\documentclass{aa}
\usepackage{graphicx}
\usepackage{amssymb}
\usepackage{natbib}
\bibpunct{(}{)}{;}{a}{}{,}
\usepackage{lscape}
\usepackage{supertabular}
\usepackage{rotating}
\usepackage{setspace}
\usepackage{epsf}
\usepackage{txfonts}
\def\vsini{$V\!\sin i$}
\def\teff{$T_{\rm eff}$}

\def\logg{$\log~g$}

\def\omc{$\Omega/\Omega_{\rm{c}}$}

\def\top{$T_{\rm eff}^{\rm o}$}
\def\gop{$\log g_{\rm o}$}

\def\vsiniap{$V\!\sin i_{\rm~\!app.}$~}
\def\vsinit{$V\!\sin i^{\rm true}$}
\def\kms{km~s$^{-1}$}
\def\tap{$T_{\rm eff}^{\rm app.}$~}
\def\gap{$\log g_{\rm app.}$~}

\def\rv{RV}
\def\ttms{$\frac{\tau}{\tau_{MS}}$~}
\begin{document}
%
\title{Effects of metallicity, star formation conditions and evolution in B and Be
stars.}
\subtitle{I: Large Magellanic Cloud, field of NGC\,2004.}

\titlerunning{B and Be stars in the LMC}
\author{
C. Martayan \inst{1}
\and Y. Fr\'emat \inst{2}
\and A.-M. Hubert \inst{1}
\and M. Floquet \inst{1}
\and J. Zorec \inst{3}
\and C. Neiner \inst{1,4}
}
\offprints {C. Martayan}
\mail{christophe.martayan@obspm.fr}
\institute{
GEPI, UMR\,8111 du CNRS, Observatoire de Paris-Meudon, 92195 Meudon Cedex, France
\and Royal Observatory of Belgium, 3 avenue circulaire, 1180 Brussels, Belgium
\and Institut d'Astrophysique de Paris (IAP), 98bis boulevard Arago, 75014 Paris,
France
\and Instituut voor Sterrenkunde, KU Leuven, Celestijnenlaan 200B, 3001 Belgium}
\date{Received /Accepted}
\abstract{
Spectroscopic observations of hot stars belonging to the young cluster LMC-NGC\,2004 and
its surrounding region were carried out with the VLT-GIRAFFE facilities in MEDUSA mode.
We determine fundamental parameters (\teff, \logg, \vsini, and radial velocity),
for all B and Be stars in the sample thanks to a code developed in our group. The effect
of fast rotation (stellar flattening and gravitational darkening) are taken into account
in this study. We also determine the age of observed clusters. We compare the mean
\vsini~obtained for field and cluster B and Be stars in the Large Magellanic Cloud (LMC)
with the ones in the Milky Way (MW). We find, in particular, that Be stars rotate
faster in the LMC than in the MW, in the field as well as in clusters. We discuss the
relations between \vsini, metallicity, star formation conditions and stellar evolution
by comparing the LMC with the MW. We conclude that Be stars begin their Main Sequence
life with an initial rotational velocity higher than the one of B stars. It is
probable that only part of the B stars, with a sufficient initial rotational velocity,
can become Be stars. This result may explain the differences in the proportion of Be
stars in clusters with similar ages.

}
\maketitle

\section{Introduction}

The study of physical properties of B stars with respect to the relative frequency of Be
stars in young open clusters and their surrounding field can provide important insights
into the origin of the Be phenomenon. Indeed, our knowledge of mass-loss conditions,
which lead to the formation of an anisotropic envelope around a fraction of B stars, is
still very poor.  Several physical processes could be involved, such as rapid rotation,
non-radial pulsations, magnetic fields, evolutionary effects, binarity...
Although the Be phenomenon has frequently been considered as a rapid rotation-related
phenomenon in OB stars, it is not yet established whether only a fraction or all of the
rapidly rotating early-type stars evolve into Be stars.

Several recent studies tackled the major question of the
influence of metallicity and evolution on rapid rotators.  The Be
phenomenon could be favoured in stars of low metallicity (Maeder
et al. 1999). Preliminary results by Royer et al. (2004) could
confirm this metallicity effect, although they use limited
stellar samples.  The appearance of the Be phenomenon could also
depend on stellar ages (Fabregat \& Torrej\'on 2000). Accounting
for the effects of fast rotation and of gravitational darkening,
Zorec et al. (2005) concluded  that Be stars spread over the
whole Main Sequence (MS) evolutionary phase. However, they find
that in massive stars the Be phenomenon tends to be present at
smaller \ttms age ratios than in less massive stars. Moreover,
Keller (2004) found that young clusters host more rapid rotators
than their surrounding field. This assessment is valid for the
LMC, as well as in the Galaxy. A similar trend was found by Gies
\& Huang (2004) from a spectroscopic survey of young galactic
clusters.

Up to now, spectroscopic surveys were only obtained with a
resolution power R$<$5000 and/or a low signal to noise ratio
(S/N). The new instrumentation FLAMES-GIRAFFE installed at the
VLT-UT2 at ESO is particularly well suited to obtain high quality
spectra of large samples needed for the study of stellar
populations. Thus, we have undertaken the determination of
fundamental parameters for a large sample of B and Be stars in
regions of different metallicity: (i) to check whether the low
metallicity favours the formation of rapid rotators and in
particular of Be stars; and (ii) to investigate the evolutionary
status of Be stars. In a first paper (Martayan et al. 2005,
hereafter Paper I), we reported the identification of numerous
B-type stars, the discovery of new Be stars and spectroscopic
binaries in the young cluster LMC-NGC\,2004 and its surrounding
field  with the help of medium resolution spectra obtained with
the FLAMES instrumentation. The present paper deals with
fundamental parameters and evolutionary status of a very large
fraction of those objects, taking into account rotational effects
(stellar flattening, gravitational darkening) when appropriate.

\section{Observations}

This work makes use of spectra obtained with the multifibre
spectrograph VLT-FLAMES in Medusa mode (132 fibres) at medium
resolution (R=6400) in setup LR02 (396.4 - 456.7 nm).
Observations  (ESO runs 72.D-0245B and 73.D-0133A) were carried
out in the young cluster LMC-NGC\,2004 and in its surrounding
field, as part of the Guaranteed Time Observation programmes  of
the Paris Observatory (P.I.: F. Hammer).  The observed field
(25\arcmin~ in diameter) is centered at $\alpha$(2000) = 05h 29m
00s and $\delta$(2000) = -67$^{\circ}$ 14\arcmin~ 00\arcsec.
Besides the young cluster NGC\,2004, this field contains several
high-density groups of stars (KMHK\,943, 971, 963, 991, 988 and
BSDL\,2001).  Spectra were obtained on November 24, 2003 and
April 12,  2004; at these dates, the heliocentric velocities are
smaller than 1.5 \kms. The strategy and conditions of
observations, as well as the spectra reduction, are described in
Paper I. A significant sample of the  B stars population (168
objects), 6 O and 2 A stars were observed during the two
observing runs. Since the V magnitude of the selected targets
ranges from 13.7 to 17.8 mag, we chose a 2-hour integration time.
This corresponds to an average S/N $\simeq$ 120, with  individual
values ranging from $\sim$20 to $\sim$150 for the fainter and
brighter stars, respectively.

\begin{figure*}[ht]
\centering
\resizebox{\hsize}{!}{\includegraphics[angle=-90]{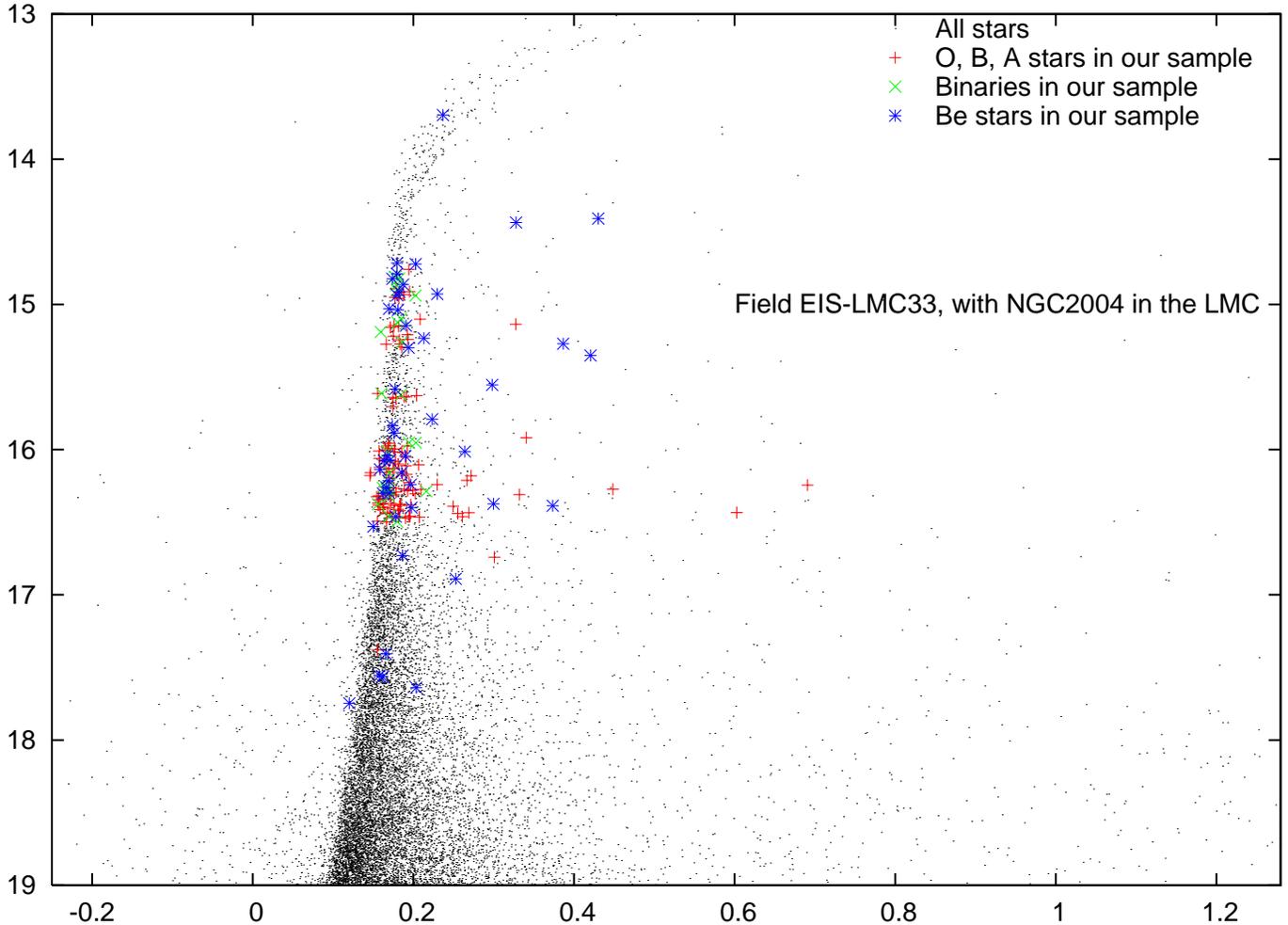}}
\caption{Instrumental V versus instrumental (B-V) colour diagram from our
photometry in the EIS LMC\,33 field. The '.' symbols correspond to all stars in
this field. '*' show the Be stars, '+' the O-B-A stars and 'x' the binaries
in the sample. Several stars in this sample have a strong reddening, they are
located mainly in clusters, and in the \ion{H}{ii} region LHA\,120-N51A.}
\label{figcoul}
\end{figure*}

\begin{figure*}[ht]
\centering
\resizebox{\hsize}{!}{\includegraphics[angle=-90]{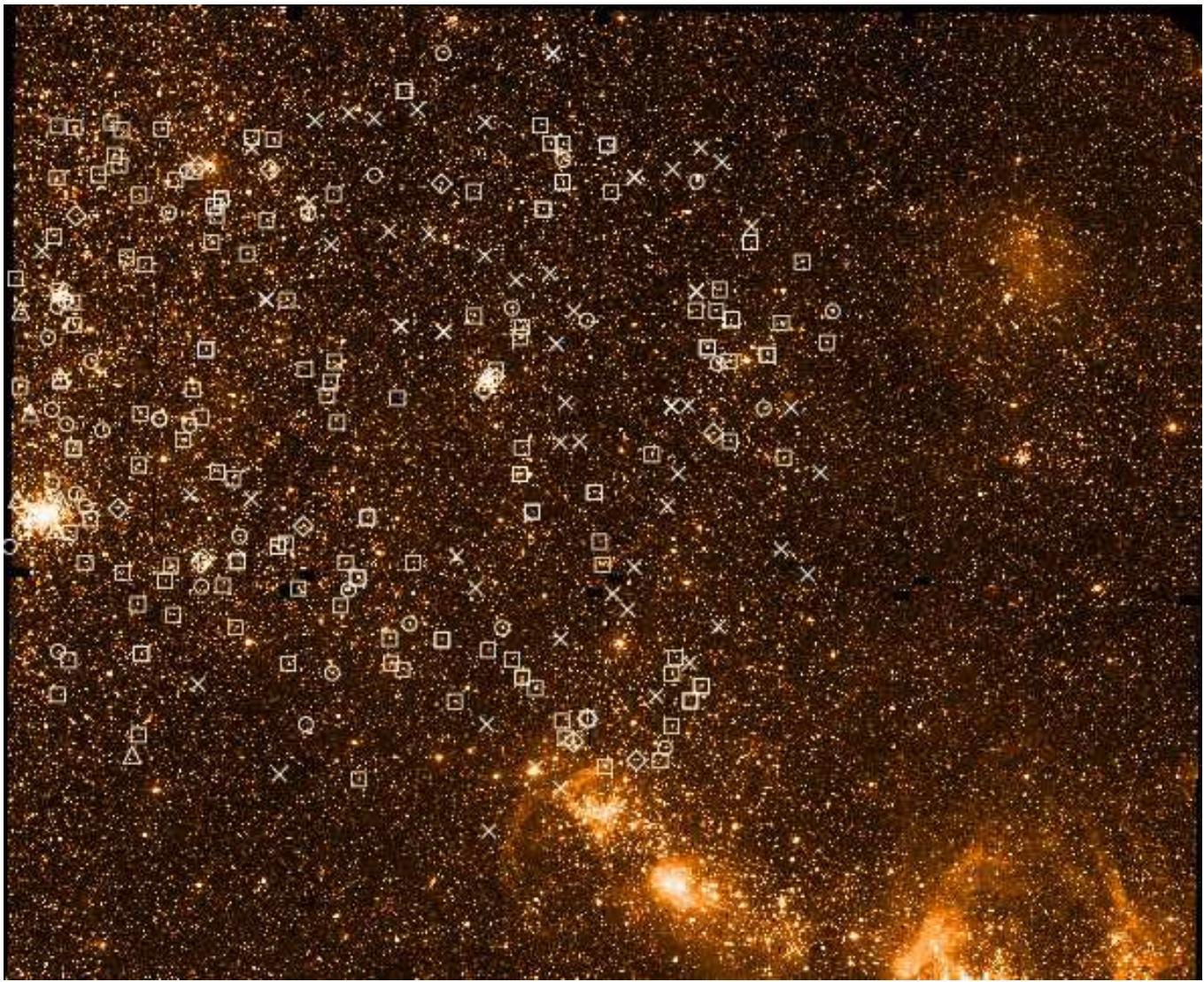}}
\caption{The LMC\,33 field from the EIS pre-FLAMES survey. Circles represent Be
stars in the sample, squares O-B-A stars, and  crosses sky fibres. Triangles
and diamonds indicate Be and B stars with a strong reddening, respectively. In
the southwest lies the \ion{H}{ii} region LHA\,120-N51A.}
\label{figure0}
\end{figure*}

The colour diagram V versus B-V  (Fig.~\ref{figcoul}), derived
from our instrumental photometry, shows the B and Be stars in our
sample  compared to all the stars in the EIS-LMC\,33 field.
Several stars present a strong reddening and are mainly located
either in the clusters NGC\,2004, KMHK\,943,  KMHK\,971,
`unknown2' and in the galactic open cluster HS\,66325, either at
the periphery of the \ion{H}{ii} region LHA\,120-N51A, or in the
field, but without explicit link between all these regions.  The
locations of the observed O, B and A-type stars are shown in the
LMC\,33 field from the EIS pre-FLAMES survey (Fig.~\ref{figure0}).

Finally, among the 124812 stars which we have listed in the
EIS LMC\,33 field, our pre-selection with photometric criteria gives 
1806 B-type stars. And we have observed 177 B-type stars among the 
1235 B-type stars which are observable in the VLT-FLAMES/GIRAFFE field.
The ratio of observed to observable B-type stars represents 14.3\%.
Consequently, this sample is statistically significant.


\section{Fundamental parameters determination}
\label{FPD}

\begin{table}[ht]
\caption{Atoms and ions treated in the computations assuming NLTE. The number
of levels taken into account for each ion is given.}
\label{tab:desc}
\centering
\begin{tabular}{@{}lll@{}}
\hline \hline\noalign{\smallskip}
Atom & Ion & Number of levels\\
\hline \noalign{\smallskip}
Hydrogen & H {\sc i} & 8 levels + 1 superlevel\\
         & H {\sc ii} & 1 level\\
Helium   & He {\sc i} & 24 levels\\
         & He {\sc ii} & 20 levels\\
         & He {\sc iii} & 1 level\\
Carbon   & C {\sc ii} & 53 levels all individual levels\\
         & C {\sc iii} & 12 levels\\
         & C {\sc iv}  & 9 levels + 4 superlevels\\
         & C {\sc v} & 1 level\\
Nitrogen & N {\sc i} & 13 levels\\
         & N {\sc ii} & 35 levels + 14 superlevels\\
         & N {\sc iii} & 11 levels\\
         & N {\sc iv} & 1 level\\
Oxygen   & O {\sc i} & 14 levels + 8 superlevels\\
         & O {\sc ii} & 36 levels + 12 superlevels\\
         & O {\sc iii} & 9 levels\\
         & O {\sc iv} & 1 level\\
Magnesium & Mg {\sc ii} & 21 levels + 4 superlevels\\
          & Mg {\sc iii} & 1 level\\
\hline
\end{tabular}
\end{table}

One important step in the analysis of the data collected with
FLAMES is the determination of the stellar fundamental
parameters. In order to derive the effective temperature (\teff),
surface gravity (\logg), projected rotational velocity (\vsini) and
radial velocity (RV) in an homogeneous and coherent way for the
whole stellar sample, we use the GIRFIT least squares procedure,
which is able to handle large datasets and was previously
developed and described by Fr\'emat et al. (2005a). GIRFIT fits
the observations with theoretical spectra interpolated in a grid
of stellar  fluxes computed with the SYNSPEC programme and from
model atmospheres calculated with TLUSTY (Hubeny \& Lanz 1995,
see references therein) or/and  with ATLAS9 (Kurucz 1993;
Castelli et al. 1997). It accounts for the instrumental
resolution through convolution of spectra with a Gaussian
function and for Doppler broadening due to rotation. Use is made
of subroutines taken from the ROTINS computer code provided with
SYNSPEC (Hubeny \& Lanz 1995).

    It is worth noting that the spectra obtained in this way do not take into account
the second order effects of fast rotation (stellar flattening and
gravitational darkening), which are expected to be strong in Be
stars.  To introduce these effects in our discussion, stellar
parameters are corrected afterwards by adopting a grid of
synthetic stellar spectra computed by Fr\'emat et al. (2005b)
with the FASTROT computer code assuming a solid-body-type
rotation. In the following sections, the terms 'apparent' and
'parent non-rotating counterpart' (pnrc) are used as defined by
Fr\'emat et al. (2005b).

We introduce in the following sections the grid of model
atmospheres (Sect.~\ref{subsec:grid}) we use, the fitting
criteria we adopt in the GIRFIT procedure
(Sect.~\ref{subsec:girfit}) and the corrections for fast rotation
we apply on the Be stars' fundamental parameters
(Sect.~\ref{subsec:effects}). The calibrations that allow us to
estimate the spectral type of each non-Be target from the
equivalent width of the hydrogen and helium lines are detailed in
Sect.~\ref{subsec:mktyp}.

\subsection{Grid of model atmospheres}
\label{subsec:grid}

The models we use to build the GIRFIT input grid of stellar fluxes are computed in two
consecutive steps. To account in the most effective way for  line-blanketing, the
temperature structure of the atmospheres is computed using the ATLAS9 computer code
(Kurucz 1993; Castelli et al. 1997).  Non-LTE level populations are then estimated for
each of the atoms we consider using TLUSTY (Hubeny \& Lanz 1995) and keeping fixed the
temperature  and density distributions obtained with ATLAS9.

Table \ref{tab:desc} lists the ions that are introduced in the computations. Except for
\ion{C}{ii}, the atomic models we use in this work were downloaded from TLUSTY's
homepage (http://tlusty.gsfc.nasa.gov) maintained by I. Hubeny and T. Lanz. \ion{C}{ii}
is treated with the MODION IDL package developed by Varosi et al. (1995) and the atomic
data (oscillator strengths, energy levels, and photoionization cross-sections) from the
TOPBASE database (Cunto et al. 1993). It reproduces the results obtained by Sigut
(1996).

In this way, and for each spectral region studied in the present
work, the specific intensity grids are computed for effective
temperatures and surface gravities ranging from 15000 K to 27000
K and from 2.5 to 5.0 dex, respectively. For \teff~$<$~15000 K
and \teff~$>$~27000 K we use LTE calculations and the OSTAR 2002
NLTE model atmospheres grid (Lanz \& Hubeny 2003),
respectively.

The metallicities of the model atmospheres are chosen to be as close as possible to the
NGC\,2004 averaged value, $[m/H]~=~-0.45$ (where $[m/H] = \log(m/H)_{\rm LMC} -
\log(m/H)_\odot$), estimated from the results by Korn et al. (2002, Table 3). The
Kurucz and OSTAR 2002 models we use are therefore those calculated for [m/H]$=-0.5$.
Finally, the complete input flux grid is built assuming the averaged element abundances
derived by Korn et al. (2002) for C,  N, O, Mg, Si, and Fe. The other elements, except
hydrogen and helium, are assumed to be underabundant by $-0.45$ dex relative to the
Sun.

\begin{figure}[ht]
\center
\resizebox{\hsize}{!}{\includegraphics[clip]{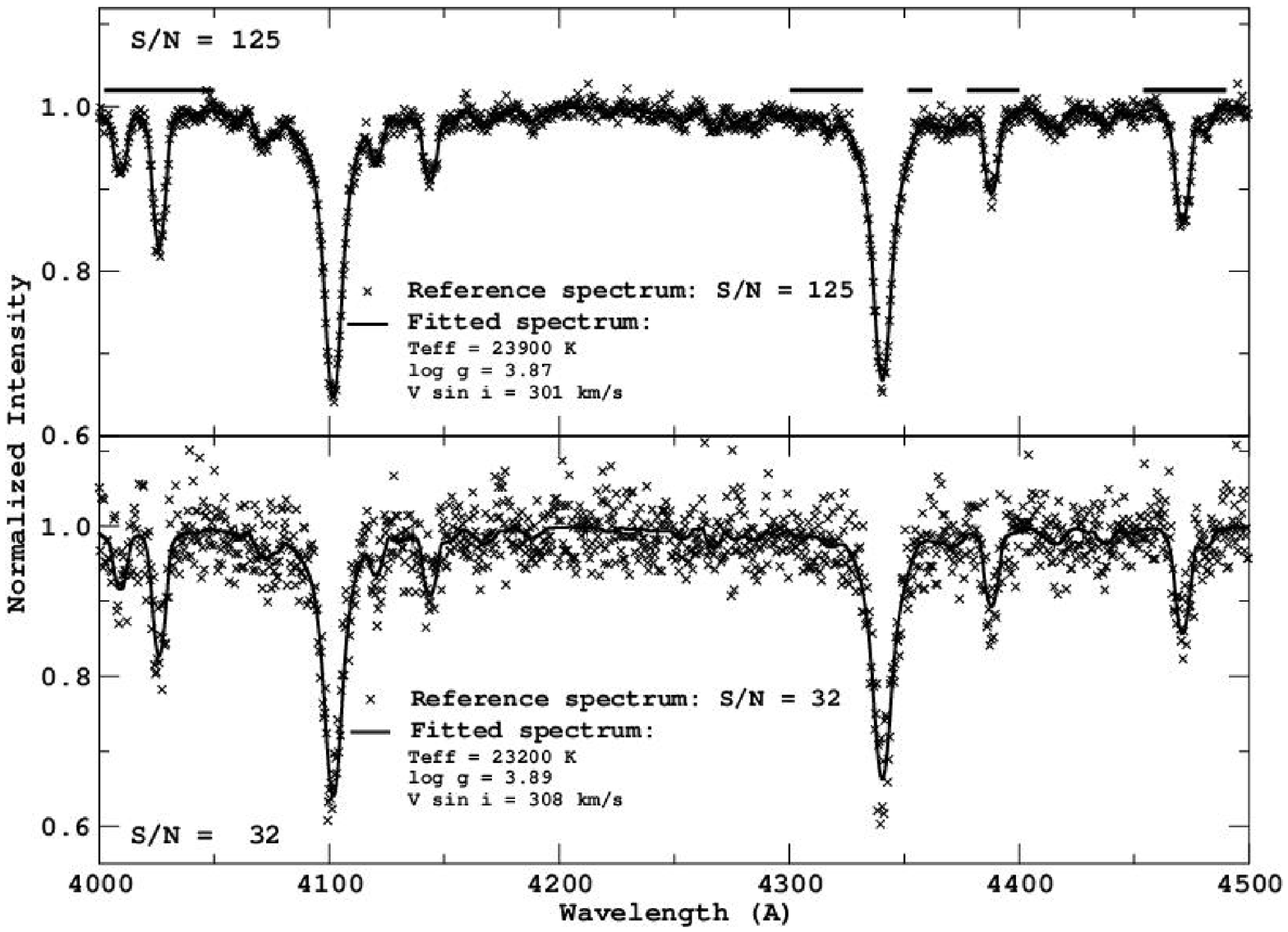}}
\caption{Reference spectra (crosses) computed for \teff~= 25000 K, \logg~= 3.87
and \vsini~= 300 \kms, for different S/N, are fitted using the {\sc girfit}
procedure. The obtained fundamental parameters are noted on the figure. }
\label{fig:sn}
\end{figure}

\subsection{The GIRFIT procedure: fitting criteria and continuum level}
\label{subsec:girfit}

The procedure we adopt to derive the stellar fundamental
parameters mainly focuses on the spectral domain ranging from
4000 to 4500 \AA, which gathers two hydrogen lines (H$\gamma$ and
H$\delta$),8 strong helium lines (He\,{\sc i}
$\lambda$ 4009, 4026, 4121, 4144, 4169, 4388, 4471 and
\ion{He}\,{\sc ii} $\lambda$ 4200) and several weak
lines of silicon and carbon. The $\chi^{2}$ parameter is computed
on different spectral zones generally centred on these
temperature-  and gravity-sensitive diagnostic features.
However, due to the moderate spectral resolution and
frequent high apparent rotational velocities of targets, other
criteria such as those based on silicon lines cannot be used. 
Furthermore It is worth noting that even the \ion{He}{\sc i}/\ion{H}{i}
line ratios used to estimate \teff~and
\logg~values are less accurate for early B-type stars (B1-B0) 
than for later types, the simultaneous fit of several hydrogen 
and helium line-profiles enables us to obtain the sought stellar 
parameters quite easily, within the error boxes given in Table \ref{tab:sn} 
(see also Fig.~1 in Fr\'emat et al. 2005b). In the most dubious cases, 
the overall agreement between observed and synthetic spectra was
checked over the complete spectral range.

In Be stars, which often display circumstellar emission/absorption in their
spectra, the zones are further defined to exclude any part of the spectral lines that
could be deformed by line emission or shell absorption (e.g. hydrogen line cores). Note
that, as the parameters derived for Be stars at this stage of the procedure do not take
into account the effects of fast rotation (see Sect.~\ref{subsec:effects}), they will be
further called apparent fundamental parameters.

During the spectra fitting procedure, 4 free parameters are considered: the effective
temperature, the surface gravity, the projected rotational velocity and the radial
velocity. To reduce as much as possible the impact of noise on the location of the
stellar continuum, we also include a fifth parameter standing for the
wavelength-independent ratio (i.e. a scaling factor) between the  mean ``flux'' level
of the normalized observed and theoretical spectra. Assuming a Poisson noise
distribution to compute a reference spectrum, Fig.~\ref{fig:sn} shows how the derived
fundamental parameters can be affected by a decrease of S/N, while Table~\ref{tab:sn}
lists the averaged absolute errors on the fundamental parameters expected for different
values of S/N.

\begin{figure}[ht]
\centering
\resizebox{8.5cm}{!}{\includegraphics[angle=-90]{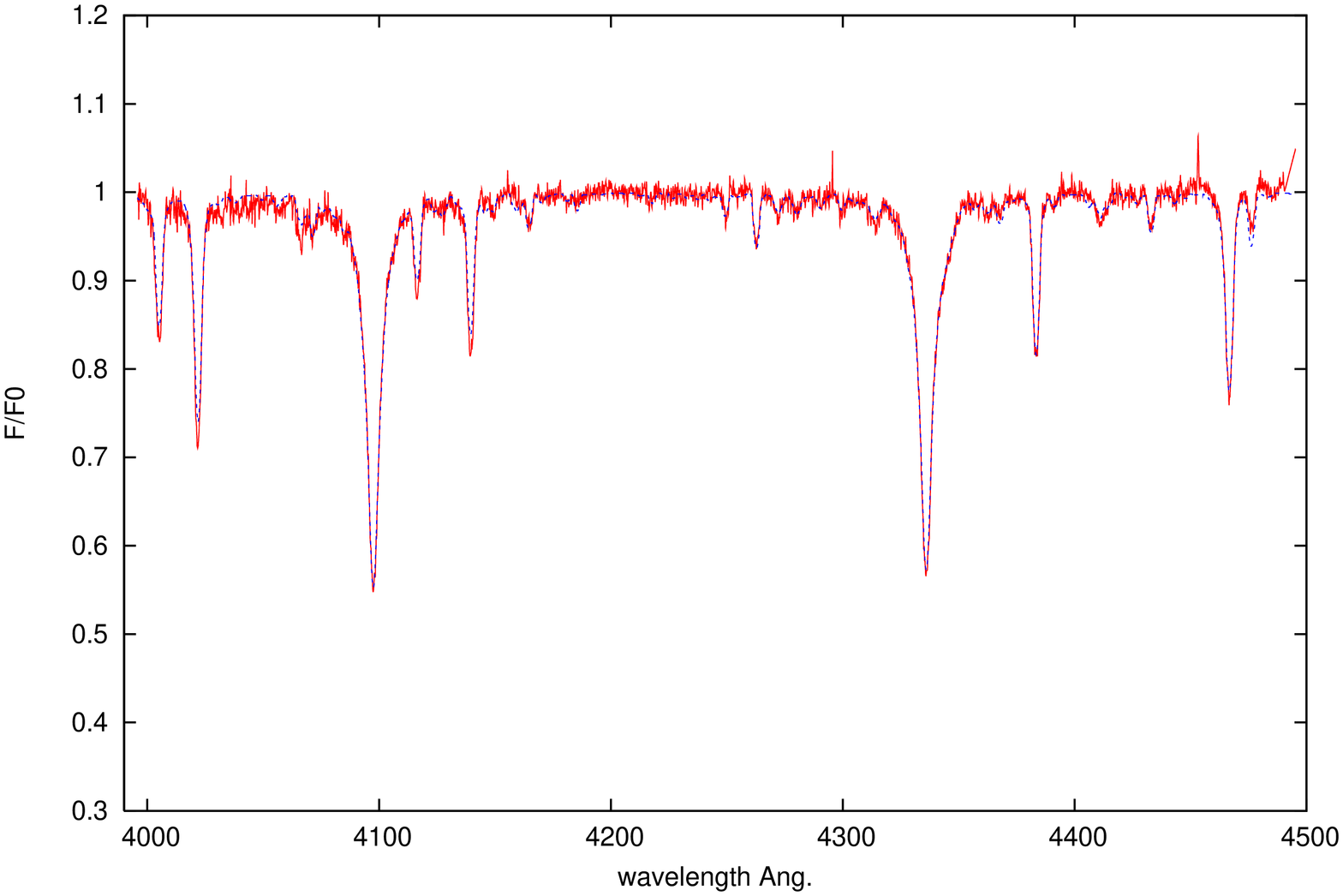}}\\
\resizebox{8.5cm}{!}{\includegraphics[angle=-90]{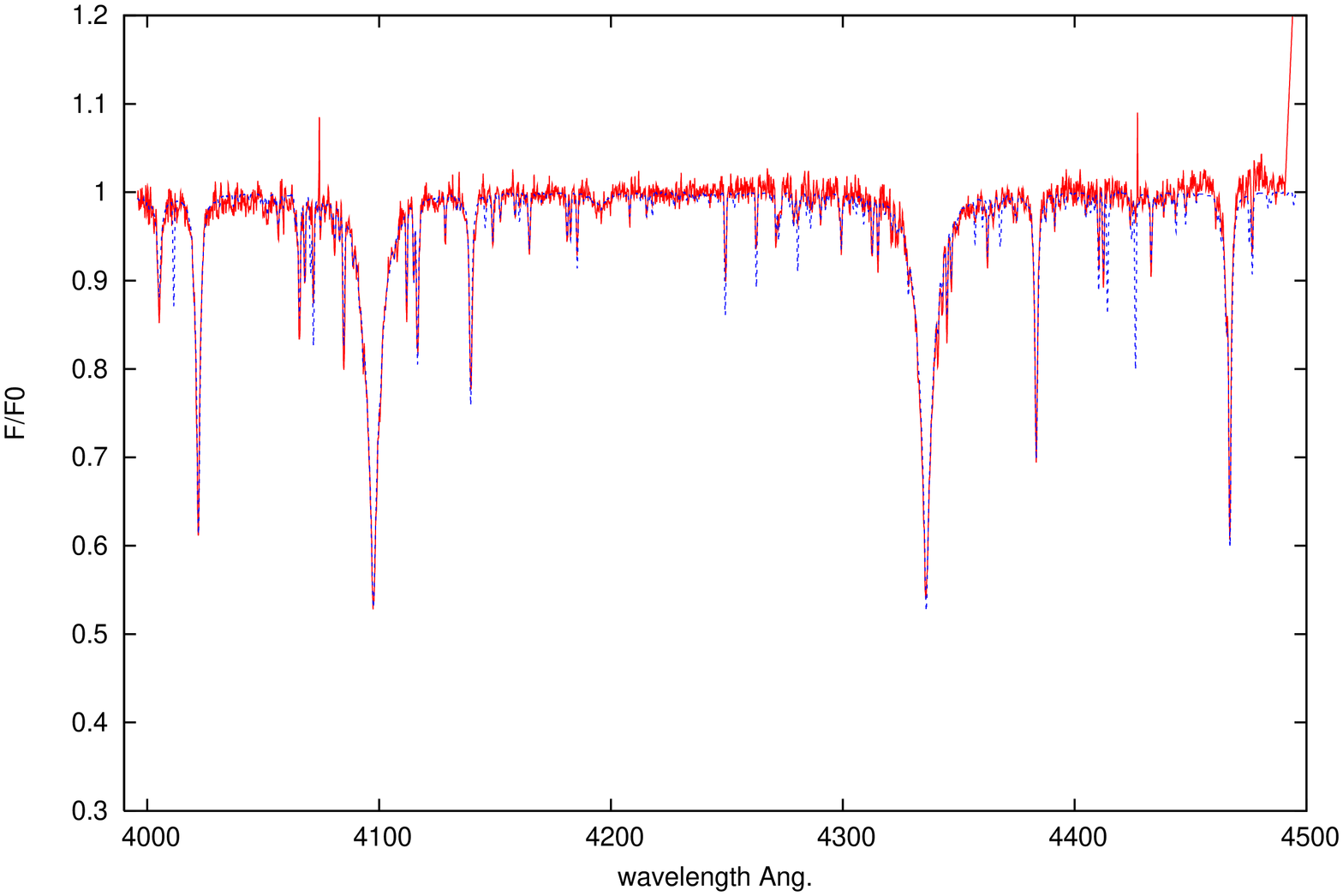}}\\
\resizebox{8.5cm}{!}{\includegraphics[angle=-90]{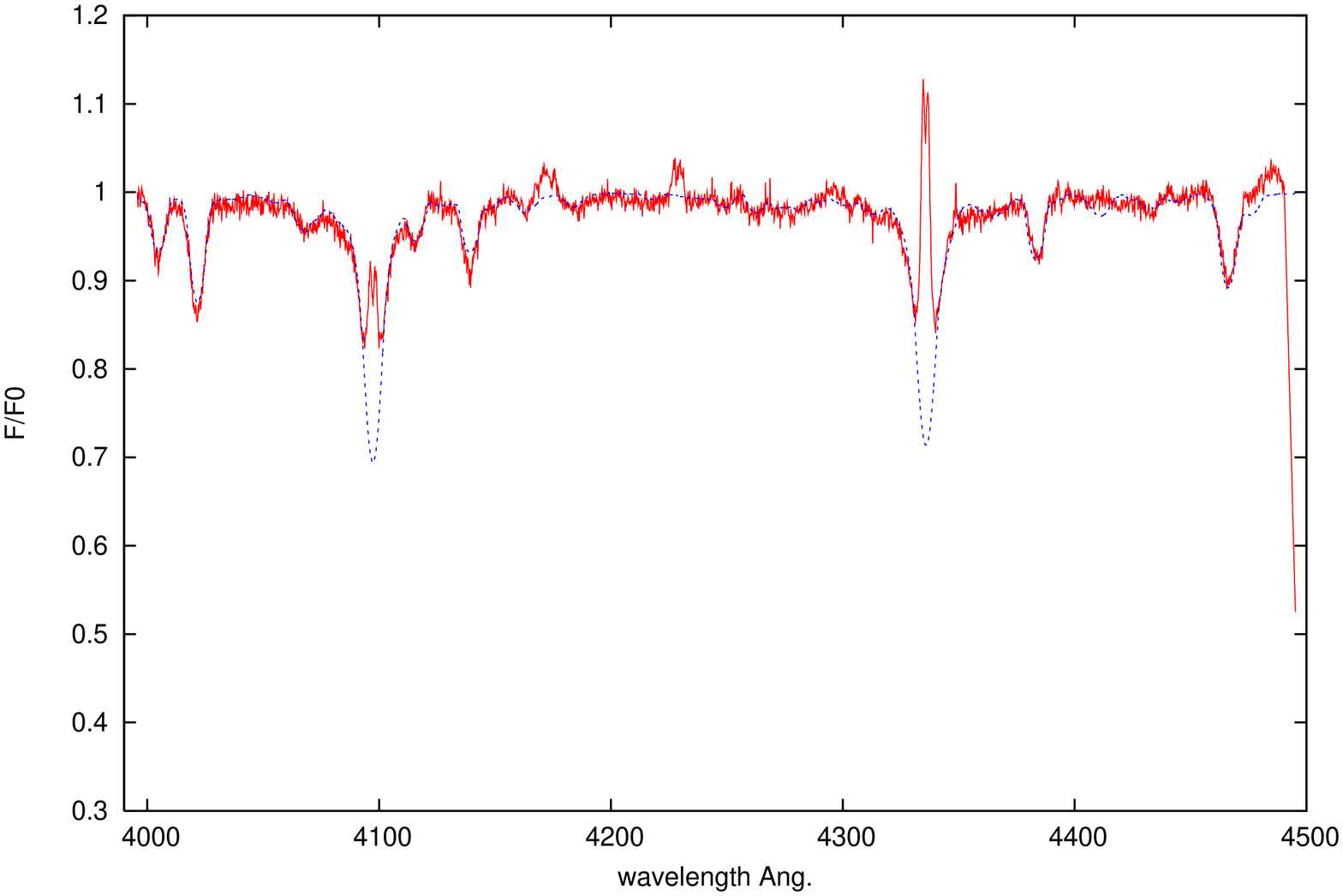}}
\caption{Example of fits (dotted lines) of observed spectra (solid lines).
Upper panel:  The MHF119707 B star.  The fit gives \teff= 21500 K, \logg=
3.6 dex, \vsini= 134 \kms, and \rv= 308 \kms.
Medium panel: The MHF67663 O-B star. The fit gives \teff= 29500 K, \logg=
4.3 dex, \vsini= 9 \kms, and \rv= 306 \kms.
Lower panel: The MHFBe136844 Be star. The fit gives \teff= 23000 K, \logg=
3.5 dex, \vsini= 348 \kms, and \rv= 315 \kms.}
\label{fits}
\end{figure}

For each star of the sample, we repeat the GIRFIT procedure
several times with different fitted zones and initial parameters
values in order to scan the complete space of solutions. The
solutions we finally select are those with the lowest $\chi^{2}$
recomputed over the same wavelength range: the whole spectrum for
O, B, and A-type stars without emission, and only the blue part
of the spectrum (4000--4250 \AA) for Be stars in order to avoid,
as much as possible, the influence of line emission in the
hydrogen lines. After a final visual check of the adjusted spectra,
the \rv~determinations obtained at the end of the GIRFIT procedure
are compared to the values directly measured on the observations.
If this ultimate verification is successful, the process
stops. Otherwise, the fitting spectral zones are modified and the
procedure is restarted. Examples of fitted spectra, for a B, an
O-B and a Be star are given in Fig.~\ref{fits}.

Following Bouret et al. (2003), who used the same atomic data, there is no difference
between fits obtained with TLUSTY-SYNSPEC (Hubeny \& Lanz 1995) and with the CMFGEN code
(Hillier \& Miller 1998), which takes into account not only NLTE effects and
line-blanketing but also a wind model with mass loss. This effect is present in O stars,
but is not critical for B-type  stars. Bouret et al. (2003) presented several fits for
their sample of O and B stars and obtained similar apparent fundamental parameters with
these two codes for early B-type stars. This study validates our choice of code for
determining the fundamental parameters of B stars.

\subsection{Spectral classification determination}
\label{subsec:mktyp}

The spectral type and luminosity class of the B-type stars we observed are
determined in two different ways. First, we use an iterative method we
developed, which has the advantage of being fast and easy to use. We estimate
the spectral type from the equivalent width of the H$\gamma$ line by assuming,
in a first step, that our sample is only composed of dwarf stars. The spectral
type is combined with the equivalent width  of the \ion{He}{i} 4471 line to
derive, in a second step, the luminosity class, which is then used to rederive
the spectral type. Several iterations are required to obtain a combination of
spectral type/luminosity class  fully coherent with the equivalent widths of the
selected lines (H$\gamma$ and \ion{He}{i} 4471). The equivalent width
calibrations we adopt in this procedure are those proposed by Azzopardi (1987)
and Jaschek \& Jaschek (1995) for H$\gamma$ and by Didelon (1982) for
the \ion{He}{i} 4471 line.

\begin{table}[ht]
\caption{Averaged absolute errors on the fundamental parameters introduced by
different S/N. For \vsini~$\le$ 50 \kms, the error is estimated to $\pm$20
\kms~(due to the intermediate resolution) and the minimum error in \vsini~is
$\pm$10 \kms~for the other cases.}
\centering
\begin{tabular}{rrrr}
\hline\hline\noalign{\smallskip}
S/N     &   $\Delta$($T_{\rm eff}$)  & $\Delta$($\log~g$)  & $\Delta$($V\!\sin i$) \\
        &   (\%)              & (\%)              & (\%)              \\
\hline\noalign{\smallskip}
 30 & 20 & 10 & 30 \\
 40 & 15 & 10 & 20 \\
 50 & 12 & 10 & 16 \\
 60 & 10 &  9 & 16 \\
 70 &  8 &  8 & 10 \\
 80 &  6 &  6 & 10 \\
 90 &  6 &  6 &  7 \\
100 &  5 &  5 &  5 \\
120 &  5 &  5 &  5 \\
$>$140 & $<$5 & $<$5 & $<$5 \\
\hline
\end{tabular}
\label{tab:sn}
\end{table}

The second method transcribes the set of fundamental parameters we derived by fitting
the observed spectra into spectral type and luminosity class, with the help of effective
temperature and surface gravity calibrations given by Gray \& Corbally (1994) and Zorec
(1986) for B stars, and by Bouret et al. (2003) for hotter stars.

The differences in the results provided by these two methods are, on average, half a
spectral subtype and half a luminosity class for stars between B0 and B5, affecting both
the equivalent width measurements and the derived stellar fundamental parameters.
However, the first method fails to give a reliable spectral classification for the few
hotter (late O) and cooler (B5-A0) stars in the sample. Moreover, for Be stars, the
spectral classification determination is only performed using the derived apparent
fundamental parameters (second method), since the emission contamination, often present
in H$\gamma$ and in several cases in the \ion{He}{i} 4471 line, makes the first method
particularly inappropriate for early Be stars.

\subsection{Effects of fast rotation}
\label{subsec:effects}

As mentioned in the introduction, Be stars are fast rotators with
angular velocities probably around 90\% of their breakup velocity
(Fr\'emat et al. 2005b). It is further expected that
solid-body-type fast rotation flattens the star, which causes a
gravitational darkening of the stellar disk due to the variation
of the temperature and density distribution from pole to equator.
For Be stars, we therefore have to account for these effects on
the stellar spectra and, consequently, on the determination of the
fundamental parameters. In the present paper, these effects are
introduced as corrections directly applied to the apparent
fundamental parameters we derived. These corrections are computed
by systematically comparing  a grid of spectra taking into
account the effects of fast rotation obtained with the FASTROT
code for different values of pnrc (i.e. parent non-rotating
counterpart) stellar parameters (\top, \gop, \vsinit) and of
angular velocity (\omc, where $\Omega$$_{c}$ is the break-up
angular velocity) to a grid of spectra computed using usual
plane-parallel model atmospheres. Adopting the same spectroscopic
criteria than those described  in Section~\ref{subsec:girfit}
(i.e. hydrogen  and helium lines), we obtain different sets of
pnrc and apparent stellar parameters (\teff, \logg, \vsini). The
corrections we apply to the apparent stellar parameters of the Be
stars in the sample (Section~\ref{subsec:girfit}) are
interpolated in this grid using an iterative procedure.
Generally, only a few iterations are needed (to reach differences
smaller than 500 K for \teff~and 0.05 dex for \logg) to obtain
the final pnrc parameters for a given \omc. The radius,
mass, and luminosity of the non-rotating stellar counterparts are
estimated by (\top; \gop)--interpolation in the theoretical
evolutionary tracks (Charbonnel et al. 1993).

\section{Results} 

In this section we present the results on stellar parameters and spectral classification
determination we obtain as described in Sect.~\ref{FPD} for non-emission line O-B-A
stars, for Be stars, and for some spectroscopic binary systems. The $\alpha$(2000) and
$\delta$(2000) coordinates, the instrumental V magnitude and the instrumental (B-V)
colour index for individual stars are extracted from EIS pre-FLAMES (LMC\,33) survey
images, as reported in Paper I. The S/N we measure in the spectra may differ for objects
with the same magnitude depending either on the position of the fibres within the
GIRAFFE field, either on transmission rate differences, or on the presence of clouds
partly obscuring the observed field (the field of GIRAFFE has a 25\arcmin~ diameter on
the sky). All these informations are given in Tables~\ref{table3}, \ref{tableBe}, and
\ref{tablebin} for the different groups of stars mentioned above, respectively.

\begin{table*}[tbph]
\caption{Parameters for O-B-A stars that need no corrections for fast rotation.
KWBBe names from Keller et al. (1999) or our MHF catalogue numbers are given in
col. 1. Coordinates ($\alpha$(2000), $\delta$(2000)) are given in col. 2 and 3.
The instrumental V magnitude and instrumental (B-V) colour index are given in
col. 4 and 5. The S/N ratio is given in col. 6. In col. 7, 8, 9 and 10,
\teff~is given in K, \logg~in dex, \vsini~and \rv~in \kms. `CFP' is the
spectral type and luminosity classification determined from fundamental
parameters (method 2), whereas 'CEW' is the spectral type and luminosity
classification determined from EW diagrams (method 1). In the last column some
complementary indications on the spectrum are given: '\ion{He}{ii}' when the
line at 4200 {\AA} is observed, 'bin' in case of suspected binary, 'not Be' for
a Be star suspected by Keller et al. (1999) but not seen as a Be star in our
study. The last column also gives the localization in clusters:
cl0 for NGC\,2004 (05h 30m 42s -67$^{\circ}$ 17$\arcmin$ 11$\arcsec$),
cl1 for KMHK\,988 (05h 30m 36.5s -67$^{\circ}$ 11$\arcmin$ 09$\arcsec$),
cl2 for KMHK\,971 (05h 29m 55s -67$^{\circ}$ 18$\arcmin$ 37$\arcsec$),
cl3 for KMHK\,930 (05h 28m 13s -67$^{\circ}$ 07$\arcmin$ 21$\arcsec$),
cl4 for KMHK\,943 (05h 28m 35s -67$^{\circ}$ 13$\arcmin$ 29$\arcsec$),
cl5 for the `unknown' cluster or association 1 (05h 30m 25s -67$^{\circ}$
13$\arcmin$ 20$\arcsec$),
cl6 for the `unknown' cluster or association 2 (05h 29m 54s -67$^{\circ}$
07$\arcmin$ 37$\arcsec$),
cl7 for the `unknown' cluster or association 3 (05h 27m 21s -67$^{\circ}$
12$\arcmin$ 52$\arcsec$),
cl8 for the association BSDL\,1930 (05h 29m 26s -67$^{\circ}$ 08$\arcmin$
54$\arcsec$),
and cl9 for the galactic open cluster HS\,66325 (05h 29m 36s -67$^{\circ}$
07$\arcmin$ 41$\arcsec$).
}
\centering
\small{
\begin{tabular}{@{\ }l@{\ \ \ }l@{\ \ \ }l@{\ \ \ }l@{\ \ \ }l@{\ \ \ }l@{\ \ \ }l@{\ \ \ }l@{\ \ \ }l@{\ \ \ }l@{\ \ \ }l@{\ \ \ }l@{\ \ \ }l@{\ }}
\hline
\hline
Star & $\alpha$ & $\delta$ & V & B-V & S/N &  \teff & \logg & \vsini & \rv & CFP & CEW & comm. \\
\hline
MHF52224 & 5 29 10.474 & -67 24 33.16 & 15.22 & 0.18 & 95 &  22000 $\pm$1300 & 3.5 $\pm$0.2  & 11 $\pm$20  & 311 $\pm$10  & B2III-IV & B1.5III & \\
MHF54275 & 5 27 43.510 & -67 23 57.52 & 16.46 & 0.19 & 90 &  23500 $\pm$1400  & 4.1 $\pm$0.2  & 37 $\pm$20  & 309 $\pm$10  & B1V & B2IV & \\
MHF54565 & 5 27 59.431 & -67 24 08.32 & 14.76 & 0.19 & 120 & 24000 $\pm$1200  & 3.6 $\pm$0.2  & 172 $\pm$10  & 318 $\pm$10 & B1IV & B0V & \\
MHF54686 & 5 27 50.434 & -67 23 57.74 & 16.21 & 0.27 & 80 &  20000 $\pm$1200  & 4.0 $\pm$0.2  & 210 $\pm$20  & 300 $\pm$10  & B2V & B3V & \\
MHF57079 & 5 28 08.580 & -67 23 28.10 & 16.39 & 0.25 & 50 &  24000 $\pm$2900  & 4.2 $\pm$0.4  & 320 $\pm$50  & 295 $\pm$10  & B1V & B2.5V&\\
MHF57428 & 5 28 10.900 & -67 23 21.10 & 16.28 & 0.20 & 70 &  25000 $\pm$2000  & 4.3 $\pm$0.3  & 53 $\pm$10  & 307 $\pm$10  & B1V & B1V& \\
MHF57975 & 5 30 13.775 & -67 23 21.25 & 16.42 & 0.18 & 70 &  20000 $\pm$1600  & 3.8 $\pm$0.3  & 345 $\pm$35  & 301 $\pm$10  & B2IV & B1.5V& \\
MHF59059 & 5 27 40.443 & -67 23 00.64 & 15.65 & 0.18 & 90 &  24500 $\pm$1500  & 4.3 $\pm$0.3  & 71 $\pm$10  & 317 $\pm$10  & B1V & B2V& \\
MHF60436 & 5 28 04.640 & -67 22 49.20 & 16.44 & 0.18 & 75 &  23000 $\pm$1800  & 4.1 $\pm$0.3  & 165 $\pm$17  & 305 $\pm$10  & B1.5V & B1.5V& \\
MHF62150 & 5 28 42.846 & -67 22 22.93 & 16.47 & 0.18 & 70 &  18500 $\pm$1500  & 4.0 $\pm$0.3  & 40 $\pm$20  & 304 $\pm$10  & B2V & B2V& \\
MHF62555 & 5 27 35.078 & -67 22 17.96 & 16.08 & 0.18 & 85 &  20500 $\pm$1600  & 4.0 $\pm$0.3  & 84 $\pm$10  & 304 $\pm$10  & B2V & B2V& \\
MHF63084 & 5 30 37.190 & -67 22 14.70 & 15.97 & 0.19 & 50 &  10500 $\pm$1200  & 2.9 $\pm$0.3  & 26 $\pm$20  & 292 $\pm$10  & B9.5III & B6.5IV&  \\
MHF63948 & 5 28 19.782 & -67 21 59.99 & 15.96 & 0.17 & 60 &  20500 $\pm$2100  & 4.0 $\pm$0.4  & 149 $\pm$24  & 313 $\pm$10  & B2V & B2V& \\
MHF65925 & 5 27 40.477 & -67 21 32.86 & 16.10 & 0.18 & 100 &  23000 $\pm$1200  & 4.0 $\pm$0.2  & 109 $\pm$10  & 322 $\pm$10  & B1.5V & B1V& \\
MHF66708 & 5 28 58.020 & -67 21 30.57 & 16.31 & 0.17 & 110 &  26500 $\pm$1300  & 4.3 $\pm$0.2  & 123 $\pm$10  & 325 $\pm$10  & B1V & B1.5IV& \\
MHF67663 & 5 29 01.263 & -67 21 19.99 & 15.63 & 0.20 & 120 &  30000 $\pm$1500  & 4.3 $\pm$0.2  & 9 $\pm$20  & 306 $\pm$10  & B0V & B1IV & HeII\\
MHF67792 & 5 28 26.590 & -67 21 12.50 & 16.44 & 0.17 & 80 &  22500 $\pm$1400  & 4.0 $\pm$0.2  & 57 $\pm$10  & 300 $\pm$10  & B1.5V & B1.5V& \\
MHF68153 & 5 30 33.960 & -67 21 17.30 & 16.13 & 0.18 & 60 &  22000 $\pm$2200  & 4.0 $\pm$0.4  & 186 $\pm$30  & 300 $\pm$10  & B1.5V & B1.5IV& \\
MHF68195 & 5 27 39.477 & -67 21 05.57 & 16.13 & 0.17 & 114 &  23000 $\pm$1200  & 4.0 $\pm$0.2  & 95 $\pm$10  & 308 $\pm$10  & B1.5V & B1.5III& \\
MHF68257 & 5 29 30.810 & -67 21 22.00 & 16.17 & 0.19 & 70 &  23000 $\pm$1800  & 4.0 $\pm$0.3  & 50 $\pm$10  & 312 $\pm$10  & B1.5V & B1III&\\
MHF69681 & 5 28 33.460 & -67 20 57.00 & 14.91 & 0.20 & 130 &  27000 $\pm$1300  & 3.8 $\pm$0.2  & 19 $\pm$20  & 308 $\pm$10  & B1IV & B0V & \\
MHF70976 & 5 29 01.686 & -67 20 39.36 & 15.64 & 0.19 & 110 &  32500 $\pm$1600  & 4.3 $\pm$0.2  & 30 $\pm$20  & 326 $\pm$10  & B0V & B0V & HeII\\
MHF72268 & 5 29 46.090 & -67 20 22.60 & 16.06 & 0.16 & 100 &  20000 $\pm$1000  & 3.8 $\pm$0.2  & 168 $\pm$10  & 266 $\pm$10  & B2IV & B1.5III & \\
MHF74015 & 5 30 03.900 & -67 20 01.50 & 16.39 & 0.16 & 70 &  22500 $\pm$1800  & 4.1 $\pm$0.3  & 118 $\pm$12  & 300 $\pm$10  & B1.5V & B1.5V & \\
MHF75373 & 5 30 14.180 & -67 19 43.90 & 16.49 & 0.16 & 65 &  20000 $\pm$2000  & 4.0 $\pm$0.4  & 104 $\pm$14  & 299 $\pm$10  & B2V & B2III & \\
MHF75553 & 5 29 15.980 & -67 19 45.65 & 15.16 & 0.17 & 140 &  19600 $\pm$1000  & 3.6 $\pm$0.2  & 33 $\pm$20  & 315 $\pm$10  & B2IV & B1III & \\
MHF77981 & 5 29 28.020 & -67 19 16.80 & 16.02 & 0.19 & 90 &  23000 $\pm$1400  & 4.0 $\pm$0.2  & 42 $\pm$20  & 296 $\pm$10  & B1.5V & B1III & \\
MHF78706 & 5 30 06.370 & -67 19 06.00 & 16.11 & 0.19 & 80 &  23000 $\pm$1400  & 4.0 $\pm$0.2  & 152 $\pm$15  & 300 $\pm$10  & B1V & B1V & \\
MHF81136 & 5 30 29.230 & -67 18 35.20 & 15.97 & 0.17 & 90 &  21000 $\pm$1300  & 4.0 $\pm$0.2  & 74 $\pm$10  & 300 $\pm$10  & B2V & B2V & \\
MHF81174 & 5 28 55.258 & -67 18 32.63 & 16.30 & 0.16 & 80 &  22000 $\pm$1300  & 4.3 $\pm$0.3  & 97 $\pm$10  & 317 $\pm$10  & B1.5V & B2IV & \\
MHF81322 & 5 28 00.860 & -67 18 32.84 & 15.24 & 0.19 & 85 &  29000 $\pm$1700  & 4.5 $\pm$0.3  & 183 $\pm$15  & 319 $\pm$10  & B0V & B0.5V & HeII\\
MHF81490 & 5 29 56.010 & -67 18 35.70 & 16.43 & 0.60 & 70 &  18000 $\pm$1500  & 4.1 $\pm$0.3  & 153 $\pm$15  & 313 $\pm$10  & B2.5V & B2.5V & cl2 \\
MHF81521 & 5 29 14.560 & -67 18 33.70 & 15.15 & 0.18 & 90 &  17000 $\pm$1000  & 3.5 $\pm$0.2  & 69 $\pm$10  & 293 $\pm$10  & B3III-IV & B2.5III & \\
MHF81807 & 5 29 45.660 & -67 18 32.90 & 16.16 & 0.17 & 80 &  20000 $\pm$1200  & 3.8 $\pm$0.2  & 80 $\pm$10  & 290 $\pm$10  & B2IV & B1.5V & \\
MHF82482 & 5 29 55.425 & -67 18 27.88 & 16.27 & 0.45 & 80 &  22000 $\pm$1300  & 4.0 $\pm$0.2  & 50 $\pm$10  & 318 $\pm$10  & B1.5V & B2V & cl2\\
MHF84042 & 5 29 31.731 & -67 18 02.40 & 16.38 & 0.18 & 100 &  20500 $\pm$1000  & 4.1 $\pm$0.2  & 194 $\pm$10  & 282 $\pm$10  & B2V & B2V & \\
MHF84176 & 5 28 01.745 & -67 17 54.83 & 16.46 & 0.20 & 75 &  27000 $\pm$2200  & 4.3 $\pm$0.3  & 102 $\pm$10  & 321 $\pm$10  & B1V & B1IV & \\
MHF85562 & 5 30 33.800 & -67 17 44.90 & 16.13 & 0.17 & 65 &  20000 $\pm$2000  & 4.0 $\pm$0.4  & 71 $\pm$12  & 305 $\pm$10  & B2V & B2V & cl0\\
MHF86995 & 5 29 26.755 & -67 17 33.49 & 15.92 & 0.34 & 105 &  18500 $\pm$900  & 3.7 $\pm$0.2  & 11 $\pm$20  & 297 $\pm$10  & B2.5IV & B2IV & \\
MHF87634 & 5 30 27.780 & -67 17 22.50 & 15.27 & 0.18 & 148 &  26000 $\pm$1300  & 3.9 $\pm$0.2  & 234 $\pm$12  & 310 $\pm$10  & B1V & B1III & cl0\\
MHF88527 & 5 28 21.190 & -67 17 06.20 & 16.47 & 0.19 & 100 &  25000 $\pm$1300  & 4.3 $\pm$0.2  & 16 $\pm$20  & 312 $\pm$10  & B1V & B1.5III & \\
MHF93347 & 5 29 46.957 & -67 16 14.79 & 16.39 & 0.18 & 102 &  24000 $\pm$1200  & 4.2 $\pm$0.2  & 191 $\pm$10  & 313 $\pm$10  & B1V & B1.5IV& \\
MHF94228 & 5 29 51.414 & -67 16 03.90 & 16.38 & 0.18 & 90 &  19500 $\pm$1200  & 3.9 $\pm$0.2  & 127 $\pm$10  & 337 $\pm$10  & B2V & B2III& \\
MHF95555 & 5 30 14.008 & -67 15 52.75 & 16.28 & 0.19 & 85 &  19500 $\pm$1200  & 3.7 $\pm$0.2  & 320 $\pm$30  & 299 $\pm$10  & B2IV & B1.5III& \\
MHF96072 & 5 27 09.328 & -67 15 31.37 & 16.37 & 0.16 & 70 &  23000 $\pm$1800  & 4.5 $\pm$0.4  & 191 $\pm$20  & 302 $\pm$10  & B1.5V & B2.5V& \\
MHF97219 & 5 27 47.076 & -67 15 29.13 & 15.10 & 0.21 & 170 & 36000 $\pm$1800  & 4.4: $\pm$0.2  & 31 $\pm$20  & 293 $\pm$10  & O8V & O5V & HeII+bin? \\
MHF97965 & 5 28 24.323 & -67 15 20.11 & 16.33 & 0.15 & 95 &  21000 $\pm$1300  & 4.3 $\pm$0.3  & 131 $\pm$10  & 300 $\pm$10  & B1.5V & B2.5V & \\
\hline
\end{tabular}
\label{table3}
}
\end{table*}

\addtocounter{table}{-1}
\begin{table*}[tbph]
\caption{continue}
\centering
\small{
\begin{tabular}{@{\ }l@{\ \ \ }l@{\ \ \ }l@{\ \ \ }l@{\ \ \ }l@{\ \ \ }l@{\ \ \ }l@{\ \ \ }l@{\ \ \ }l@{\ \ \ }l@{\ \ \ }l@{\ \ \ }l@{\ \ \ }l@{\ }}
\hline
MHF98622 & 5 27 24.867 & -67 15 05.31 & 15.99 & 0.17 & 120 &  19000 $\pm$900  & 3.8 $\pm$0.2  & 21 $\pm$20  & 298 $\pm$10  & B3IV & B2IV & \\
MHF98629 & 5 30 01.360 & -67 15 10.60 & 16.50 & 0.17 & 70 &  21000 $\pm$1700  & 3.8 $\pm$0.3  & 340 $\pm$35  & 315 $\pm$10  & B2IV & B1III & \\
MHF100069 & 5 27 29.545 & -67 14 52.44 & 16.18 & 0.27 & 95 &  19000 $\pm$1100  & 4.1 $\pm$0.2  & 204 $\pm$15  & 301 $\pm$10  & B2V & B2.5V & \\
MHF101934 & 5 29 17.200 & -67 14 40.60 & 15.99 & 0.17 & 90 & 15000 $\pm$900  & 3.4 $\pm$0.2  & 72 $\pm$10  & 285 $\pm$10  & B5III-IV & B2.5III & \\
MHF105436 & 5 29 20.133 & -67 13 55.71 & 15.64 & 0.17 & 116 &  19500 $\pm$1000  & 3.6 $\pm$0.2  & 134 $\pm$10  & 292 $\pm$10  & B2III-IV & B2.5V & \\
MHF106600 & 5 30 47.920 & -67 13 42.60 & 16.18 & 0.15 & 80 &  20500 $\pm$1200  & 4.1 $\pm$0.2  & 227 $\pm$22  & 291 $\pm$10  & B2V & B2IV & \\
MHF106613 & 5 28 35.209 & -67 13 45.45 & 16.31 & 0.33 & 91 &  17500 $\pm$1100  & 3.6 $\pm$0.2  & 324 $\pm$25  & 300 $\pm$10  & B3III-IV & B2.5V & cl4\\
MHF106692 & 5 29 58.573 & -67 13 47.01 & 14.89 & 0.18 & 145 &  15000 $\pm$700  & 3.1 $\pm$0.1  & 33 $\pm$20  & 295 $\pm$10  & B5III & B5 & \\
MHF107300 & 5 29 19.200 & -67 13 34.20 & 16.07 & 0.17 & 70 &  25000 $\pm$2000  & 4.0 $\pm$0.3  & 24 $\pm$20  & 292 $\pm$10  & B1IV-V & B1V & \\
MHF107458 & 5 30 36.450 & -67 13 33.70 & 16.27 & 0.17 & 70 &  19500 $\pm$1600  & 3.7 $\pm$0.3  & 343 $\pm$35  & 312 $\pm$10  & B2III-IV & B1III & \\
MHF109280 & 5 29 26.489 & -67 13 12.50 & 16.43 & 0.17 & 125 &  19300 $\pm$1000  & 3.8 $\pm$0.2  & 57 $\pm$10  & 270 $\pm$10  & B2IV-V & B2V & \\
MHF110170 & 5 29 18.147 & -67 13 00.25 & 16.32 & 0.26 & 110 &  20500 $\pm$1000  & 4.0 $\pm$0.2  & 137 $\pm$10  & 298 $\pm$10  & B2IV-V & B2IV & \\
MHF112935 & 5 27 31.570 & -67 12 30.60 & 15.29 & 0.19 & 130 &  29000 $\pm$1500  & 4.2 $\pm$0.2  & 178 $\pm$10  & 306 $\pm$10  & B0.5V & B0.5V & HeII\\
MHF113982 & 5 28 25.150 & -67 12 17.33 & 16.00 & 0.17 & 108 &  21500 $\pm$1100  & 3.8 $\pm$0.2  & 51 $\pm$10  & 301 $\pm$10  & B2IV& B1.5IV&\\
MHF115761 & 5 30 32.530 & -67 12 00.40 & 16.12 & 0.16 & 74 &  23000 $\pm$1900  & 4.0 $\pm$0.3  & 23 $\pm$20  & 307 $\pm$10  & B1.5V & B1V & \\
MHF115844 & 5 28 24.445 & -67 12 00.22 & 16.34 & 0.17 & 100 &  18500 $\pm$1000  & 3.9 $\pm$0.2  & 103 $\pm$10  & 294 $\pm$10  & B2.5IV & B2.5V & \\
MHF116094 & 5 27 10.510 & -67 11 47.70 & 16.28 & 0.17 & 67 &  21000 $\pm$2100  & 3.8 $\pm$0.3  & 374 $\pm$60  & 300 $\pm$10  & B2IV & B2V & \\
MHF117096 & 5 28 38.272 & -67 11 41.86 & 16.26 & 0.18 & 70 &  20000 $\pm$1600  & 3.9 $\pm$0.3  & 19 $\pm$20  & 305 $\pm$10  & B2V & B1V & \\
MHF117930 & 5 27 29.079 & -67 11 29.27 & 16.29 & 0.16 & 80 &  26000 $\pm$1600  & 4.4 $\pm$0.3  & 191 $\pm$20  & 373 $\pm$10  & B1V & B1.5V & \\
MHF117946 & 5 27 34.980 & -67 11 28.60 & 16.32 & 0.16 & 85 &  20000 $\pm$1200  & 4.1 $\pm$0.2  & 129 $\pm$13  & 299 $\pm$10  & B2V & B2V& \\
MHF119603 & 5 29 31.746 & -67 11 17.65 & 15.61 & 0.16 & 118 &  21500 $\pm$1100  & 3.9 $\pm$0.2  & 49 $\pm$15  & 300 $\pm$10  & B2V & B1.5V & \\
MHF119707 & 5 30 32.940 & -67 11 22.20 & 15.15 & 0.18 & 120 &  21500 $\pm$1100  & 3.6 $\pm$0.2  & 134 $\pm$10  & 308 $\pm$10  & B2III-IV & B1IV & cl1\\
MHF120461 & 5 30 36.730 & -67 11 02.90 & 16.16 & 0.15 & 60 &  21000 $\pm$2100  & 4.0 $\pm$0.4  & 131 $\pm$21  & 297 $\pm$10  & B2V & B1.5III& cl1\\
MHF121339 & 5 27 28.317 & -67 10 53.78 & 14.95 & 0.18 & 120 &  25000 $\pm$1200  & 3.9 $\pm$0.2  & 19 $\pm$20  & 300 $\pm$10  & B1V & B1V & \\
MHF122794 & 5 30 49.220 & -67 10 43.70 & 16.30 & 0.20 & 50 &  22000 $\pm$2600  & 4.0 $\pm$0.4  & 53 $\pm$20  & 298 $\pm$10  & B2V & B1V & \\
MHF124760 & 5 30 12.170 & -67 10 19.50 & 15.28 & 0.17 & 110 &  24500 $\pm$1200  & 3.8 $\pm$0.2  & 172 $\pm$10  & 329 $\pm$10  & B1V & B0V & \\
MHF124844 & 5 27 04.836 & -67 10 05.99 & 16.23 & 0.20 & 85 &  16000 $\pm$1000  & 3.5 $\pm$0.2  & 166 $\pm$15  & 300 $\pm$10  & B3III-IV & B2.5III & \\
MHF125614 & 5 30 17.430 & -67 10 07.90 & 16.38 & 0.20 & 60 &  20500 $\pm$2100  & 3.9 $\pm$0.4  & 157 $\pm$25  & 287 $\pm$10  & B2V & B1.5IV & \\
MHF126078 & 5 29 43.232 & -67 10 00.82 & 16.02 & 0.18 & 86 & 21000 $\pm$1300  & 3.8 $\pm$0.2  & 38 $\pm$20  & 312 $\pm$10  & B2IV & B1.5V & \\
MHF128212 & 5 29 53.238 & -67 9 42.49 & 14.93 & 0.19 & 160 &  21000 $\pm$1000  & 3.8 $\pm$0.2  & 64 $\pm$10  & 322 $\pm$10  & B2IV & B1.5V & \\
MHF131188 & 5 29 37.665 & -67 9 05.80 & 15.24 & 0.18 & 140 &  30000 $\pm$1500  & 4.3 $\pm$0.2  & 325 $\pm$16  & 299 $\pm$10  & B0V & B0.5V & HeII\\
MHF131563 & 5 30 31.970 & -67 8 58.80 & 16.46 & 0.26 & 55 &  17500 $\pm$1900  & 4.2 $\pm$0.4  & 140 $\pm$22  & 318 $\pm$10  & B3V & B3V & \\
MHF131570 & 5 29 51.860 & -67 8 57.40 & 16.41 & 0.16 & 70 &  19500 $\pm$1600  & 4.0 $\pm$0.3  & 122 $\pm$12  & 298 $\pm$10  & B2V & B2.5V & \\
MHF132507 & 5 28 18.853 & -67 8 43.75 & 16.41 & 0.18 & 80 &  19500 $\pm$1200  & 4.0 $\pm$0.2  & 13 $\pm$20  & 286 $\pm$10  & B2V & B2V & \\
MHF133049 & 5 29 52.694 & -67 8 42.79 & 16.10 & 0.21 & 70 &  21000 $\pm$1700  & 4.0 $\pm$0.3  & 35 $\pm$20  & 287 $\pm$10  & B2V & B1.5V & \\
MHF134545 & 5 30 14.008 & -67 8 23.97 & 16.31 & 0.18 & 80 &  20000 $\pm$1200  & 3.9 $\pm$0.2  & 78 $\pm$10  & 319 $\pm$10  & B2V & B2V & \\
MHF134864 & 5 29 18.600 & -67 8 22.35 & 16.37 & 0.17 & 85 &  18000 $\pm$1100  & 3.7 $\pm$0.2  & 3 $\pm$20  & 313 $\pm$10  & B2.5IV & B2.5IV&\\
MHF135232 & 5 28 38.670 & -67 8 14.50 & 16.01 & 0.16 & 95 &  24500 $\pm$1500  & 4.1 $\pm$0.2  & 222 $\pm$16  & 352 $\pm$10 & B1V & B1V & \\
MHF136076 & 5 28 47.904 & -67 8 02.56 & 16.44 & 0.27 & 70 &  22000 $\pm$1800  & 4.1 $\pm$0.3  & 23 $\pm$20  & 307 $\pm$10  & B2V & B2V & \\
MHF136846 & 5 30 03.662 & -67 7 58.97 & 16.19 & 0.17 & 80 &  19500 $\pm$1200  & 3.6 $\pm$0.2  & 248 $\pm$25  & 300 $\pm$10  & B2III-IV & B1V & \\
MHF136943 & 5 30 37.390 & -67 7 56.00 & 16.24 & 0.23 & 60 &  19500 $\pm$2000  & 4.1 $\pm$0.4  & 205 $\pm$33  & 299 $\pm$10  & B2V & B2III & \\
MHF137534 & 5 30 25.610 & -67 7 50.70 & 16.17 & 0.19 & 63 &  21000 $\pm$2100  & 3.9 $\pm$0.4  & 107 $\pm$17  & 313 $\pm$10  & B2V & B1V & \\
MHF137890 & 5 29 58.370 & -67 7 44.37 & 16.32 & 0.17 & 85 &  21000 $\pm$1300  & 4.1 $\pm$0.2  & 123 $\pm$11  & 295 $\pm$10  & B2V & B1V & cl6\\
MHF138223 & 5 29 36.708 & -67 7 40.89 & 16.24 & 0.69 & 65 &  14500 $\pm$1500  & 4.1 $\pm$0.4  & 83 $\pm$14  & 281 $\pm$10  & B5V & B6V & cl9\\
MHF139231 & 5 30 19.580 & -67 7 34.20 & 14.92 & 0.18 & 75 &  21000 $\pm$1700  & 3.6 $\pm$0.3  & 168 $\pm$17  & 297 $\pm$10  & B2III & B0.5IV & \\
MHF140653 & 5 30 21.138 & -67 7 17.83 & 15.21 & 0.19 & 75 &  20500 $\pm$1600  & 3.9 $\pm$0.3  & 52 $\pm$20  & 272 $\pm$10  & B2IV & B2IV & \\
MHF141004 & 5 29 57.130 & -67 7 35.50 & 15.14 & 0.33 & 115 & 32500 $\pm$1600  & 4.4 $\pm$0.2  & 21 $\pm$20  & 296 $\pm$10  & O9V & O9V & HeII, cl6 \\
MHF141834 & 5 28 17.049 & -67 6 54.57 & 16.38 & 0.16 & 90 &  22500 $\pm$1300  & 4.0 $\pm$0.2  & 129 $\pm$10  & 311 $\pm$10  & B2V & B2IV & cl3\\
MHF142249 & 5 28 13.500 & -67 6 54.80 & 16.46 & 0.21 & 80 &  18500 $\pm$1100  & 3.9 $\pm$0.2  & 87 $\pm$10  & 312 $\pm$10  & B2.5V & B2.5IV & cl3\\
MHF142489 & 5 29 35.870 & -67 6 51.20 & 16.35 & 0.16 & 50 &   9600 $\pm$1200  & 3.2 $\pm$0.3  & 1 $\pm$20  & 288 $\pm$10  & A0.5III & B9III & \\
MHF142798 & 5 29 41.935 & -67 6 47.67 & 16.47 & 0.17 & 70 &  19500 $\pm$1600  & 4.1 $\pm$0.3  & 149 $\pm$15  & 303 $\pm$10  & B2V & B2.5V & \\
MHF144083 & 5 30 07.760 & -67 6 34.71 & 16.27 & 0.19 & 80 &  22000 $\pm$1300  & 4.0 $\pm$0.2  & 241 $\pm$25  & 300 $\pm$10  & B2V & B1V & \\
MHF144186 & 5 30 19.070 & -67 6 36.00 & 16.05 & 0.19 & 63 &  19000 $\pm$1900  & 3.7 $\pm$0.3  & 169 $\pm$30  & 290 $\pm$10  & B2III-IV & B1III & \\
MHF144562 & 5 28 19.926 & -67 6 24.61 & 16.40 & 0.16 & 90 &  13500 $\pm$800  & 3.6 $\pm$0.2  & 167 $\pm$12  & 299 $\pm$10  & B5III & B3III & \\
MHF144608 & 5 30 37.355 & -67 6 29.83 & 15.64 & 0.19 & 100 &  21000 $\pm$1100  & 3.7 $\pm$0.2  & 18 $\pm$20  & 309 $\pm$10  & B2IV & B1V & \\
MHF144637 & 5 30 32.420 & -67 6 32.40 & 14.96 & 0.18 & 85 & 21000 $\pm$1300  & 3.5 $\pm$0.2  & 202 $\pm$20  & 248 $\pm$10  & B2III-IV & B1IV & \\
KWBBe0554 & 5 30 19.794 & -67 17 05.02 & 16.74 & 0.30 & 56 &  18500 $\pm$2200  & 4.2 $\pm$0.4  & 47 $\pm$20  & 310 $\pm$10  & B2.5V & B2.5V& not Be\\
KWBBe0993 & 5 30 18.581 & -67 18 51.61 & 17.38 & 0.16 & 60 &  19500 $\pm$2000  & 4.2 $\pm$0.4  & 33 $\pm$20  & 307 $\pm$10  & B2V & B2.5V& not Be\\
KWBBe1169 & 5 30 45.160 & -67 17 18.30 & 16.62 & 1.33 & 40 &  15500 $\pm$2300  & 4.1 $\pm$0.4  & 283 $\pm$57  & 300 $\pm$10  & B4V & B6III& not Be, cl0\\
\hline
\end{tabular}
\label{table3}
}
\end{table*}

\subsection{O-B-A stars}
\subsubsection{Fundamental parameters of O-B-A stars}

Early-type stars that do not show intrinsic emission lines in
their spectrum and have not been detected as spectroscopic
binaries are listed in Table~\ref{table3} sorted by their MHF
catalogue number. Moreover, three stars with a KWBBe name,
reported as Be stars by Keller et al. (1999) but not confirmed or
in a temporary B phase at epochs of VLT/FLAMES observations, are
added at the end of the Table.  The fundamental parameters \teff,
\logg, \vsini, and \rv~obtained by fitting the observed spectra,
as well as the spectral classification deduced on one hand from
\teff--\logg~plane calibration (CFP determination, method 2) and
on the other hand from equivalent width diagrams (CEW
determination, method 1), are reported in columns 7, 8, 9 and 10,
respectively. As the heliocentric velocities are smaller than 1.5
\kms~ and the mean error on \rv~are 9-10 km~s$^{-1}$, we do not
correct \rv~from the heliocentric velocity.

\subsubsection{Luminosity, mass, and radius for O-B-A stars}
\label{subsec:HRtracks}

Once the fundamental parameters of O, B, and A stars are
known, to derive their luminosity, mass and radius, we
interpolate in the HR-diagram grids calculated for the LMC
metallicity (Z = 0.004; Korn et al. 2002, Rolleston et
al. 1996) and for stars without rotation in Charbonnel et al.
(1993).

In order to justify the use of non-rotating models, we
estimate  the mean radius, mean mass and mean \vsini~ in
various mass bins (e.g. 5 $<$ M $<$ 7 M$\odot$, 7 $<$ M $<$ 9
M$\odot$, etc). We then obtain a mean equatorial
velocity for a random angle distribution using:\\
\begin{equation}
V_{e} = \frac{4}{\pi} <V\!\sin i_{\rm~\!app.}>,
\label{eqVevsini}
\end{equation}
where $<$\vsiniap$>$ is the mean \vsiniap.

We calculate the critical velocity with the classical formula:\\
\begin{equation}
V_{c} \simeq 436.7 \left(\frac{<M>}{<R>}\right)^{1/2},
\label{veqc}
\end{equation}
where $<$M$>$ and $<$R$>$ are the mean mass in
M$_{\odot}$ and mean radius in R$_{\odot}$.

\begin{table*}[]
\small{
\caption{Parameters $\log(L/L_{\odot}$), $M/M_{\odot}$ and $R/R_{\odot}$
interpolated or calculated for our sample of O-B stars and for several SB1 from HR diagrams taken
from Charbonnel et al. (1993).}
\centering
\begin{tabular}{lllllllll}
\hline
\hline
Star & $\log(L/L_{\odot})$ & $M/M_{\odot}$ & $R/R_{\odot}$ & \vline & Star & $\log(L/L_{\odot})$ & $M/M_{\odot}$ & $R/R_{\odot}$ \\
\hline
MHF52224 & 4.2$\pm$0.3 & 10.0$\pm$1.0 & 9.2$\pm$1.0 & \vline & MHF107300 & 4.0$\pm$0.3 & 9.7$\pm$0.5 & 5.3$\pm$1.0 \\
MHF54275 & 3.7$\pm$0.3 & 8.4$\pm$0.5 & 4.5$\pm$0.5 & \vline & MHF107458 & 3.8$\pm$0.3 & 7.4$\pm$0.5 & 6.8$\pm$1.0 \\
MHF54565 & 4.4$\pm$0.3 & 11.3$\pm$1.0 & 9.1$\pm$1.0 & \vline & SBMHF109251 & 3.5$\pm$0.3 & 6.9$\pm$0.5 & 4.7$\pm$0.5 \\
MHF54686 & 3.3$\pm$0.3 & 6.1$\pm$0.5 & 3.9$\pm$0.5 & \vline & MHF109280 & 3.5$\pm$0.3 & 6.5$\pm$0.5 & 5.4$\pm$1.0 \\
MHF57079 & 3.6$\pm$0.3 & 8.1$\pm$0.5 & 3.6$\pm$0.5 & \vline & MHF110170 & 3.5$\pm$0.3 & 6.7$\pm$0.5 & 4.3$\pm$0.5 \\
MHF57428 & 3.6$\pm$0.3 & 8.4$\pm$0.5 & 3.2$\pm$0.5 & \vline & SBMHF110467 & 3.9$\pm$0.3 & 10.1$\pm$1.0 & 4.5$\pm$0.5 \\
MHF57975 & 3.7$\pm$0.3 & 7.0$\pm$0.5 & 5.9$\pm$1.0 & \vline & MHF112935 & 4.1$\pm$0.3 & 12.0$\pm$1.0 & 4.6$\pm$0.5 \\
MHF59059 & 3.5$\pm$0.3 & 8.2$\pm$0.5 & 3.2$\pm$0.5 & \vline & SBMHF113048 & 3.4$\pm$0.3 & 5.8$\pm$0.5 & 5.5$\pm$1.0 \\
MHF60436 & 3.7$\pm$0.3 & 8.0$\pm$0.5 & 4.6$\pm$0.5 & \vline & MHF113982 & 3.8$\pm$0.3 & 8.2$\pm$0.5 & 5.8$\pm$1.0 \\
MHF62150 & 3.2$\pm$0.3 & 5.8$\pm$0.5 & 4.1$\pm$0.5 & \vline & MHF115761 & 3.8$\pm$0.3 & 8.5$\pm$0.5 & 5.1$\pm$1.0 \\
MHF62555 & 3.5$\pm$0.3 & 6.7$\pm$0.5 & 4.2$\pm$0.5 & \vline & MHF115844 & 3.4$\pm$0.3 & 6.1$\pm$0.5 & 4.9$\pm$0.5 \\
MHF63084 & 3.3$\pm$0.3 & 5.2$\pm$0.5 & 14.3$\pm$1.5 & \vline & MHF116094 & 3.8$\pm$0.3 & 8.0$\pm$0.5 & 5.8$\pm$1.0 \\
MHF63948 & 3.4$\pm$0.3 & 6.6$\pm$0.5 & 4.1$\pm$0.5 & \vline & MHF117096 & 3.5$\pm$0.3 & 6.8$\pm$0.5 & 4.8$\pm$0.5 \\
MHF65925 & 3.8$\pm$0.3 & 8.5$\pm$0.5 & 4.9$\pm$0.5 & \vline & MHF117930 & 3.6$\pm$0.3 & 8.2$\pm$0.5 & 3.0$\pm$0.5 \\
MHF66708 & 3.8$\pm$0.3 & 9.6$\pm$0.5 & 3.7$\pm$0.5 & \vline & MHF117946 & 3.4$\pm$0.3 & 6.4$\pm$0.5 & 3.9$\pm$0.5 \\
MHF67663 & 4.1$\pm$0.3 & 12.5$\pm$1.0 & 4.3$\pm$0.5 & \vline & MHF119603 & 3.6$\pm$0.3 & 7.6$\pm$0.5 & 4.9$\pm$0.5 \\
MHF67792 & 3.7$\pm$0.3 & 8.0$\pm$0.5 & 4.7$\pm$0.5 & \vline & MHF119707 & 4.1$\pm$0.3 & 9.0$\pm$0.5 & 8.2$\pm$1.0 \\
MHF68153 & 3.7$\pm$0.3 & 7.8$\pm$0.5 & 4.6$\pm$0.5 & \vline & MHF120461 & 3.5$\pm$0.3 & 6.9$\pm$0.5 & 4.2$\pm$0.5 \\
MHF68195 & 3.8$\pm$0.3 & 8.4$\pm$0.5 & 4.7$\pm$0.5 & \vline & MHF121339 & 4.1$\pm$0.3 & 10.4$\pm$1.0 & 6.0$\pm$1.0 \\
MHF68257 & 3.7$\pm$0.3 & 8.1$\pm$0.5 & 4.6$\pm$0.5 & \vline & MHF122794 & 3.6$\pm$0.3 & 7.6$\pm$0.5 & 4.7$\pm$0.5 \\
MHF69681 & 4.4$\pm$0.3 & 12.9$\pm$1.0 & 7.3$\pm$1.0 & \vline & MHF124760 & 4.1$\pm$0.3 & 10.7$\pm$1.0 & 6.7$\pm$1.0 \\
MHF70976 & 4.3$\pm$0.3 & 15.0$\pm$1.0 & 4.8$\pm$0.5 & \vline & MHF124844 & 3.5$\pm$0.3 & 6.1$\pm$0.5 & 7.4$\pm$1.0 \\
MHF72268 & 3.6$\pm$0.3 & 7.2$\pm$0.5 & 5.6$\pm$1.0 & \vline & MHF125614 & 3.6$\pm$0.3 & 7.0$\pm$0.5 & 4.8$\pm$0.5 \\
MHF74015 & 3.6$\pm$0.3 & 7.9$\pm$0.5 & 4.4$\pm$0.5 & \vline & MHF126078 & 3.8$\pm$0.3 & 8.3$\pm$0.5 & 6.1$\pm$1.0 \\
MHF75373 & 3.4$\pm$0.3 & 6.6$\pm$0.5 & 4.2$\pm$0.5 & \vline & MHF128212 & 3.7$\pm$0.3 & 7.8$\pm$0.5 & 5.7$\pm$1.0 \\
MHF75553 & 3.9$\pm$0.3 & 8.1$\pm$0.5 & 8.1$\pm$1.0 & \vline & SBMHF128963 & 3.4$\pm$0.3 & 6.6$\pm$0.5 & 4.2$\pm$0.5 \\
MHF77981 & 3.7$\pm$0.3 & 8.2$\pm$0.5 & 4.7$\pm$0.5 & \vline & MHF131188 & 4.1$\pm$0.3 & 12.4$\pm$1.0 & 4.0$\pm$0.5 \\
MHF78706 & 3.7$\pm$0.3 & 8.3$\pm$0.5 & 4.6$\pm$0.5 & \vline & MHF131563 & 2.8$\pm$0.2 & 4.5$\pm$0.5 & 2.8$\pm$0.5 \\
MHF81807 & 3.6$\pm$0.3 & 7.2$\pm$0.5 & 5.6$\pm$0.5 & \vline & MHF131570 & 3.3$\pm$0.3 & 6.1$\pm$0.5 & 4.0$\pm$0.5 \\
MHF81136 & 3.5$\pm$0.3 & 6.9$\pm$0.5 & 4.4$\pm$0.5 & \vline & MHF132507 & 3.4$\pm$0.3 & 6.4$\pm$0.5 & 4.4$\pm$0.5 \\
MHF81174 & 3.3$\pm$0.3 & 6.7$\pm$0.5 & 3.1$\pm$0.5 & \vline & MHF133049 & 3.5$\pm$0.3 & 7.0$\pm$0.5 & 4.4$\pm$0.5 \\
MHF81322 & 3.7$\pm$0.3 & 7.0$\pm$0.5 & 2.7$\pm$0.5 & \vline & MHF134545 & 3.5$\pm$0.3 & 6.7$\pm$0.5 & 4.6$\pm$0.5 \\
MHF81490 & 3.1$\pm$0.3 & 5.2$\pm$0.5 & 3.4$\pm$0.5 & \vline & MHF134864 & 3.5$\pm$0.3 & 6.3$\pm$0.5 & 5.9$\pm$1.0 \\
MHF81521 & 3.6$\pm$0.3 & 6.6$\pm$0.5 & 7.8$\pm$1.0 & \vline & MHF135232 & 3.8$\pm$0.3 & 9.1$\pm$0.5 & 4.6$\pm$0.5 \\
MHF82482 & 3.6$\pm$0.3 & 7.4$\pm$0.5 & 4.4$\pm$0.5 & \vline & MHF136076 & 3.6$\pm$0.3 & 7.4$\pm$0.5 & 4.1$\pm$0.5 \\
MHF84042 & 3.3$\pm$0.3 & 6.4$\pm$0.5 & 3.6$\pm$0.5 & \vline & MHF136846 & 3.8$\pm$0.3 & 7.3$\pm$0.5 & 7.0$\pm$1.0 \\
MHF84176 & 3.8$\pm$0.3 & 10.1$\pm$1.0 & 3.7$\pm$0.5 & \vline & MHF136943 & 3.3$\pm$0.3 & 6.2$\pm$0.5 & 3.9$\pm$0.5 \\
MHF85562 & 3.4$\pm$0.3 & 6.5$\pm$0.5 & 4.2$\pm$0.5 & \vline & MHF137534 & 3.6$\pm$0.3 & 7.4$\pm$0.5 & 5.0$\pm$0.5 \\
MHF86995 & 3.5$\pm$0.3 & 6.5$\pm$0.5 & 6.0$\pm$1.0 & \vline & MHF137890 & 3.5$\pm$0.3 & 6.9$\pm$0.5 & 4.1$\pm$0.5 \\
MHF87634 & 4.1$\pm$0.3 & 11.1$\pm$1.0 & 5.9$\pm$1.0 & \vline & MHF138223 & 2.5$\pm$0.2 & 3.6$\pm$0.5 & 2.7$\pm$0.5 \\
MHF88527 & 3.7$\pm$0.3 & 8.5$\pm$0.5 & 3.3$\pm$0.5 & \vline & MHF139231 & 4.0$\pm$0.3 & 8.8$\pm$0.5 & 8.3$\pm$1.0 \\
MHF93347 & 3.6$\pm$0.3 & 8.1$\pm$0.5 & 3.8$\pm$0.5 & \vline & MHF140653 & 3.6$\pm$0.3 & 7.1$\pm$0.5 & 5.0$\pm$0.5 \\
MHF94228 & 3.4$\pm$0.3 & 6.6$\pm$0.5 & 4.6$\pm$0.5 & \vline & MHF141004 & 4.1$\pm$0.3 & 8.9$\pm$0.5 & 3.2$\pm$0.5 \\
MHF95555 & 3.7$\pm$0.3 & 7.1$\pm$0.5 & 6.4$\pm$1.0 & \vline & MHF141834 & 3.7$\pm$0.3 & 7.8$\pm$0.5 & 4.5$\pm$0.5 \\
MHF96072 & 3.1$\pm$0.3 & 4.8$\pm$0.5 & 2.2$\pm$0.5 & \vline & SBMHF141891 & 3.5$\pm$0.3 & 6.8$\pm$0.5 & 4.5$\pm$0.5 \\
MHF97219 & 4.4$\pm$0.3 & 11.9$\pm$1.0 & 3.8$\pm$0.5 & \vline & MHF142249 & 3.3$\pm$0.3 & 6.0$\pm$0.5 & 4.5$\pm$0.5 \\
MHF97965 & 3.2$\pm$0.3 & 6.2$\pm$0.5 & 2.9$\pm$0.5 & \vline & MHF142489 & 2.6$\pm$0.2 & 3.4$\pm$0.5 & 7.6$\pm$1.0 \\
MHF98622 & 3.5$\pm$0.3 & 6.5$\pm$0.5 & 5.6$\pm$1.0 & \vline & MHF142798 & 3.2$\pm$0.3 & 5.9$\pm$0.5 & 3.4$\pm$0.5 \\
MHF98629 & 3.8$\pm$0.3 & 8.0$\pm$0.5 & 6.0$\pm$1.0 & \vline & MHF144083 & 3.6$\pm$0.3 & 7.6$\pm$0.5 & 4.3$\pm$0.5 \\
MHF100069 & 3.1$\pm$0.3 & 5.6$\pm$0.5 & 3.6$\pm$0.5 & \vline & MHF144186 & 3.7$\pm$0.3 & 7.1$\pm$0.5 & 6.7$\pm$1.0 \\
MHF101934 & 3.3$\pm$0.3 & 5.4$\pm$0.5 & 7.4$\pm$1.0 & \vline & MHF144562 & 3.0$\pm$0.3 & 4.3$\pm$0.5 & 5.7$\pm$1.0 \\
SBMHF102053 & 3.3$\pm$0.3 & 5.5$\pm$0.5 & 5.6$\pm$1.0 & \vline & MHF144608 & 3.9$\pm$0.3 & 8.2$\pm$0.5 & 6.5$\pm$1.0 \\
SBMHF103207 & 4.0$\pm$0.3 & 9.8$\pm$0.5 & 6.4$\pm$1.0 & \vline & MHF144637 & 4.2$\pm$0.3 & 10.0$\pm$0.5 & 10.2$\pm$1.0 \\
MHF105436 & 3.9$\pm$0.3 & 8.0$\pm$0.5 & 8.1$\pm$1.0 & \vline & KWBBe0554 & 3.0$\pm$0.3 & 5.2$\pm$0.5 & 3.2$\pm$0.5 \\
MHF106600 & 3.4$\pm$0.3 & 6.6$\pm$0.5 & 4.0$\pm$0.5 & \vline & KWBBe0993 & 3.0$\pm$0.3 & 5.5$\pm$0.5 & 2.9$\pm$0.5 \\
MHF106613 & 3.6$\pm$0.3 & 6.6$\pm$0.5 & 7.2$\pm$1.0 & \vline & KWBBe1169 & 2.7$\pm$0.2 & 4.0$\pm$0.5 & 2.9$\pm$0.5 \\
MHF106692 & 3.8$\pm$0.3 & 7.5$\pm$0.5 & 13.3$\pm$1.5 & \vline &  & & & \\
\hline
\end{tabular}
\label{tabLMR}
}
\end{table*}

\begin{figure*}[!ht]
\centering
\includegraphics[width=12 cm, height=17 cm, angle=-90]{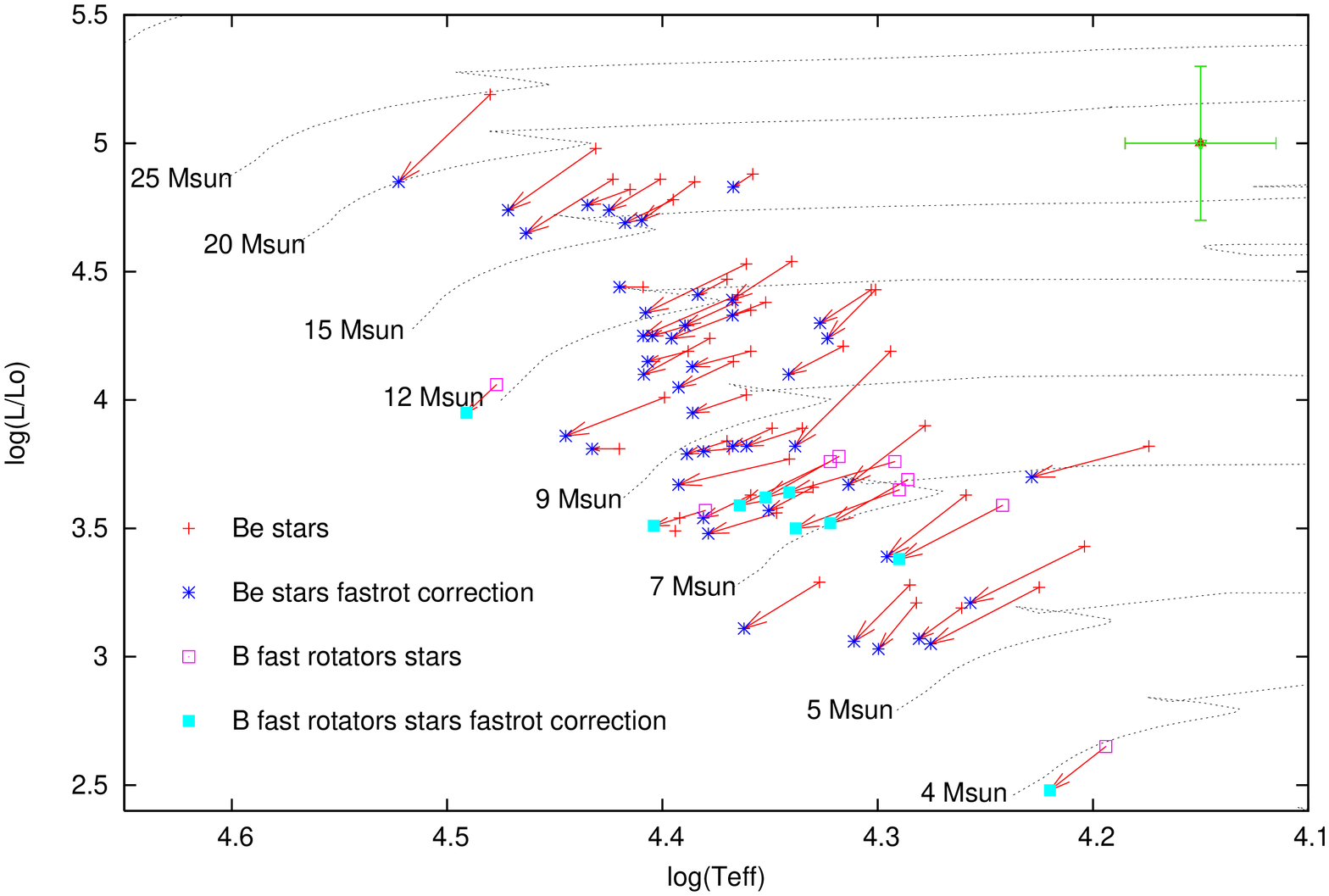}
\includegraphics[width=12 cm, height=17 cm, angle=-90]{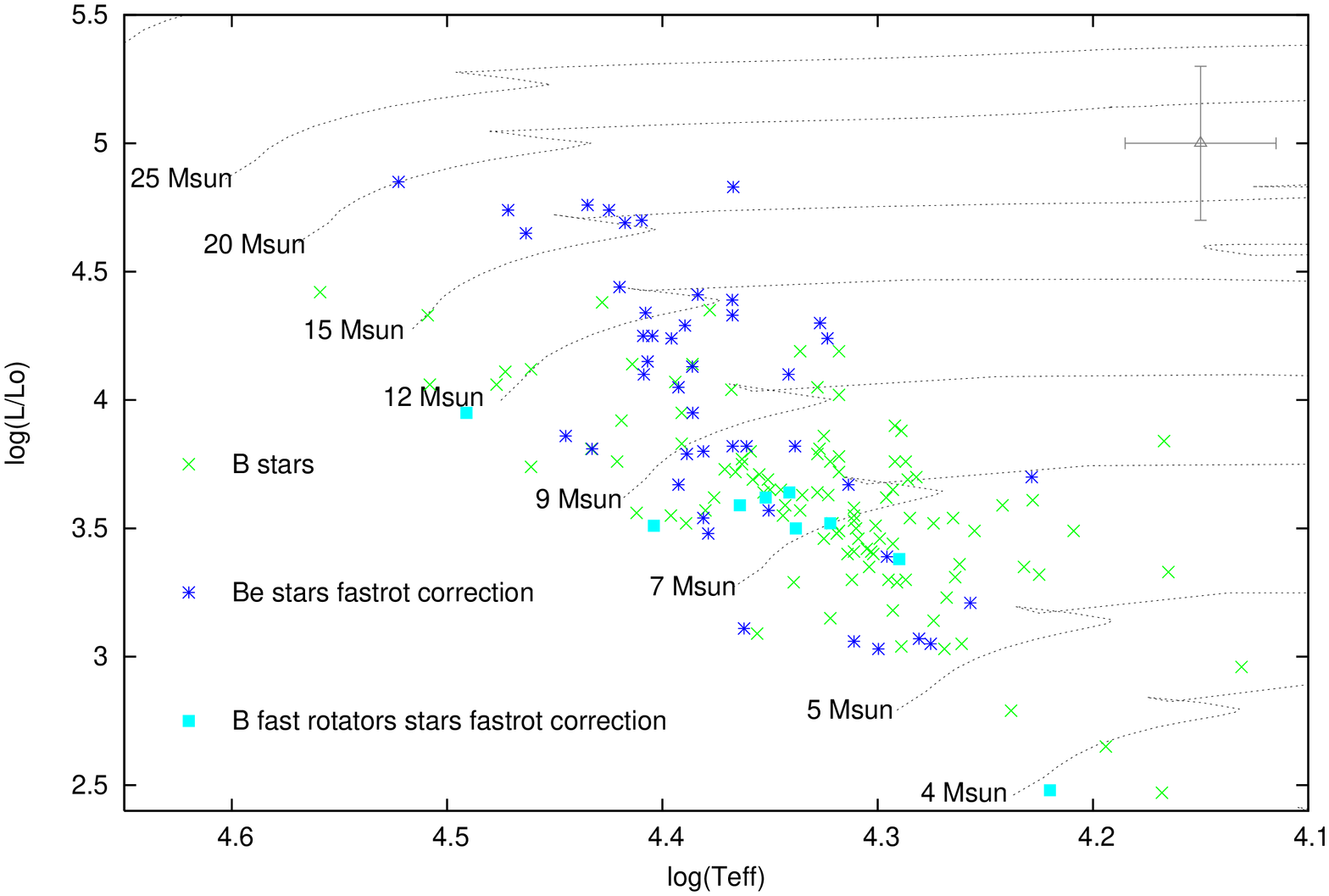}
\caption{HR diagrams for the studied B stars, including rapid rotators,
and Be stars. Top: The effects of fast rotation are taken into account with \omc~=
85\%  for Be stars and rapidly rotating B stars. Bottom: B stars and fast
rotators (Be and B stars) corrected for their fast rotation. Common: The adopted
metallicity for the LMC comes from Korn et al. (2002), Rolleston et al. (1996).
Green 'x' represent B stars, red '+' represent Be stars with their apparent
parameters, and blue '*' Be stars corrected with FASTROT with  \omc~= 85\%. Pink
empty squares represent rapidly rotating B stars with their apparent parameters,
and filled blue squares, rapidly rotating B stars corrected with FASTROT with
\omc~= 85\%. Typical error bars are shown in the upper right corner of the
figure.
}
\label{hr2}
\end{figure*}

This gives the $V_{e}/V_{c}$ ratio, and we can then obtain \omc~thanks to formulae taken
from Chauville et al. (2001):

\begin{equation}
$\omc$ = \frac{1}{0.724} V_{e}/V_{c} [1 - 0.276 (V_{e}/V_{c})^{2}].
\label{omc}
\end{equation}

\begin{table}[]
\caption{Corrections for \omc~= 85\% for rapidly rotating B stars in the
sample.  The units are K for \top, dex for \gop, and \kms~for \vsinit.}
\centering
\begin{tabular}{llll}
\hline
\hline
Star & \omc  & = 85\% & \\
     & \top  & \gop & \vsinit \\
\hline
MHF57079 & 25000$\pm$2900 & 4.4$\pm$0.4 & 327$\pm$50 \\
MHF57975 & 21500$\pm$1600 & 4.1$\pm$0.3 & 357$\pm$35 \\
MHF95555 & 21000$\pm$1200 & 4.0$\pm$0.2 & 331$\pm$30 \\
MHF98629 & 22500$\pm$1700 & 4.1$\pm$0.3 & 348$\pm$35 \\
MHF106613 & 19500$\pm$1100 & 4.0$\pm$0.2 & 338$\pm$25 \\
MHF107458 & 22000$\pm$1600 & 4.0$\pm$0.3 & 350$\pm$35 \\
MHF116094 & 23000$\pm$2100 & 4.1$\pm$0.3 & 384$\pm$60 \\
MHF131188 & 31000$\pm$1500 & 4.4$\pm$0.2 & 333$\pm$16 \\
KWBBe1169 & 16500$\pm$2400 & 4.4$\pm$0.4 & 294$\pm$57 \\
\hline
\end{tabular}
\label{tabBnfastrot}
\end{table}

For B stars $<$\vsiniap$>$ is close to 110 km~s$^{-1}$, thus $V_{e}/V_{c}$ $\simeq$ 27\% and
\omc $\simeq$ 37\%. As the effects of fast rotation appear for \omc $>$ 50\% (Zorec et
al. 2005), we do not need to correct B stars for fast rotation effects except for 9 of
them which have a strong \vsini.

We obtain in this way the luminosity, mass and radius of most O, B and A stars of the
sample (see Table~\ref{tabLMR}). The position of these stars in the HR diagram is shown
in Fig.~\ref{hr2}.

\subsection{Corrections for rapidly rotating B-type stars}
\label{Bnstars}

The 9 B-type stars MHF57079, MHF57975, MHF95555, MHF98629,
MHF106613, MHF107458, MHF116094, MHF131188, and KWBBe1169 have a
high rotational velocity, although they do not show emission
lines as Be stars. The star KWBBe1169 was previously observed like
a Be star by Keller (1999) but, in our observations, it
does not show any emission. This could be due to the transient
nature of the Be phenomenon. Results on fundamental parameters
taking into account fast rotation effects are given in
Table~\ref{tabBnfastrot} and Fig.~\ref{hr2}.

\begin{figure}[ht]
\centering
\resizebox{\hsize}{!}{\includegraphics[angle=-90,clip]{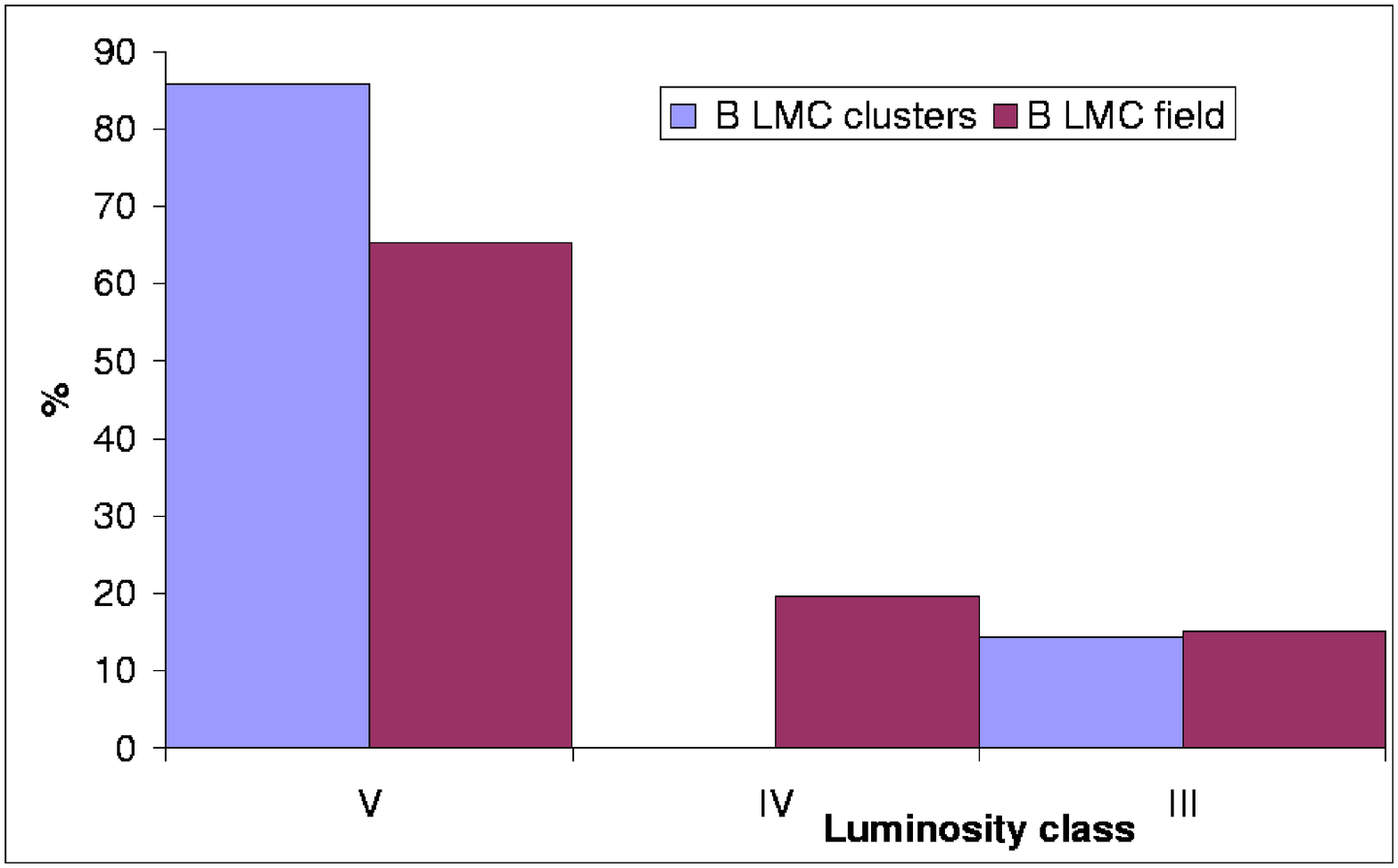}}\\
\resizebox{\hsize}{!}{\includegraphics[angle=-90,clip]{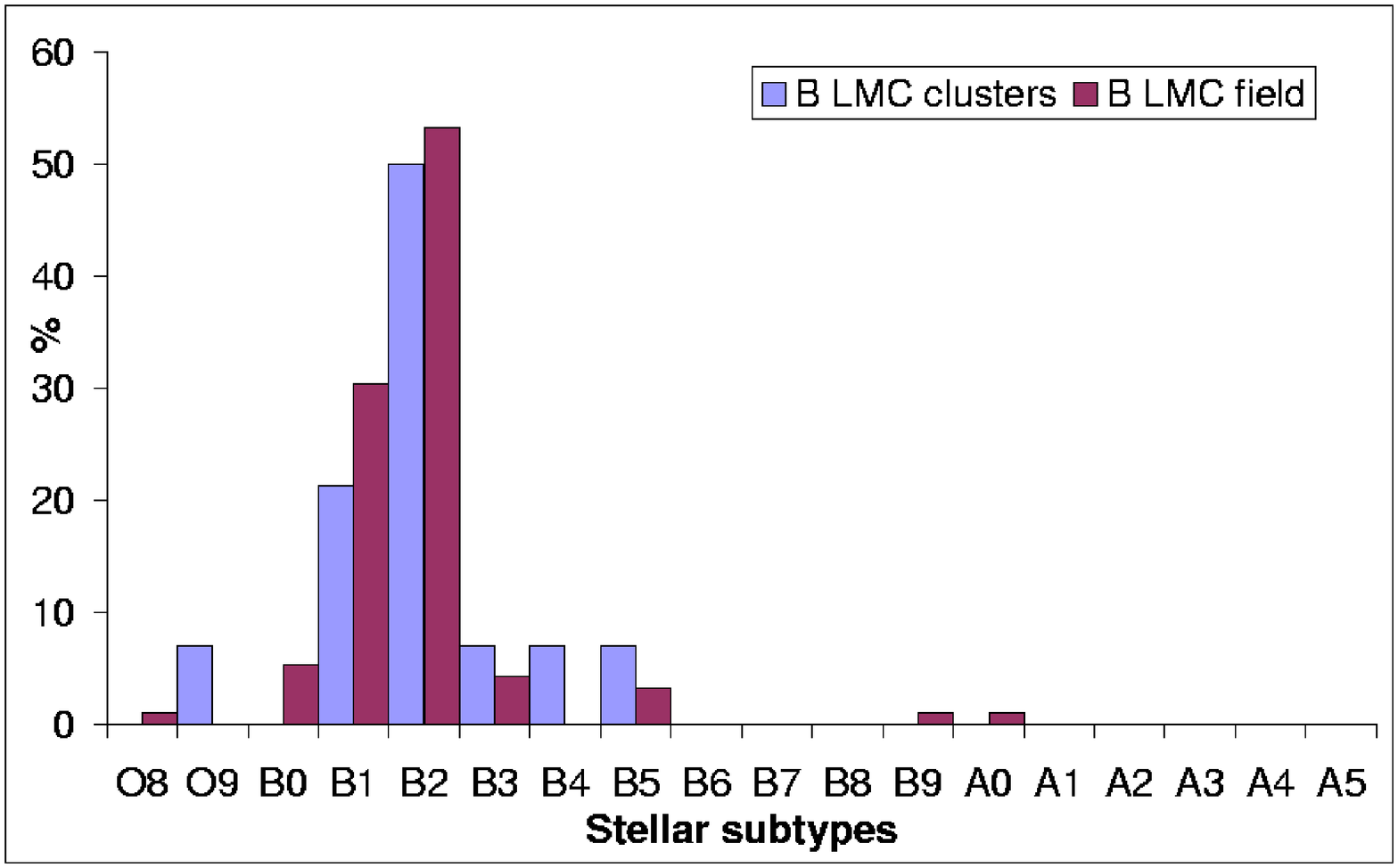}}
\caption{Luminosity class (upper panel) and spectral type (lower panel)
distributions of B-type stars in the sample in the LMC.}
\label{LCSTBLMC}
\end{figure}

\subsection{Be stars}
\subsubsection{Fundamental parameters of Be stars}

The total number of Be stars in the sample is 47. It includes 22 known Be stars, called
KWBBe in Keller et al. (1999), for which the H$\alpha$ emissive character has been
confirmed, and 25 new Be stars reported in Paper I and called MHFBe. For a description
of the emission line characteristics of these stars we refer to Paper I. The apparent
fundamental parameters (\tap, \gap, \vsiniap, and \rv) we derive in a first step for
these stars are reported in Table~\ref{tableBe}. The spectral classification derived
from apparent fundamental parameters is also given in the last column of the Table.
Without correction for fast rotation nearly all Be stars would have a sub-giant or
giant luminosity class. The apparent position of Be stars in the HR diagram compared to
B stars is also shown in Fig.~\ref{hr2}.

\begin{figure}[ht]
\centering
\resizebox{\hsize}{!}{\includegraphics[angle=-90,clip]{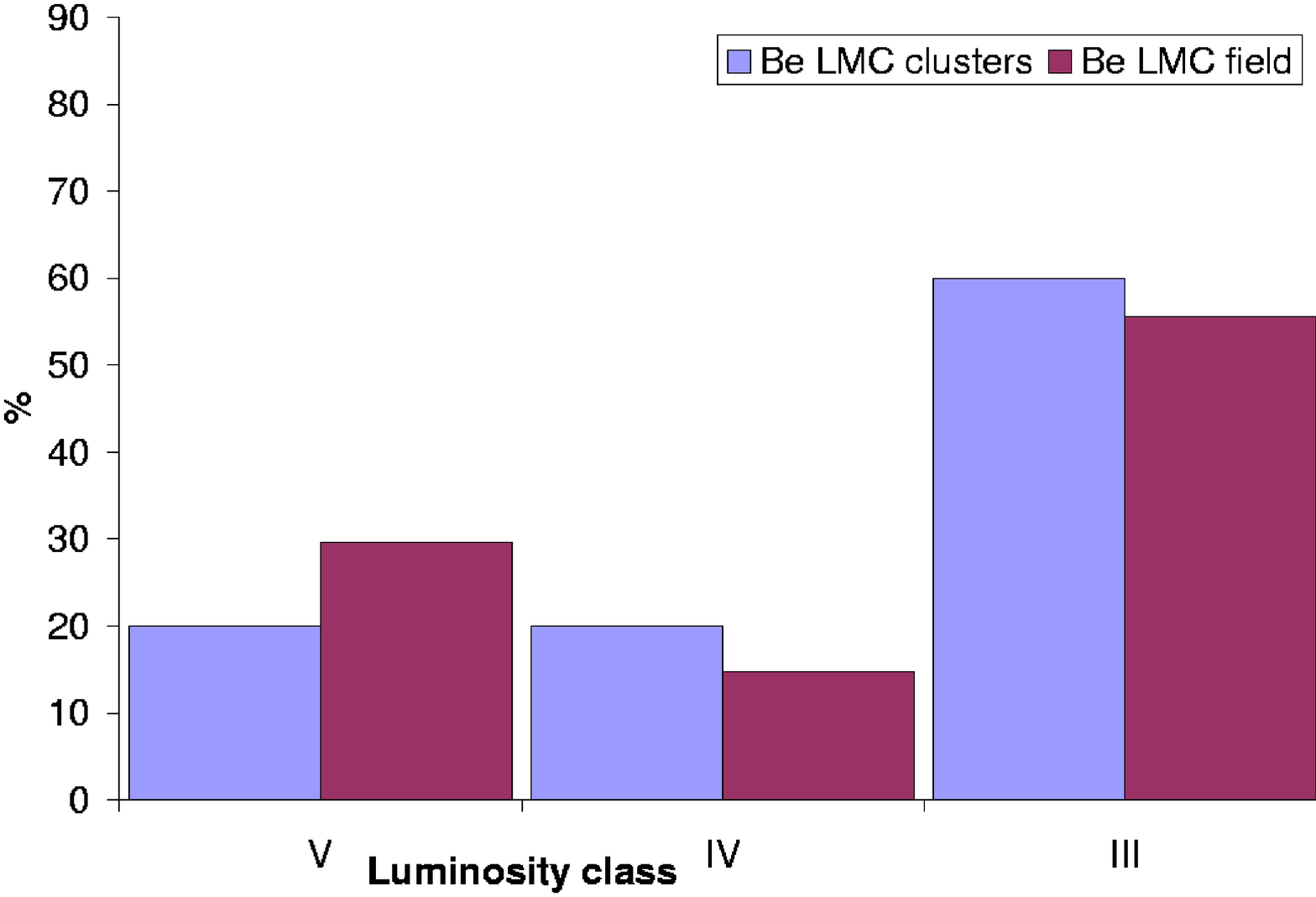}}\\
\resizebox{\hsize}{!}{\includegraphics[angle=-90,clip]{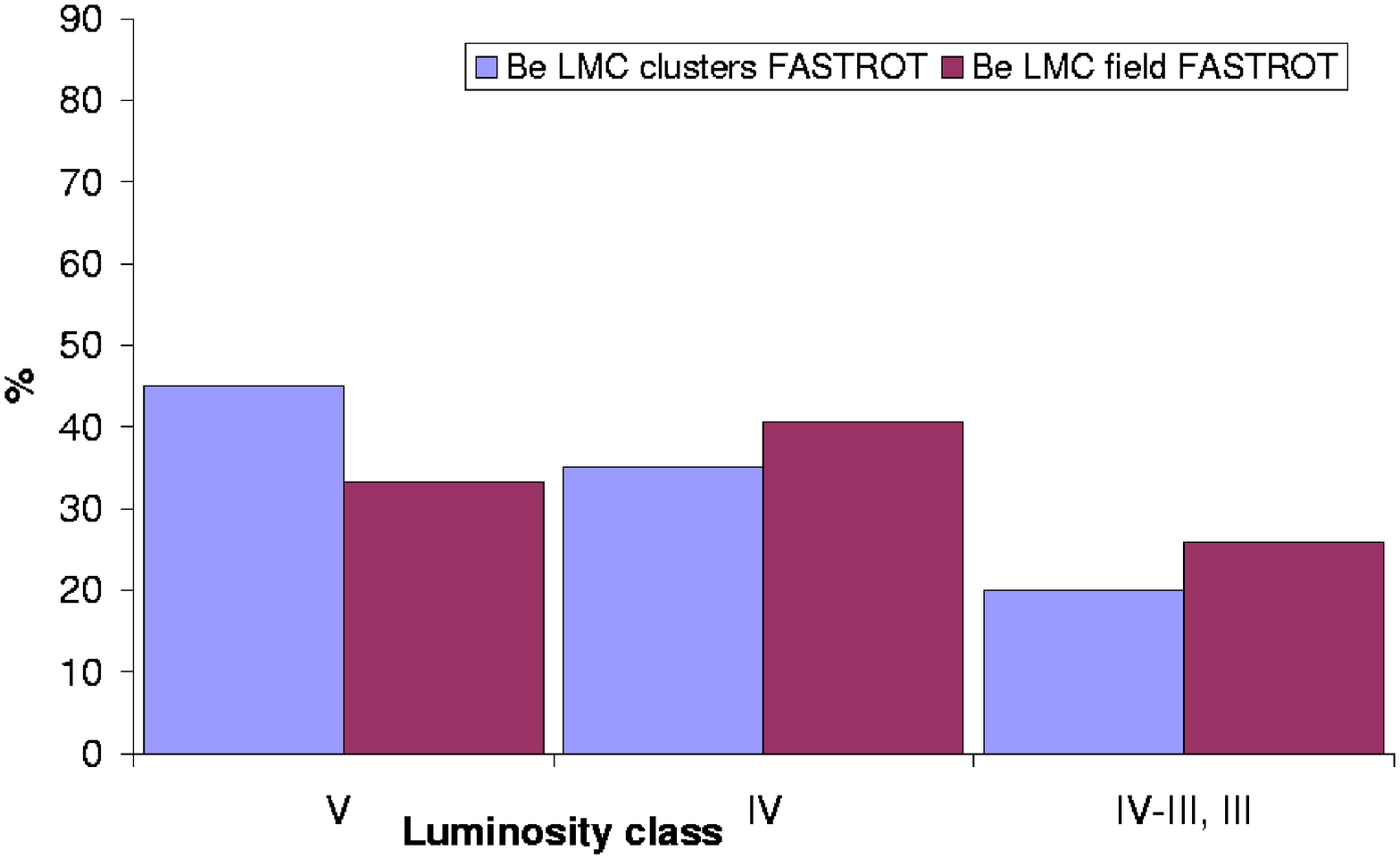}}
\caption{Percentages of Be stars in the sample in the field and
in clusters in the LMC versus luminosity class. Upper
panel: without correction for fast rotation. Lower panel: with
FASTROT corrections.} \label{LCBeLMC}
\end{figure}

\subsubsection{Rapid rotation corrections for Be stars}
\label{befastcalculs}

The pnrc (i.e. parent non rotating counterpart) fundamental parameters (\top, \gop,
\vsinit) we obtain after correction with FASTROT (see Sect. 3.4) are given in
Table~\ref{tabBefastrot} for different rotation rates \omc. We estimate the rotation
rate \omc~to be used for the selection of the probable most suitable pnrc fundamental
parameters of Be stars in the LMC thanks to the equations mentioned above. We obtain
$V_{e}/V_{c}$~$\simeq$~70\% and \omc~$\simeq$~85\% on average. As for O-B-A stars and
with the pnrc fundamental parameters corresponding to the rotation rate \omc~=~85\%,
we derive $\log(L/L_{\odot}$), $M/M_{\odot}$, and $R/R_{\odot}$ for Be stars. These
parameters are given in Table~\ref{tabLMRBe}. After correction for rapid rotation, Be
stars globally shift in the HR diagram towards lower luminosity and higher temperature,
as illustrated in Fig.\ref{hr2}. It clearly demonstrates that Be stars are less evolved
than their apparent fundamental parameters could indicate.

\subsection{Spectroscopic binaries}

The determination of fundamental parameters was undertaken for 15 spectroscopic binaries
that show a single line spectrum (SB1, see Table 2 in Paper I). However, since we
usually obtained only one spectrum for each suspected SB1 binary, the influence of the
secondary component on the spectrum is not known. Results are given in
Table~\ref{tablebin}. We note a fair agreement between spectral classifications derived
from fundamental parameters and from the equivalent width of the H$\gamma$ and
\ion{He}{I} 4471 lines, except for one star (MHF91603). The mean \vsini~for these
binaries is 100 \kms.

\begin{figure}[htbp]
\centering
\resizebox{\hsize}{!}{\includegraphics[angle=-90,clip]{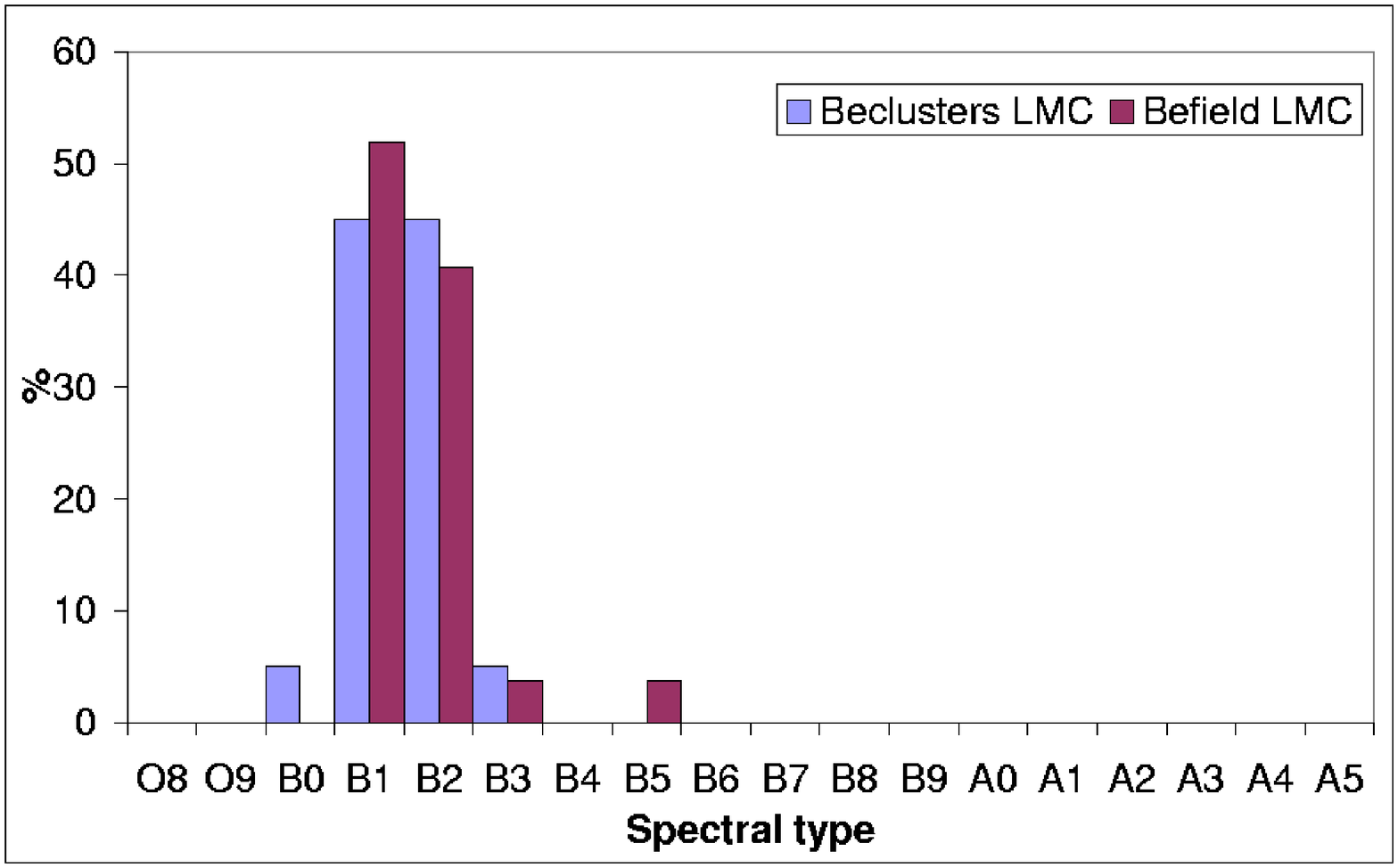}}\\
\resizebox{\hsize}{!}{\includegraphics[angle=-90,clip]{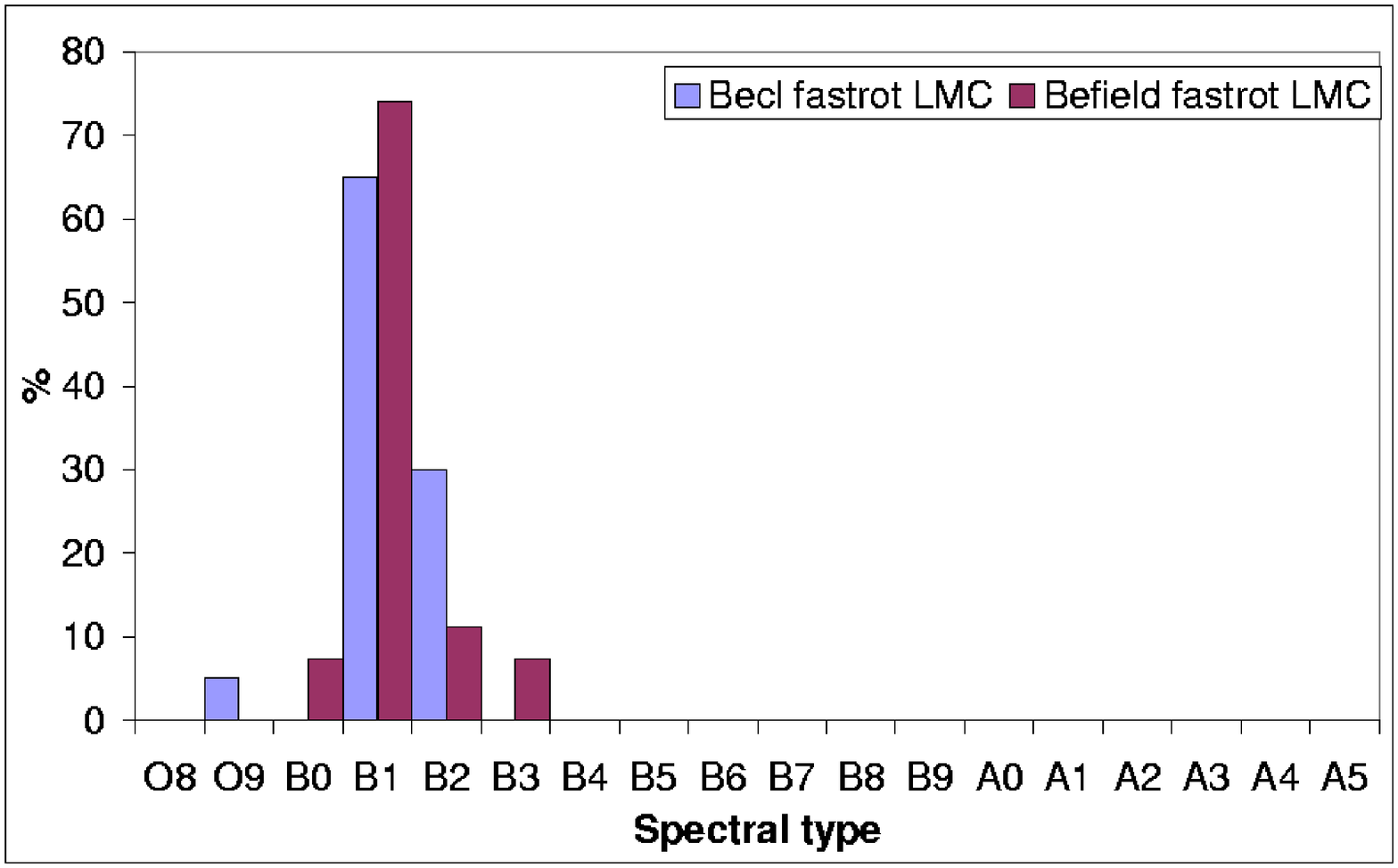}}
\caption{Same as in Fig.~\ref{LCBeLMC},versus spectral type.}
\label{STBeLMC}
\end{figure}

\begin{table*}[ht]
\caption{Same as Table~\ref{table3} for spectroscopic binaries.}
\centering
\small{
\begin{tabular}{@{\ }l@{\ \ \ }l@{\ \ \ }l@{\ \ \ }l@{\ \ \ }l@{\ \ \ }l@{\ \ \ }l@{\ \ \ }l@{\ \ \ }l@{\ \ \ }l@{\ \ \ }l@{\ \ \ }l@{\ \ \ }l@{\ }}
\hline
\hline
Star & $\alpha$ (2000) & $\delta$ (2000) & V & B-V & S/N & \teff  & \logg  & \vsini & \rv & CFP & CEW & comm.\\
\hline
MHF64847 & 5 27 32.040 & -67 21 53.20 & 15.13 & 0.21 & 110 & 26000 $\pm$1300  & 4.1 $\pm$0.2  & 137 $\pm$10  & 347 $\pm$10  & B1V & B1V & \\
MHF65587 & 5 28 23.460 & -67 21 41.30 & 16.01 & 0.23 & 120 &  21500 $\pm$1100  & 3.8 $\pm$0.2  & 173 $\pm$10  & 336 $\pm$10  & B1.5IV & B1V & \\
MHF71137 & 5 28 46.650 & -67 20 40.90 & 15.62 & 0.24 & 107 &  21500 $\pm$1100  & 4.0 $\pm$0.2  & 117 $\pm$10  & 321  $\pm$10  & B1.5V & B1.5IV & \\
MHF79301 & 5 29 11.050 & -67 18 58.90 & 15.61 & 0.23 & 85 &  18000 $\pm$1100  & 3.5 $\pm$0.2  & 39 $\pm$20  & 319 $\pm$10  & B2III-IV & B2III & \\
MHF91603 & 5 28 03.400 & -67 16 33.20 & 15.19 & 0.24 & 140 &  19000 $\pm$1000  & 3.5 $\pm$0.2  & 54 $\pm$10  & 322  $\pm$10  & B2III-IV & B1IV & \\
MHF98013 & 5 30 32.650 & -67 15 25.70 & 14.82 & 0.25 & 120 &  19500 $\pm$1000  & 3.5 $\pm$0.2  & 89 $\pm$10  & 277  $\pm$10  & B2III-IV & B0.5III & \\
MHF102053 & 5 29 56.242 & -67 14 32.95 & 16.26 & 0.31 & 85 &  17000 $\pm$1000  & 3.7 $\pm$0.2  & 188 $\pm$19  & 340 $\pm$10  & B3IV & B3V & \\
MHF103207 & 5 30 13.560 & -67 14 27.10 & 14.94 & 0.23 & 140 &  23500 $\pm$1200  & 3.8 $\pm$0.2  & 122 $\pm$10  & 221 $\pm$10  & B1IV & B1III& \\
MHF109251 & 5 28 31.950 & -67 13 11.90 & 15.95 & 0.25 & 108 &  20500 $\pm$1000  & 3.9  $\pm$0.2 & 5 $\pm$20  & 284 $\pm$10  & B2IV-V & B2IV & cl4\\
MHF110467 & 5 27 25.310 & -67 12 52.50 & 15.70 & 0.17 & 123 &  26500 $\pm$1300  & 4.1 $\pm$0.2  & 31 $\pm$20  & 264  $\pm$10  & B1V & B1V & cl7\\
MHF111340 & 05 27 14.43 & -67 12 40.90 & 16.16 & 0.21 & 90 & & & & & SB2 & SB2 & cl7 \\
MHF113048 & 5 26 57.527 & -67 12 18.95 & 16.38 & 0.21 & 90 &  17000 $\pm$1000  & 3.7 $\pm$0.2  & 75 $\pm$10  & 283  $\pm$10  & B3IV & B2IV & \\
MHF128963 & 5 30 38.290 & -67 9 32.70 & 15.95 & 0.25 & 75 &  20000 $\pm$1600  & 4.0 $\pm$0.3  & 97 $\pm$10  & 314  $\pm$10  & B2V & B2.5V & \\
MHF133975 & 5 29 50.610 & -67 8 30.00 & 16.27 & 0.25 & 80 &  21000 $\pm$1300  & 4.0 $\pm$0.2  & 79 $\pm$10  & 233  $\pm$10  & B2V & B2IV & \\
MHF141891 & 5 28 00.770 & -67 6 55.50 & 16.29 & 0.28 & 90 &  20500 $\pm$1200  & 4.0 $\pm$0.2  & 165 $\pm$12  & 266  $\pm$10  & B2V & B2III & \\
MHF149652 & 5 28 58.672 & -67 5 29.28 & 16.50 & 0.24 & 90 &  20000 $\pm$1200  & 3.9 $\pm$0.2  & 122 $\pm$12  & 341  $\pm$10  & B2IV & B2IV & \\
\hline
\end{tabular}
\label{tablebin}
}
\end{table*}

\subsection{Characteristics of the sample}
\label{charsample}

To characterize the sample of stars, we study the distribution in spectral types,
luminosity classes and masses for stars in clusters and in the field. These results are
presented in the following subsections.

\subsubsection{O-B-A stars}
\label{charBstars}

We present in Fig.~\ref{LCSTBLMC} the distribution of O-B-A stars with respect to
spectral type and luminosity class. The classification used here is the one obtained
from the fundamental parameters determination.

Fig.~\ref{LCSTBLMC} indicates that the B stars in the sample are essentially early
B-type stars (B0 to B3) and are mainly dwarfs (class V), in the field as well as in
clusters.

\begin{figure}[ht]
\centering
\resizebox{\hsize}{!}{\includegraphics[angle=-90,clip]{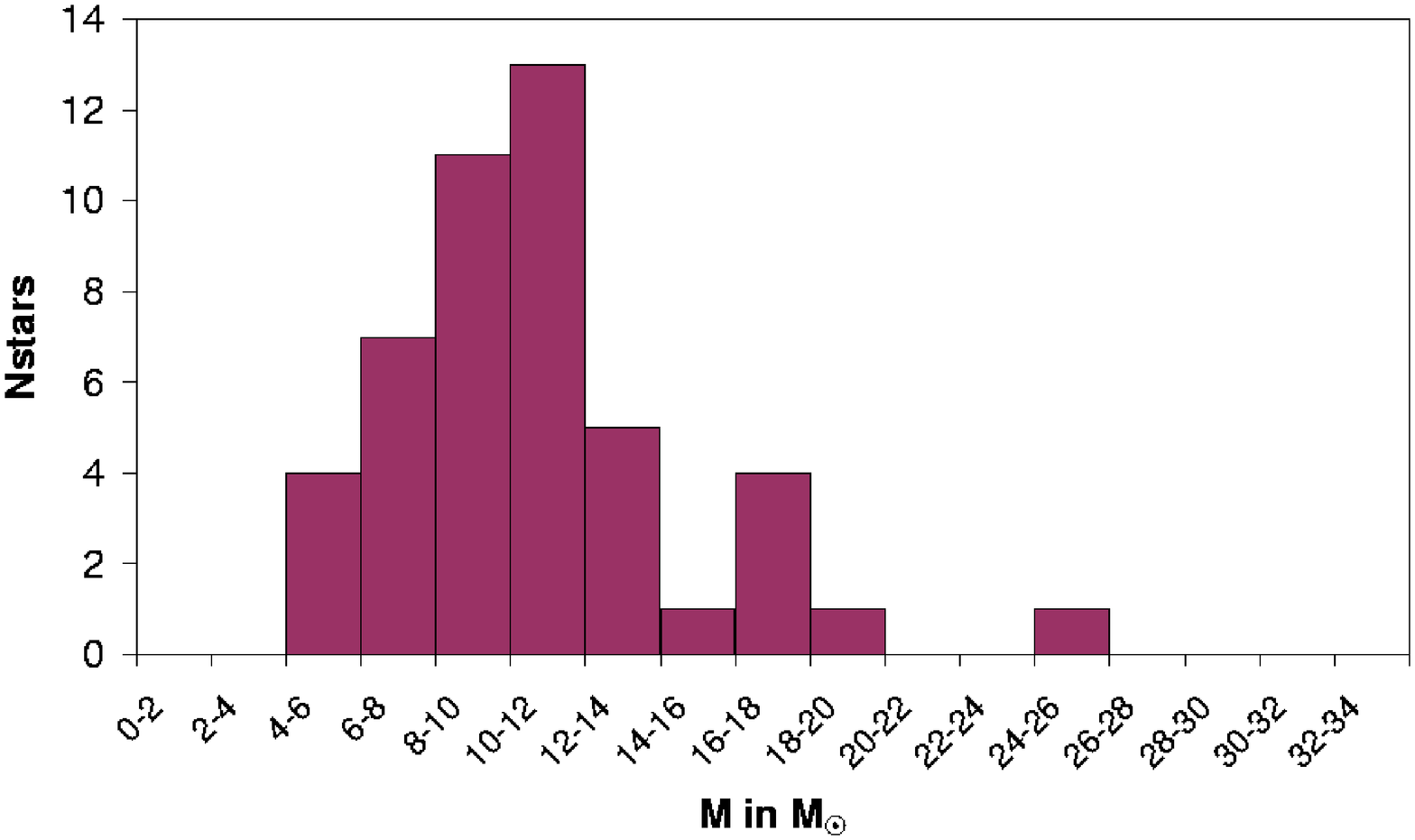}}\\
\resizebox{\hsize}{!}{\includegraphics[angle=-90,clip]{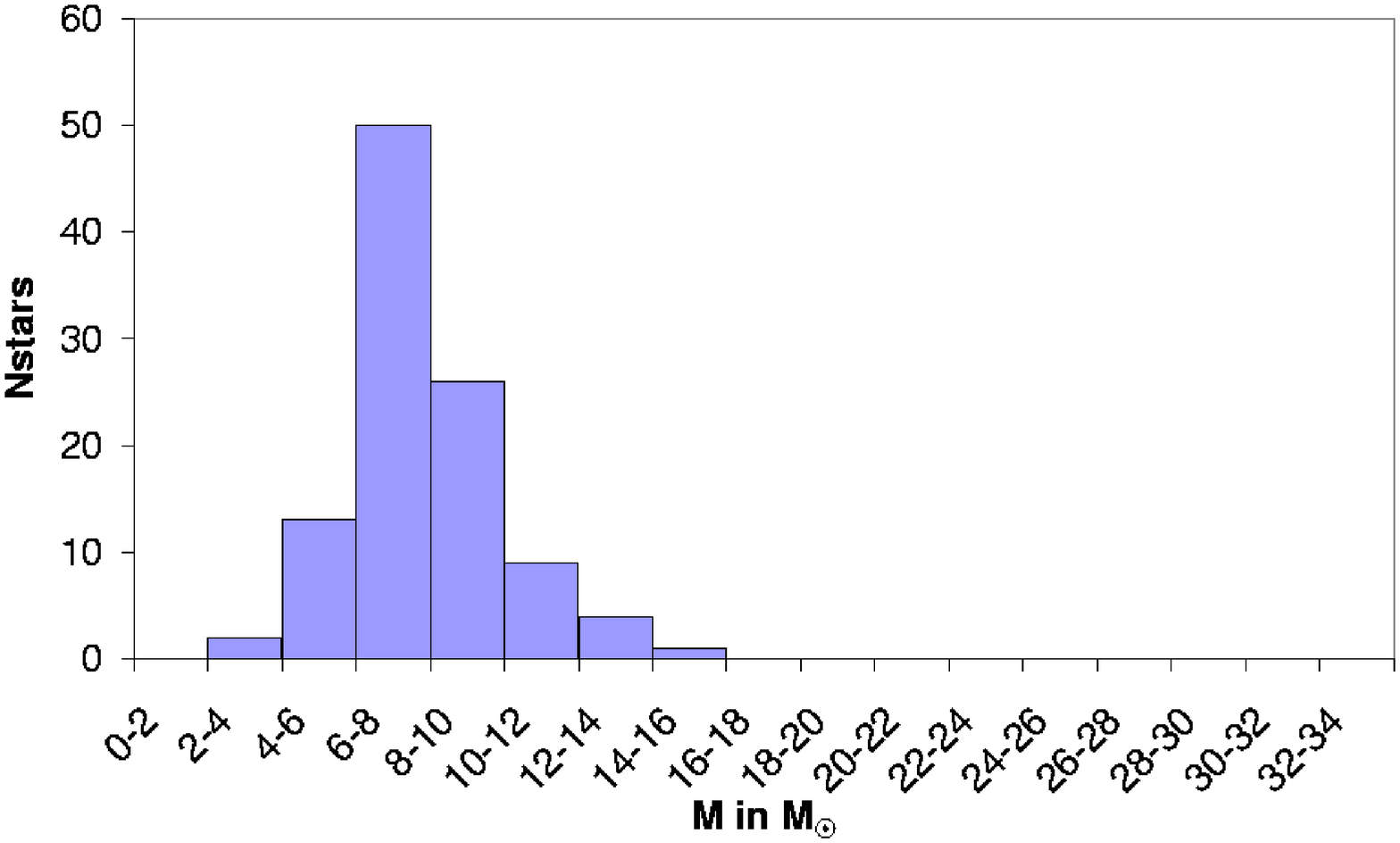}}
\caption{Mass distribution of Be (upper panel) and B stars (lower panel) in the
sample in the LMC.}
\label{fmasse}
\end{figure}

\subsubsection{Be stars}

As in Sect.~\ref{charBstars}, we present the distribution of Be stars with respect to
luminosity class and spectral type. Again the classification used here is the one
obtained from the fundamental parameters determination.  We also compare the
distribution obtained before and after correction of fast rotation effects.

Fig.~\ref{LCBeLMC} shows that Be stars after fast rotation
treatment appear less evolved than apparent parameters would
suggest: some stars in classes III and IV are redistributed in
classes IV and V, but about 60\% of the Be stars still appear as
giants and subgiants.

Fig.~\ref{STBeLMC} presents the distribution in spectral types for Be stars before and
after fast rotation treatment. The stars corrected for rotation effects appear hotter
than apparent fundamental parameters would suggest. In particular there are more B1
types. Nevertheless, in both distributions the sample is composed of early-type stars
(B0 to B3).

\subsubsection{Masses}
\label{masses}

In addition, we investigate the mass distribution of B and Be stars
(Fig.~\ref{fmasse}). The sample shows a distribution peaking around 7
and 10 M$_{\odot}$ for B and Be stars, respectively. Among
the stars with $M$ $\le$ 9 $M_{\odot}$ 18\% are Be stars, whereas
among the stars with $M$ $>$  9 $M_{\odot}$ 62\% are Be stars. This
is probably due to a bias effect in the target selection
procedure, since the sample includes 61\% of Be stars among
stars brighter than V=15.

\begin{table*}[!ht]
\caption{Fundamental parameters of Be stars in the observed sample. Col. 1
gives the name of the star. Coordinates ($\alpha$(2000), $\delta$(2000)) are
given in col. 2 and 3. The instrumental V magnitude and instrumental (B-V)
colour index are given in col. 4 and 5. The S/N ratio is given in col. 6. Col.
from 7 to 10 give the apparent \tap in K, apparent \gap in dex, apparent
\vsiniap in \kms~ and \rv~in \kms. `CFP' (col. 11) gives the spectral type and
luminosity class derived from the fundamental parameters. The last column gives
some complementary remarks about the localization in clusters:
cl0 for NGC\,2004 (05h 30m 42s -67$^{\circ}$ 17$\arcmin$ 11$\arcsec$),
cl1 for KMHK\,988 (05h 30m 36.5s -67$^{\circ}$ 11$\arcmin$ 09$\arcsec$),
cl2 for KMHK\,971 (05h 29m 55s -67$^{\circ}$ 18$\arcmin$ 37$\arcsec$),
cl3 for KMHK\,930 (05h 28m 13s -67$^{\circ}$ 07$\arcmin$ 21$\arcsec$),
cl4 for KMHK\,943 (05h 28m 35s -67$^{\circ}$ 13$\arcmin$ 29$\arcsec$),
cl5 for the `unknown' cluster or association 1 (05h 30m 25s -67$^{\circ}$
13$\arcmin$ 20$\arcsec$),
cl6 for the `unknown' cluster or association 2 (05h 29m 54s -67$^{\circ}$
07$\arcmin$ 37$\arcsec$),
cl7 for the `unknown' cluster or association 3 (05h 27m 21s -67$^{\circ}$
12$\arcmin$ 52$\arcsec$),
cl8 for the association BSDL\,1930 (05h 29m 26s -67$^{\circ}$ 08$\arcmin$
54$\arcsec$)
and cl9 for the galactic open cluster HS\,66325 (05h 29m 36s -67$^{\circ}$
07$\arcmin$ 41$\arcsec$).
}
\centering
\small{
\begin{tabular}{@{\ }l@{\ \ \ }l@{\ \ \ }ll@{\ \ \ }lll@{\ \ \ }ll@{\ \ \ }l@{\ \ \ }l@{\ \ \ }l@{\ }}
\hline
\hline
Star & $\alpha$ (2000) & $\delta$ (2000) & V & B-V & S/N & \tap  & \gap  & \vsiniap & RV & CFP & comm.\\
\hline
KWBBe0044 & 5 30 45.049 & -67 14 26.14 & 13.70 & 0.24 & 140 & 23000 $\pm$1100  & 3.2 $\pm$0.2  & 111 $\pm$10  & 319 $\pm$10  & B2III & \\
KWBBe0075 & 5 30 37.690 & -67 17 39.50 & 14.41 & 0.43 & 90 & 23500 $\pm$1400  & 3.5 $\pm$0.2  & 185 $\pm$13  & 312 $\pm$10  & B1III & cl0\\
KWBBe0091 & 5 30 44.284 & -67 17 22.72 & 14.44 & 0.33 & 140 & 20000 $\pm$1000  & 3.2 $\pm$0.2  & 201 $\pm$10  & 301 $\pm$10  & B2III & cl0\\
KWBBe0152 & 5 30 24.391 & -67 14 55.50 & 15.58 & 0.18 & 103 & 23000 $\pm$1200  & 3.4 $\pm$0.2  & 379 $\pm$19  & 300 $\pm$10  & B2III & \\
KWBBe0171 & 5 30 36.340 & -67 16 51.00 & 15.55 & 0.30 & 110 & 23000 $\pm$1100  & 3.6 $\pm$0.2  & 270 $\pm$14  & 290 $\pm$10  & B2III & cl0\\
KWBBe0177 & 5 30 39.582 & -67 16 49.63 & 15.27 & 0.39 & 83 & 30000 $\pm$1800  & 3.5 $\pm$0.2  & 491: $\pm$50  & 301 $\pm$10  & B0III & cl0\\
KWBBe0203 & 5 30 48.700 & -67 16 49.30 & 15.35 & 0.42 & 80 & 24500 $\pm$1500  & 3.3 $\pm$0.2  & 256 $\pm$25  & 301 $\pm$10  & B1III & cl0\\
KWBBe0276 & 5 29 59.358 & -67 14 46.49 & 15.79 & 0.22 & 119 & 22000 $\pm$1100  & 3.3 $\pm$0.2  & 246 $\pm$12  & 273 $\pm$10  & B2III & \\
KWBBe0287 & 5 30 08.070 & -67 14 36.20 & 15.83 & 0.17 & 90 & 20000 $\pm$1200  & 3.2 $\pm$0.2  & 221 $\pm$15  & 275 $\pm$10  & B2III & \\
KWBBe0323 & 5 30 31.976 & -67 16 40.26 & 16.21 & 0.17 & 90 & 23500 $\pm$1400  & 3.7 $\pm$0.2  & 274 $\pm$19  & 302 $\pm$10  & B1.5III & cl0\\
KWBBe0342 & 5 30 38.630 & -67 16 23.00 & 16.04 & 0.19 & 80 & 23000 $\pm$1400  & 4.1 $\pm$0.2  & 309 $\pm$31  & 323 $\pm$10  & B1.5V & cl0\\
KWBBe0344 & 5 30 38.859 & -67 14 21.13 & 15.89 & 0.18 & 100 & 22500 $\pm$1100  & 3.8 $\pm$0.2  & 216 $\pm$11  & 301 $\pm$10  & B2IV & \\
KWBBe0347 & 5 30 39.900 & -67 12 21.01 & 16.04 & 0.17 & 95 & 22500 $\pm$1400  & 3.5 $\pm$0.2  & 334 $\pm$23  & 302 $\pm$10  & B2III & \\
KWBBe0374 & 5 30 47.799 & -67 11 36.43 & 16.37 & 0.30 & 60 & 22000 $\pm$2200  & 4.1 $\pm$0.4  & 337 $\pm$54  & 339 $\pm$10  & B2V & \\
KWBBe0579 & 5 30 27.610 & -67 13 00.21 & 16.53 & 0.15 & 55 & 19000 $\pm$2300  & 3.5 $\pm$0.4  & 310 $\pm$50  & 301 $\pm$10  & B2III & cl5\\
KWBBe0622 & 5 30 36.223 & -67 13 27.71 & 16.89 & 0.25 & 60 & 16000 $\pm$1600  & 3.5 $\pm$0.3  & 299 $\pm$48  & 302 $\pm$10  & B3III & \\
KWBBe0624 & 5 30 37.040 & -67 21 02.62 & 16.73 & 0.19 & 54 & 19670 $\pm$2400  & 3.4 $\pm$0.3  & 343 $\pm$55  & 275 $\pm$10  & B2III & \\
KWBBe0874 & 5 29 55.846 & -67 19 14.27 & 17.56 & 0.16 & 45 & 19000 $\pm$2900  & 4.1 $\pm$0.4  & 262 $\pm$52  & 300 $\pm$10  & B2V & \\
KWBBe1055 & 5 30 29.265 & -67 17 01.65 & 17.41 & 0.17 & 50 & 22000 $\pm$2600  & 3.9 $\pm$0.4  & 419 $\pm$67  & 300 $\pm$10  & B2IV & cl0\\
KWBBe1108 & 5 30 37.238 & -67 11 29.31 & 17.64 & 0.20 & 50 & 19500 $\pm$2300  & 4.0 $\pm$0.4  & 316 $\pm$51  & 300 $\pm$10  & B2V & cl1\\
KWBBe1175 & 5 30 46.960 & -67 17 39.30 & 17.75 & 0.12 & 25 & 21500 $\pm$4300  & 4.2 $\pm$0.4  & 345 $\pm$100  & 297 $\pm$10  & B2V & cl0\\
KWBBe1196 & 5 30 50.893 & -67 18 08.51 & 17.56 & 0.16 & 45 & 17000 $\pm$2200  & 3.8 $\pm$0.4  & 359 $\pm$65  & 299 $\pm$10  & B3IV & cl0\\
MHFBe55075 & 5 30 15.660 & -67 23 54.60 & 16.01 & 0.26 & 50 & 21500 $\pm$2600  & 3.8 $\pm$0.4  & 266 $\pm$43  & 301 $\pm$10  & B2IV & \\
MHFBe55920 & 5 27 42.256 & -67 23 36.58 & 16.08 & 0.16 & 80 & 23500 $\pm$1400  & 4.0 $\pm$0.2  & 120 $\pm$12  & 304 $\pm$10  & B1V & \\
MHFBe59721 & 5 29 25.670 & -67 23 02.40 & 16.46 & 0.18 & 85 & 18000 $\pm$1090  & 4.0 $\pm$0.2  & 196 $\pm$18  & 299 $\pm$10  & B2V & \\
MHFBe66252 & 5 29 18.200 & -67 21 37.10 & 16.26 & 0.17 & 80 & 25000 $\pm$1500  & 4.0 $\pm$0.2  & 446 $\pm$45  & 274 $\pm$10  & B1V & \\
MHFBe72704 & 5 28 29.430 & -67 20 20.10 & 14.86 & 0.19 & 150 & 23000 $\pm$1100  & 3.5 $\pm$0.2  & 108 $\pm$10  & 304 $\pm$10  & B1.5III-IV & \\
MHFBe73013 & 5 28 55.994 & -67 20 14.55 & 16.16 & 0.19 & 100 & 25000 $\pm$120  & 4.3 $\pm$0.2  & 402 $\pm$20  & 302 $\pm$10  & B1V & \\
MHFBe77796 & 5 29 13.609 & -67 19 18.70 & 14.91 & 0.18 & 90 & 25500 $\pm$1500  & 3.7 $\pm$0.2  & 120 $\pm$10  & 311 $\pm$10  & B1III-IV & \\
MHFBe85028 & 5 29 45.000 & -67 17 50.50 & 16.07 & 0.17 & 100 & 23000 $\pm$1200  & 3.8 $\pm$0.2  & 269 $\pm$13  & 316 $\pm$10  & B1.5IV & \\
MHFBe101350 & 5 30 34.640 & -67 14 45.30 & 15.30 & 0.19 & 120 & 24500 $\pm$1200  & 3.8 $\pm$0.2  & 224 $\pm$11  & 285 $\pm$10  & B1IV & \\
MHFBe103914 & 5 27 15.216 & -67 14 10.15 & 15.03 & 0.17 & 130 & 26500 $\pm$1300  & 3.4 $\pm$0.2  & 368 $\pm$18  & 300 $\pm$10  & B1III & \\
MHFBe107771 & 5 28 34.651 & -67 13 23.97 & 16.39 & 0.37 & 84 & 18000 $\pm$1100  & 3.6 $\pm$0.2  & 319 $\pm$29  & 300 $\pm$10  & B2.5III-IV & cl4\\
MHFBe107877 & 5 28 34.260 & -67 13 25.90 & 15.23 & 0.21 & 90 & 20500 $\pm$1200  & 3.4 $\pm$0.2  & 208  $\pm$15 & 319 $\pm$10  & B2III & cl4\\
MHFBe108272 & 5 28 32.127 & -67 13 23.68 & 16.40 & 0.20 & 100 & 21500 $\pm$1100  & 3.9 $\pm$0.2  & 226 $\pm$11  & 320 $\pm$10  & B2IV & cl4\\
MHFBe110827 & 5 27 28.582 & -67 12 56.81 & 14.93 & 0.23 & 130 & 26500 $\pm$1300  & 4.2 $\pm$0.2  & 195 $\pm$14  & 305 $\pm$10  & B1V & cl7\\
MHFBe116297 & 5 28 06.021 & -67 11 47.44 & 16.14 & 0.16 & 110 & 23500 $\pm$1200  & 4.0 $\pm$0.2  & 226 $\pm$11  & 319 $\pm$10  & B1V & \\
MHFBe118313 & 5 26 55.945 & -67 11 27.10 & 15.04 & 0.18 & 85 & 25000 $\pm$1500  & 3.4 $\pm$0.2  & 197 $\pm$20  & 301 $\pm$10  & B1III & \\
MHFBe118784 & 5 28 27.565 & -67 11 28.50 & 14.79 & 0.18 & 185 & 27000 $\pm$1400  & 3.4 $\pm$0.2  & 388 $\pm$19  & 296 $\pm$10  & B1III & \\
MHFBe119521 & 5 30 35.252 & -67 11 19.10 & 16.30 & 0.17 & 94 & 24000 $\pm$1400  & 3.7 $\pm$0.2  & 321 $\pm$22  & 341 $\pm$10  & B1III & cl1 \\
MHFBe132079 & 5 30 05.496 & -67 8 53.45 & 16.24 & 0.20 & 80 & 25000 $\pm$1500  & 4.4 $\pm$0.3  & 318 $\pm$32  & 305 $\pm$10  & B1V & \\
MHFBe132205 & 5 29 25.872 & -67 8 53.93 & 14.72 & 0.18 & 134 & 26000 $\pm$1300  & 3.4 $\pm$0.2  & 129 $\pm$10  & 301 $\pm$10  & B1III & cl8\\
MHFBe136844 & 5 27 35.241 & -67 7 54.44 & 15.15 & 0.19 & 110 & 23000 $\pm$1200  & 3.5 $\pm$0.2  & 348 $\pm$17  & 315 $\pm$10  & B1III-IV & \\
MHFBe137325 & 5 29 07.009 & -67 7 49.28 & 16.30 & 0.16 & 110 & 15000 $\pm$800  & 3.1 $\pm$0.2  & 252 $\pm$13  & 320 $\pm$10  & B5III & \\
MHFBe138610 & 5 30 00.332 & -67 7 44.02 & 14.82 & 0.17 & 129 & 25000 $\pm$1200  & 3.7 $\pm$0.2  & 171 $\pm$10  & 314 $\pm$10  & B1IV & cl6\\
MHFBe140012 & 5 28 13.281 & -67 7 20.77 & 14.94 & 0.18 & 130 & 23000 $\pm$1200  & 3.5 $\pm$0.2  & 246 $\pm$12  & 297 $\pm$10  & B1.5III & cl3\\
MHFBe155603 & 5 28 47.760 & -67 4 25.70 & 14.72 & 0.20 & 100 & 25000 $\pm$1300  & 3.4 $\pm$0.2  & 249 $\pm$12  & 308 $\pm$10  & B1III & \\
\hline
\end{tabular}
\label{tableBe}
}
\end{table*}

\begin{table*}[ht]
\caption{Stellar parameters of the LMC Be stars corrected for the effects of
fast rotation assuming different rotation rates (\omc). The probably most
suitable corrections are those corresponding to \omc~= 85\%. The units are K
for \top, dex for \gop, and \kms~for \vsinit.}
\centering
\small{
\begin{tabular}{l@{\ \ }lll@{\ }l@{\ \ }lll@{\ }l@{\ \ }lll}
\hline
\hline
Star & \multicolumn{3}{c}{\omc = 85\%} & \vline & \multicolumn{3}{c}{\omc = 90\%} & \vline & \multicolumn{3}{c}{\omc = 95\%} \\
     & \top  & \gop & \vsinit & \vline & \top  & \gop & \vsinit & \vline &\top  & \gop & \vsinit \\
\hline
KWBBe0044 & 23000$\pm$1100 & 3.2$\pm$0.2 & 119$\pm$10 & \vline & 23390$\pm$1100 & 3.3$\pm$0.2 & 121$\pm$10 & \vline & 23300$\pm$1100 & 3.3$\pm$0.2 & 121$\pm$10 \\
KWBBe0075 & 24000$\pm$1400 & 3.6$\pm$0.2 & 193$\pm$13 & \vline & 24450$\pm$1400 & 3.6$\pm$0.2 & 195$\pm$13 & \vline & 23950$\pm$1400 & 3.6$\pm$0.2 & 198$\pm$13 \\
KWBBe0091 & 21000$\pm$1000 & 3.4$\pm$0.2 & 211$\pm$10 & \vline & 21175$\pm$1000 & 3.4$\pm$0.2 & 211$\pm$10 & \vline & 21310$\pm$1000 & 3.5$\pm$0.2 & 217$\pm$10 \\
KWBBe0152 & 25500$\pm$1200 & 3.8$\pm$0.2 & 394$\pm$19 & \vline & 25900$\pm$1200 & 3.8$\pm$0.2 & 388$\pm$19 & \vline & 25490$\pm$1200 & 3.8$\pm$0.2 & 400$\pm$19 \\
KWBBe0171 & 24500$\pm$1200 & 3.8$\pm$0.2 & 275$\pm$14 & \vline & 24180$\pm$1200 & 3.9$\pm$0.2 & 277$\pm$14 & \vline & 24155$\pm$1200 & 3.8$\pm$0.2 & 284$\pm$14 \\
KWBBe0177 & 33500$\pm$1800 & 3.9$\pm$0.2 & 508:$\pm$50 & \vline & 33400$\pm$1800 & 3.9$\pm$0.2 & 516:$\pm$50 & \vline & 33740$\pm$1800 & 4.0$\pm$0.2 & 527:$\pm$50 \\
KWBBe0203 & 25500$\pm$1500 & 3.6$\pm$0.2 & 268$\pm$25 & \vline & 26350$\pm$1500 & 3.5$\pm$0.2 & 268$\pm$25 & \vline & 25970$\pm$1500 & 3.5$\pm$0.2 & 269$\pm$25 \\
KWBBe0276 & 23500$\pm$1100 & 3.5$\pm$0.2 & 253$\pm$12 & \vline & 23360$\pm$1100 & 3.5$\pm$0.2 & 257$\pm$12 & \vline & 23120$\pm$1100 & 3.5$\pm$0.2 & 261$\pm$12 \\
KWBBe0287 & 21000$\pm$1200 & 3.4$\pm$0.2 & 230$\pm$15 & \vline & 21190$\pm$1200 & 3.5$\pm$0.2 & 234$\pm$15 & \vline & 21250$\pm$1200 & 3.5$\pm$0.2 & 237$\pm$15 \\
KWBBe0323 & 25000$\pm$1400 & 3.9$\pm$0.2 & 282$\pm$19 & \vline & 24535$\pm$1400 & 3.9$\pm$0.2 & 281$\pm$19 & \vline & 24610$\pm$1400 & 3.9$\pm$0.2 & 283$\pm$19 \\
KWBBe0342 & 24000$\pm$1400 & 4.3$\pm$0.2 & 316$\pm$31 & \vline & 23450$\pm$1400 & 4.3$\pm$0.2 & 322$\pm$31 & \vline & 24545$\pm$1400 & 4.3$\pm$0.2 & 321$\pm$31 \\
KWBBe0344 & 23500$\pm$1100 & 4.0$\pm$0.2 & 225$\pm$11 & \vline & 23450$\pm$1100 & 4.0$\pm$0.2 & 225$\pm$11 & \vline & 23370$\pm$1100 & 4.0$\pm$0.2 & 228$\pm$11 \\
KWBBe0347 & 25000$\pm$1300 & 3.8$\pm$0.2 & 344$\pm$23 & \vline & 25240$\pm$1300 & 3.8$\pm$0.2 & 343$\pm$23 & \vline & 25470$\pm$1300 & 3.9$\pm$0.2 & 349$\pm$23 \\
KWBBe0374 & 24000$\pm$2200 & 4.3$\pm$0.4 & 345$\pm$54 & \vline & 23880$\pm$2200 & 4.3$\pm$0.4 & 347$\pm$54 & \vline & 24080$\pm$2200 & 4.4$\pm$0.4 & 352$\pm$54 \\
KWBBe0579 & 20500$\pm$2300 & 3.8$\pm$0.4 & 320$\pm$50 & \vline & 21000$\pm$2300 & 3.9$\pm$0.4 & 323$\pm$50 & \vline & 21060$\pm$2300 & 3.9$\pm$0.4 & 332$\pm$50 \\
KWBBe0622 & 18000$\pm$1600 & 4.0$\pm$0.3 & 315$\pm$48 & \vline & 17240$\pm$1600 & 3.9$\pm$0.3 & 314$\pm$48 & \vline & 17700$\pm$1600 & 3.9$\pm$0.3 & 335$\pm$48 \\
KWBBe0624 & 22000$\pm$2400 & 3.8$\pm$0.3 & 356$\pm$55 & \vline & 22690$\pm$2400 & 3.8$\pm$0.3 & 353$\pm$55 & \vline & 22880$\pm$2400 & 3.8$\pm$0.3 & 359$\pm$55 \\
KWBBe0874 & 20000$\pm$2900 & 4.3$\pm$0.4 & 270$\pm$52 & \vline & 20380$\pm$2900 & 4.3$\pm$0.4 & 275$\pm$52 & \vline & 20610$\pm$2900 & 4.4$\pm$0.4 & 280$\pm$52 \\
KWBBe1055 & 25000$\pm$2600 & 4.2$\pm$0.4 & 433$\pm$67 & \vline & 24600$\pm$2600 & 4.3$\pm$0.4 & 441$\pm$67 & \vline & 24710$\pm$2600 & 4.3$\pm$0.4 & 441$\pm$67 \\
KWBBe1108 & 20500$\pm$2300 & 4.3$\pm$0.4 & 326$\pm$51 & \vline & 20860$\pm$2300 & 4.3$\pm$0.4 & 332$\pm$51 & \vline & 20970$\pm$2300 & 4.4$\pm$0.4 & 338$\pm$51 \\
KWBBe1175 & 23000$\pm$4300 & 4.5$\pm$0.4 & 354$\pm$100 & \vline & 23030$\pm$4300 & 4.5$\pm$0.4 & 359$\pm$100 & \vline & 23260$\pm$4300 & 4.5$\pm$0.4 & 364$\pm$100 \\
KWBBe1196 & 19000$\pm$2200 & 4.2$\pm$0.4 & 378$\pm$65 & \vline & 18980$\pm$2200 & 4.2$\pm$0.4 & 386$\pm$65 & \vline & 19120$\pm$2200 & 4.2$\pm$0.4 & 394$\pm$65 \\
MHFBe55075 & 23000$\pm$2600 & 3.9$\pm$0.4 & 274$\pm$43 & \vline & 23100$\pm$2600 & 4.0$\pm$0.4 & 276$\pm$43 & \vline & 23165$\pm$2600 & 4.0$\pm$0.4 & 282$\pm$43 \\
MHFBe55920 & 24000$\pm$1400 & 4.1$\pm$0.2 & 129$\pm$12 & \vline & 23980$\pm$1400 & 4.1$\pm$0.2 & 128$\pm$12 & \vline & 23930$\pm$1400 & 4.1$\pm$0.2 & 129$\pm$12 \\
MHFBe59721 & 19000$\pm$1100 & 4.2$\pm$0.2 & 208$\pm$18 & \vline & 19190$\pm$1100 & 4.2$\pm$0.2 & 212$\pm$18 & \vline & 19240$\pm$1100 & 4.2$\pm$0.2 & 214$\pm$18 \\
MHFBe66252 & 28000$\pm$1500 & 4.3$\pm$0.2 & 459$\pm$45 & \vline & 27880$\pm$1500 & 4.3$\pm$0.2 & 465$\pm$45 & \vline & 28100$\pm$1500 & 4.4$\pm$0.2 & 473$\pm$45 \\
MHFBe72704 & 23500$\pm$1100 & 3.6$\pm$0.2 & 114$\pm$10 & \vline & 23380$\pm$1100 & 3.6$\pm$0.2 & 117$\pm$10 & \vline & 23370$\pm$1100 & 3.6$\pm$0.2 & 117$\pm$10 \\
MHFBe73013 & 27000$\pm$1200 & 4.6:$\pm$0.2 & 411$\pm$20 & \vline & 27100$\pm$1200 & 4.6:$\pm$0.2 & 422$\pm$20 & \vline & 27250$\pm$1200 & 4.6:$\pm$0.2 & 422$\pm$20 \\
MHFBe77796 & 26500$\pm$1500 & 3.7$\pm$0.2 & 126$\pm$10 & \vline & 26050$\pm$1500 & 3.7$\pm$0.2 & 130$\pm$10 & \vline & 26060$\pm$1500 & 3.7$\pm$0.2 & 129$\pm$10 \\
MHFBe85028 & 24500$\pm$1200 & 4.0$\pm$0.2 & 277$\pm$13 & \vline & 24240$\pm$1200 & 4.0$\pm$0.2 & 279$\pm$13 & \vline & 24350$\pm$1200 & 4.0$\pm$0.2 & 282$\pm$13 \\
MHFBe101350 & 25500$\pm$1200 & 3.9$\pm$0.2 & 230$\pm$11 & \vline & 25270$\pm$1200 & 3.9$\pm$0.2 & 232$\pm$11 & \vline & 25475$\pm$1200 & 3.9$\pm$0.2 & 234$\pm$11 \\
MHFBe103914 & 29000$\pm$1300 & 3.8$\pm$0.2 & 382$\pm$18 & \vline & 27270$\pm$1300 & 3.6$\pm$0.2 & 385$\pm$18 & \vline & 28930$\pm$1300 & 3.8$\pm$0.2 & 394$\pm$18 \\
MHFBe107771 & 20000$\pm$1100 & 4.0$\pm$0.2 & 331$\pm$29 & \vline & 19680$\pm$1100 & 4.0$\pm$0.2 & 333$\pm$29 & \vline & 19890$\pm$1100 & 4.0$\pm$0.2 & 341$\pm$29 \\
MHFBe107877 & 22000$\pm$1200 & 3.6$\pm$0.2 & 216$\pm$15 & \vline & 22040$\pm$1200 & 3.7$\pm$0.2 & 219$\pm$15 & \vline & 22130$\pm$1200 & 3.7$\pm$0.2 & 222$\pm$15 \\
MHFBe108272 & 22500$\pm$1100 & 4.1$\pm$0.2 & 235$\pm$11 & \vline & 22420$\pm$1100 & 4.1$\pm$0.2 & 236$\pm$11 & \vline & 22550$\pm$1100 & 4.1$\pm$0.2 & 237$\pm$11 \\
MHFBe110827 & 27000$\pm$1300 & 4.3$\pm$0.2 & 201$\pm$14 & \vline & 26620$\pm$1300 & 4.3$\pm$0.2 & 205$\pm$14 & \vline & 27400$\pm$1300 & 4.3$\pm$0.2 & 205$\pm$14 \\
MHFBe116297 & 25000$\pm$1200 & 4.1$\pm$0.2 & 233$\pm$11 & \vline & 24310$\pm$1200 & 4.1$\pm$0.2 & 234$\pm$11 & \vline & 24310$\pm$1200 & 4.1$\pm$0.2 & 237$\pm$11 \\
MHFBe118313 & 26000$\pm$1500 & 3.5$\pm$0.2 & 207$\pm$20 & \vline & 26380$\pm$1500 & 3.6$\pm$0.2 & 208$\pm$20 & \vline & 27060$\pm$1500 & 3.6$\pm$0.2 & 209$\pm$20 \\
MHFBe118784 & 30000$\pm$1400 & 3.8$\pm$0.2 & 404$\pm$19 & \vline & 28400$\pm$1400 & 3.7$\pm$0.2 & 406$\pm$19 & \vline & 30470$\pm$1400 & 3.8$\pm$0.2 & 416$\pm$19 \\
MHFBe119521 & 25500$\pm$1400 & 4.0$\pm$0.2 & 330$\pm$22 & \vline & 24370$\pm$1400 & 3.9$\pm$0.2 & 334$\pm$22 & \vline & 25830$\pm$1400 & 4.0$\pm$0.2 & 337$\pm$22 \\
MHFBe132079 & 25500$\pm$1500 & 4.5:$\pm$0.3 & 326$\pm$32 & \vline & 26985$\pm$1500 & 4.6:$\pm$0.3 & 335$\pm$32 & \vline & 26495$\pm$1500 & 4.6:$\pm$0.3 & 332$\pm$32 \\
MHFBe132205 & 27000$\pm$1300 & 3.6$\pm$0.2 & 137$\pm$10 & \vline & 26950$\pm$1300 & 3.6$\pm$0.2 & 136$\pm$10 & \vline & 28145$\pm$1300 & 3.7$\pm$0.2 & 139$\pm$10 \\
MHFBe136844 & 25500$\pm$1200 & 3.8$\pm$0.2 & 360$\pm$17 & \vline & 26240$\pm$1200 & 3.9$\pm$0.2 & 362$\pm$17 & \vline & 25320$\pm$1200 & 3.8$\pm$0.2 & 365$\pm$17 \\
MHFBe137325 & 17000$\pm$800 & 3.4$\pm$0.2 & 263$\pm$13 & \vline & 16550$\pm$800 & 3.5$\pm$0.2 & 271$\pm$13 & \vline & 16330$\pm$800 & 3.4$\pm$0.2 & 276$\pm$13 \\
MHFBe138610 & 25500$\pm$1200 & 3.8$\pm$0.2 & 178$\pm$10 & \vline & 25210$\pm$1200 & 3.8$\pm$0.2 & 181$\pm$10 & \vline & 25560$\pm$1200 & 3.8$\pm$0.2 & 183$\pm$10 \\
MHFBe140012 & 24500$\pm$1200 & 3.7$\pm$0.2 & 252$\pm$12 & \vline & 24400$\pm$1200 & 3.7$\pm$0.2 & 254$\pm$12 & \vline & 24470$\pm$1200 & 3.7$\pm$0.2 & 257$\pm$12 \\
MHFBe155603 & 26600$\pm$1300 & 3.5$\pm$0.2 & 261$\pm$12 & \vline & 27220$\pm$1300 & 3.6$\pm$0.2 & 261$\pm$12 & \vline & 27525$\pm$1300 & 3.6$\pm$0.2 & 263$\pm$12 \\
\hline
\end{tabular}
\label{tabBefastrot}
}
\end{table*}

\section{Rotational velocity and metallicity: results and discussion}

In the following subsections, we first give some preliminary remarks about the study by
Gies \& Huang (2004) on B-type stars in clusters of the Milky Way (MW), which are used
as a central thread in our study. Then we summarize results on \vsini~we obtained for B
and Be stars in the LMC. Finally, we compare these results with previous studies in the
LMC and in the MW, and we discuss the effect of age and metallicity on rotational
velocity.

\begin{table*}[ht]
\small{
\caption{Apparent parameters $\log(L/L_{\odot}$), $M/M_{\odot}$, and $R/R_{\odot}$
interpolated or calculated for the sample of Be stars from HR diagrams taken
from Charbonnel et al. (1993).}
\centering
\begin{tabular}{@{\ }lllllllll@{\ }}
\hline
\hline
Star & $\log(L/L_{\odot})$ & $M/M_{\odot}$ & $R/R_{\odot}$ & \vline & Star & $\log(L/L_{\odot})$ & $M/M_{\odot}$ & $R/R_{\odot}$ \\
\hline
KWBBe0044 & 4.9$\pm$0.3 & 16.8$\pm$1.0 & 18.0$\pm$2.0 & \vline & MHFBe59721 & 3.2$\pm$0.3 & 5.6$\pm$0.5 & 4.0$\pm$0.5 \\
KWBBe0075 & 4.5$\pm$0.3 & 12.3$\pm$1.0 & 11.0$\pm$1.5 & \vline & MHFBe66252 & 4.0$\pm$0.3 & 10.1$\pm$1.0 & 5.4$\pm$1.0 \\
KWBBe0091 & 4.4$\pm$0.3 & 11.7$\pm$1.0 & 14.1$\pm$1.5 & \vline & MHFBe72704 & 4.4$\pm$0.3 & 11.3$\pm$1.0 & 10.1$\pm$1.0 \\
KWBBe0152 & 4.5$\pm$0.3 & 12.9$\pm$1.0 & 12.2$\pm$1.5 & \vline & MHFBe73013 & 3.5$\pm$0.3 & 8.3$\pm$0.5 & 3.3$\pm$0.5 \\
KWBBe0171 & 4.2$\pm$0.3 & 10.3$\pm$1.0 & 8.1$\pm$1.0 & \vline & MHFBe77796 & 4.4$\pm$0.3 & 12.4$\pm$1.0 & 8.5$\pm$1.0 \\
KWBBe0177 & 5.2$\pm$0.3 & 24.4$\pm$2.0 & 14.5$\pm$1.5 & \vline & MHFBe85028 & 4.0$\pm$0.3 & 9.6$\pm$0.5 & 6.5$\pm$1.0 \\
KWBBe0203 & 4.6$\pm$0.3 & 13.6$\pm$1.0 & 10.3$\pm$1.5 & \vline & MHFBe101350 & 4.2$\pm$0.3 & 11.0$\pm$1.0 & 7.1$\pm$1.0 \\
KWBBe0276 & 4.5$\pm$0.3 & 12.8$\pm$1.0 & 13.5$\pm$1.5 & \vline & MHFBe103914 & 4.9$\pm$0.3 & 16.8$\pm$1.0 & 13.0$\pm$1.5 \\
KWBBe0287 & 4.3$\pm$0.3 & 10.6$\pm$1.0 & 10.8$\pm$1.5 & \vline & MHFBe107771 & 3.6$\pm$0.3 & 6.7$\pm$0.5 & 6.9$\pm$1.0 \\
KWBBe0323 & 4.2$\pm$0.3 & 10.5$\pm$1.0 & 7.4$\pm$1.0 & \vline & MHFBe107877 & 4.2$\pm$0.3 & 10.1$\pm$1.0 & 10.5$\pm$1.5 \\
KWBBe0342 & 3.6$\pm$0.3 & 7.9$\pm$0.5 & 4.2$\pm$0.5 & \vline & MHFBe108272 & 3.7$\pm$0.3 & 7.7$\pm$0.5 & 5.0$\pm$0.5 \\
KWBBe0344 & 3.9$\pm$0.3 & 8.8$\pm$0.5 & 6.0$\pm$1.0 & \vline & MHFBe110827 & 3.8$\pm$0.3 & 9.7$\pm$0.5 & 3.9$\pm$0.5 \\
KWBBe0347 & 4.4$\pm$0.3 & 11.4$\pm$1.0 & 10.8$\pm$1.5 & \vline & MHFBe116297 & 3.8$\pm$0.3 & 8.8$\pm$0.5 & 5.1$\pm$0.5 \\
KWBBe0374 & 3.6$\pm$0.3 & 7.5$\pm$0.5 & 4.1$\pm$0.5 & \vline & MHFBe118313 & 4.8$\pm$0.3 & 15.6$\pm$1.0 & 13.6$\pm$1.5 \\
KWBBe0579 & 3.9$\pm$0.3 & 8.1$\pm$0.5 & 8.7$\pm$1.0 & \vline & MHFBe118784 & 5.0$\pm$0.3 & 18.6$\pm$1.0 & 14.4$\pm$1.5 \\
KWBBe0622 & 3.4$\pm$0.3 & 5.9$\pm$0.5 & 7.1$\pm$1.0 & \vline & MHFBe119521 & 4.2$\pm$0.3 & 10.7$\pm$1.0 & 7.8$\pm$1.0 \\
KWBBe0624 & 4.2$\pm$0.3 & 9.9$\pm$0.5 & 11.3$\pm$1.5 & \vline & MHFBe132079 & 3.5$\pm$0.3 & 8.2$\pm$0.5 & 3.1$\pm$0.5 \\
KWBBe0874 & 3.2$\pm$0.3 & 5.8$\pm$0.5 & 3.7$\pm$0.5 & \vline & MHFBe132205 & 4.8$\pm$0.3 & 16.3$\pm$1.0 & 13.0$\pm$1.5 \\
KWBBe1055 & 3.8$\pm$0.3 & 8.2$\pm$0.5 & 5.3$\pm$1.0 & \vline & MHFBe136844 & 4.4$\pm$0.3 & 11.7$\pm$1.0 & 10.5$\pm$1.5 \\
KWBBe1108 & 3.3$\pm$0.3 & 6.0$\pm$0.5 & 4.0$\pm$0.5 & \vline & MHFBe137325 & 3.8$\pm$0.3 & 7.4$\pm$0.5 & 12.6$\pm$1.5 \\
KWBBe1175 & 3.3$\pm$0.3 & 6.5$\pm$0.5 & 3.3$\pm$0.5 & \vline & MHFBe138610 & 4.3$\pm$0.3 & 11.6$\pm$1.0 & 7.9$\pm$1.0 \\
KWBBe1196 & 3.3$\pm$0.3 & 5.6$\pm$0.5 & 5.2$\pm$1.0 & \vline & MHFBe140012 & 4.4$\pm$0.3 & 11.5$\pm$1.0 & 10.0$\pm$1.0 \\
MHFBe55075 & 3.9$\pm$0.3 & 8.2$\pm$0.5 & 6.3$\pm$1.0 & \vline & MHFBe155603 & 4.9$\pm$0.3 & 16.7$\pm$1.0 & 14.3$\pm$1.5 \\
MHFBe55920 & 3.8$\pm$0.3 & 8.7$\pm$0.5 & 4.9$\pm$0.5 \\
\hline
\end{tabular}
\label{tabLMRBe}
}
\end{table*}

\subsection{Ages, \vsini, and Be stars}

Gies \& Huang (2004, hereafter GH04) studied the link between
rotational velocity and age in clusters of the MW. They noted a
good agreement for the rotational velocity between data
and predictions by Meynet \& Maeder (2000) for a 12 M$_{\odot}$
star and for clusters with $\log(t)$ $\le$ 7. However,
clusters with $\log(t)$ $ >$ 7 seem to rotate faster than
predicted. According to GH04, there could be several
explanations: binarity, initial spin rates, and rotational
velocity dependence on mass. For $\log(t)$ $\le$ 7 the
mean fraction of binaries calculated from the fraction
given by GH04 for each cluster is $\frac{SB}{all}$ = 15\%,
whereas for $\log(t)$ $>$ 7, $\frac{SB}{all}$ = 19\%. Therefore,
binaries do not seem to be at the origin of the difference in
rotational velocity.

As GH04 merged B and Be stars in their sample and as Be stars are
fast rotators, a possible explanation of the differences in
rotational velocity may be the proportion of Be stars in the
clusters. We therefore searched in the WEBDA
database\footnote{The WEBDA database is maintained by J.C.
Mermilliod. See http://obswww.unige.ch/webda/navigation.html} for
the amount of Be stars in the clusters studied by GH04 and
calculated the average  percentages of Be stars for clusters
younger or older than $\log(t)$ = 7 in their sample. We obtained
that for $\log(t)$ $\le$ 7  the fraction of Be stars is
$\frac{Be}{all}$ = 4.8\%, whereas for $\log(t)$ $>$ 7,
$\frac{Be}{all}$ = 22\%.Thus, the
high proportion of Be stars in the clusters with $\log(t)$ $>$ 7
and their high rotational velocities may explain the discrepancy
between the observed and predicted \vsini~for a 12 M$_{\odot}$
star of the same clusters. The high proportion of Be stars found
in clusters with $\log(t)$ $>$ 7 is also in agreement  with
results by Fabregat \& Torrej\'on (2000), who found that Be stars
reached the  maximun abundance between 13 and 25 Myr.

Another explanation is related to the mass function of the sample stars of GH04. They
used the prediction in rotational velocity across the main sequence for a 12
M$_{\odot}$. However, the behaviour in \vsini~during evolution is not identical for a 12
M$_{\odot}$ and a 7 M$_{\odot}$ star.  In the latter case, following Meynet \& Maeder
(2000), the mass loss is much weaker than for massive stars (M $\ge$ 10 M$_{\odot}$) and
the evolutionary track presents an increase of velocity instead of a decrease. This may
explain why B-type stars are more likely to increase their rotational velocity during
their MS life than the more massive O-type stars.

\subsection{Results for the LMC and comparison with the MW}

In the following sub-sections we compare the obtained apparent fundamental parameters
to previous studies, which generally did not take fast rotation effects into account in
the determination of fundamental parameters.

For all stars in the LMC stellar sample studied in this paper, we find the following mean
\vsini~(in \kms) for B and Be stars in the field and clusters, where the given errors are the
mean errors and the number in brackets is the number of stars used in the average:

\begin{tabular}{ll}
Be$_{field+clusters}$:  & \vsini~= 272 $\pm$20 (47),\\
B$_{field+clusters}$:   & \vsini~= 119 $\pm$20 (106)\\
Be$_{field}$:           & \vsini~= 268 $\pm$20 (27),\\
B$_{field}$:            & \vsini~= 117 $\pm$20 (93),\\
(B+Be)$_{clusters}$:    & \vsini~= 223 $\pm$20 (33),\\
Be$_{clusters}$:        & \vsini~= 278 $\pm$20 (20),\\
B$_{clusters}$:         & \vsini~= 140 $\pm$20 (13).
\end{tabular}

\begin{table*}
\caption{Comparison of mean rotational velocities for B and Be stars with
spectral types B1-B3 and luminosity classes from V to III in the LMC and the
MW. The values in brackets represent the number of stars in the samples.}
\centering
\begin{tabular}{lllll}
\hline
\hline
                  &  Field B stars          &  Field Be stars         & Clusters B stars     & Clusters Be stars \\
 \hline
LMC this study & 121 $\pm$ 10 (81) & 268 $\pm$ 30 (26)  &144 $\pm$ 20 (10) & 266$ \pm$ 30  (19)\\
LMC Keller (2004) & 112 $\pm$ 50 (51) &                              & 146 $\pm$ 50 (49) &                     \\
\hline
MW Glebocki et al. (2000) & 124 $\pm$ 10 (449)&   204 $\pm$ 20 (48)   &                             &                          \\
MW Levato et al. (2004) & 108 $\pm$ 10 (150) &                                &                            &                          \\
MW  Yudin (2001) &                                 &  207 $\pm$ 30 (254)  &                           &                          \\
MW Chauville et al. (2001) &                 &  231 $\pm$ 20 (56)  &                     &                                 \\
\hline
MW WEBDA $\log(t)$ $<$ 7 &                    &                                     & 127 $\pm$ 20 (44)  & 199 $\pm$ 20 (8)     \\
MW WEBDA $\log(t)$ $>$ 7 &                     &                                     & 149 $\pm$ 20 (59)  &   208 $\pm$ 20 (45)   \\
\hline
\end{tabular}
\label{Vsinifield}
\end{table*}

\begin{table*}
\footnotesize{
\caption{Results of the Student's t-test comparisons of the mean rotational velocities for B and
Be stars with spectral types B1-B3 and luminosity classes from V to III in the LMC and the MW. The first
column gives the compared studies, the 2nd column gives the result of the test, the 3rd column gives the $\alpha$
coefficient and the 4th column gives the probability of differences `P'. The last column gives a comment about the
result according to the convention. The comment ``limit'' means that the number of stars in samples is low
and the value of mean \vsini~may be affected by the distribution of the inclination angles.}
\centering
\begin{tabular}{lllll}
\hline
Comparison & result & $\alpha$ & P & comment\\
\hline
\hline
 & & Be & & \\
Be LMC field/ Be MW Glebocki et al. (2000) & 10.81 & 0.05 & 90-95 \% & significant difference\\
Be LMC field/ Be MW Yudin (2001) & 9.84 & 0.05 & 90-95\% & significant difference\\
Be LMC field/ Be MW Chauville et al. (2001) & 6.52 & 0.05 & 90-95\% & significant difference\\
\hline
Be LMC clusters/ Be MW WEBDA $\log(t)$ $<$ 7 & 5.58 & 0.1 & 80-90 \% & limit, slight difference\\
Be LMC clusters/ Be MW WEBDA $\log(t)$ $>$ 7 & 8.91 & 0.05 & 90-95\% & significant difference\\
\hline
Be MW WEBDA $\log(t)$ $<$ 7/ Be MW WEBDA $\log(t)$ $>$ 7 & 1.15 & 0.3 & 50-70\% & limit, no significant difference\\
\hline
\hline
 & & B & & \\
B LMC field/ B LMC field Keller (2004) & 1.56 & 0.3 & 50-70\% & no significant difference\\
B LMC field/ B MW Glebocki et al. (2000) & 2.48 & 0.2 & 70-80\% & no significant difference\\
B LMC field/B MW Levato et al. (2004) & 9.4 & 0.05 & 90-95\% & significant difference\\
\hline
B LMC clusters/ B LMC clusters Keller (2004) & 0.12 & 0.9 & $<$10\% & limit, no difference\\
B LMC clusters/ B MW WEBDA $\log(t)$ $<$ 7 & 2.38 & 0.2 & 70-80\% & limit, no significant difference\\
B LMC clusters/B MW WEBDA $\log(t)$ $>$ 7 & 0.72 & 0.5 & 10-50 \% & limit, no difference\\
\hline
B MW WEBDA $\log(t)$ $<$ 7/B MW WEBDA $\log(t)$ $>$ 7 & 5.47 & 0.1 & 80-90 \% & slight difference\\
\hline
\hline
 & & Fields / clusters & & \\
Be LMC field/ Be LMC clusters & 0.22 & 0.5 & 10-50 \% & no difference\\
B LMC field/ B LMC clusters & 5.88 & 0.1 & 80-90 \% & limit, slight difference\\
\hline
Be MW Yudin (2001)/ Be MW WEBDA $\log(t)$ $<$ 7 & 0.70 & 0.5 & 10-50\% & no difference\\
Be MW Yudin (2001)/ Be MW WEBDA $\log(t)$ $>$ 7 & 0.21 & 0.5 & 10-50\% & no difference\\
B MW Glebocki et al. (2000)/ B MW WEBDA $\log(t)$ $<$ 7 & 1.68 & 0.3 & 50-70 \%  & no difference\\
B MW Glebocki et al. (2000)/ B MW WEBDA $\log(t)$ $>$ 7 & 15.50 & 0.02 & 95-98 \% & significant difference\\
\hline
\hline
\end{tabular}
\label{testStudent1}
}
\end{table*}

These mean \vsini~values cannot be compared directly with values in the MW, because they
are affected by ages and evolution, mass function of samples, etc. We must therefore
select B and Be stars in the same range of spectral types  and luminosity classes or of
masses (when they are known) and ages for samples in the LMC and in the MW.
Then, to investigate the effect of metallicity and age on the rotational velocity, we
first compare the mean \vsini~of the B and Be stars in the LMC to the ones in the MW.

    We calculate rotational velocities evolutionary tracks for different 
initial velocities and for
a 7 M$_{\odot}$ star, which corresponds to the maximum of the
mass function for the B-type stars sample. 
We have obtained these curves by interpolation thanks to the Figure 5
published in Meynet  \& Maeder (2002) for the tracks of a 7 M$_{\odot}$ with an 
initial velocity at the ZAMS=300 \kms. And we have used the Figure 12 published
in Meynet \& Maeder (2000), and more particularly the Figure 5 from Maeder \&
Meynet (2001) in order to obtain tracks  for different  initial rotational
velocities (V0=100, 200, 300, 400, 500 \kms). The use of tracks with a
metallicity Z=0.00001 for a 7 M$_{\odot}$ is justified because the tracks
for a star with 9 M$_{\odot}$ at metallicity Z=0.004 or Z=0.00001 are quasi identical,
then we expect that the tracks for a 7 M$_{\odot}$ are identical at Z=0.00001
and at Z=0.004.

Let us note that due to fast internal angular momentum
redistribution in the first $\simeq$ 10$^{4}$ years in the ZAMS,
the surface rotational velocities decrease 0.8 times their
initial value. Then, for the comparison sake with our
observational data, the values plotted are not $V$ but are
average \vsini$=(\pi/4)V$ (see eq.~\ref{eqVevsini}). For example,
for an initial rotational velocity equals to 300 \kms, the angular
momentum redistribution leads roughly to V$_{ZAMS}=$ 240 \kms, which
corresponds to \vsini$=(\pi/4) \times 240$ $\simeq$ 190 \kms. The
curves are only slightly affected by mass loss effects. Due to
the low metallicity LMC environment, the mass-loss dependent
effects are even less noticeable.

In the present work, the purpose of these curves is only to give a 
rough interpretation of the behavior observed of $<$\vsini$>$.

We determine the ages of stars of the field and of
several clusters or associations in our observations. For this purpose, we use HR
evolutionary tracks (for non-rotating stars) for the stars of the sample  unaffected by
rapid rotation and for Be stars corrected for the effects of fast rotation with
\omc~=~85\%. For the cluster NGC\,2004, we obtain $\log(t)$ = 7.40, which is close to
the value obtained in previous studies: $\log(t)$ = 7.30 (Keller 1999) and 7.40 (Maeder
et al. 1999). This comparison validates our method to determine ages for clusters.

\subsubsection{Field B and Be stars}
\label{BBefield}

\begin{figure*}[]
\centering
\resizebox{\hsize}{!}{\includegraphics[angle=-90]{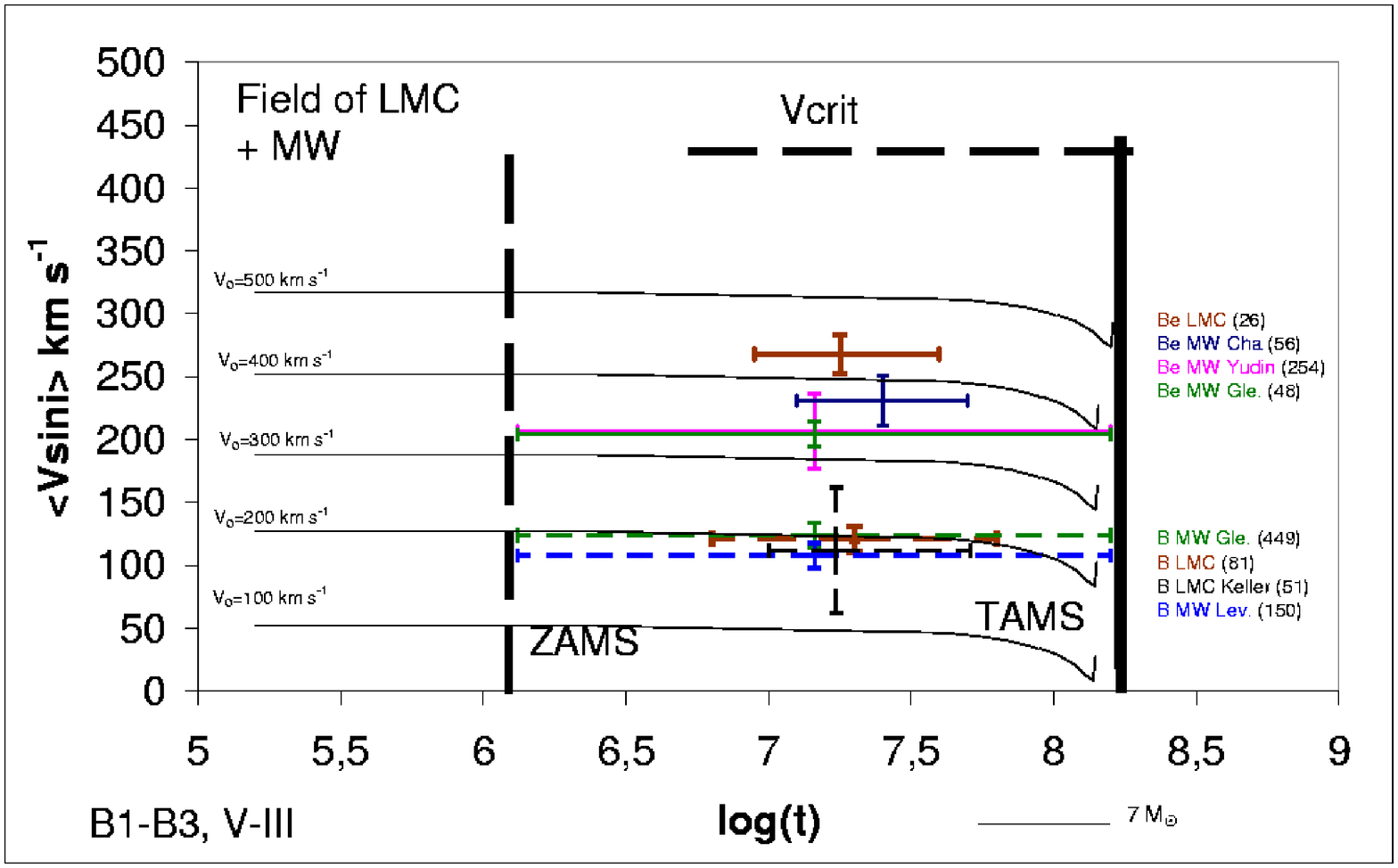}}\\
\resizebox{\hsize}{!}{\includegraphics[angle=-90]{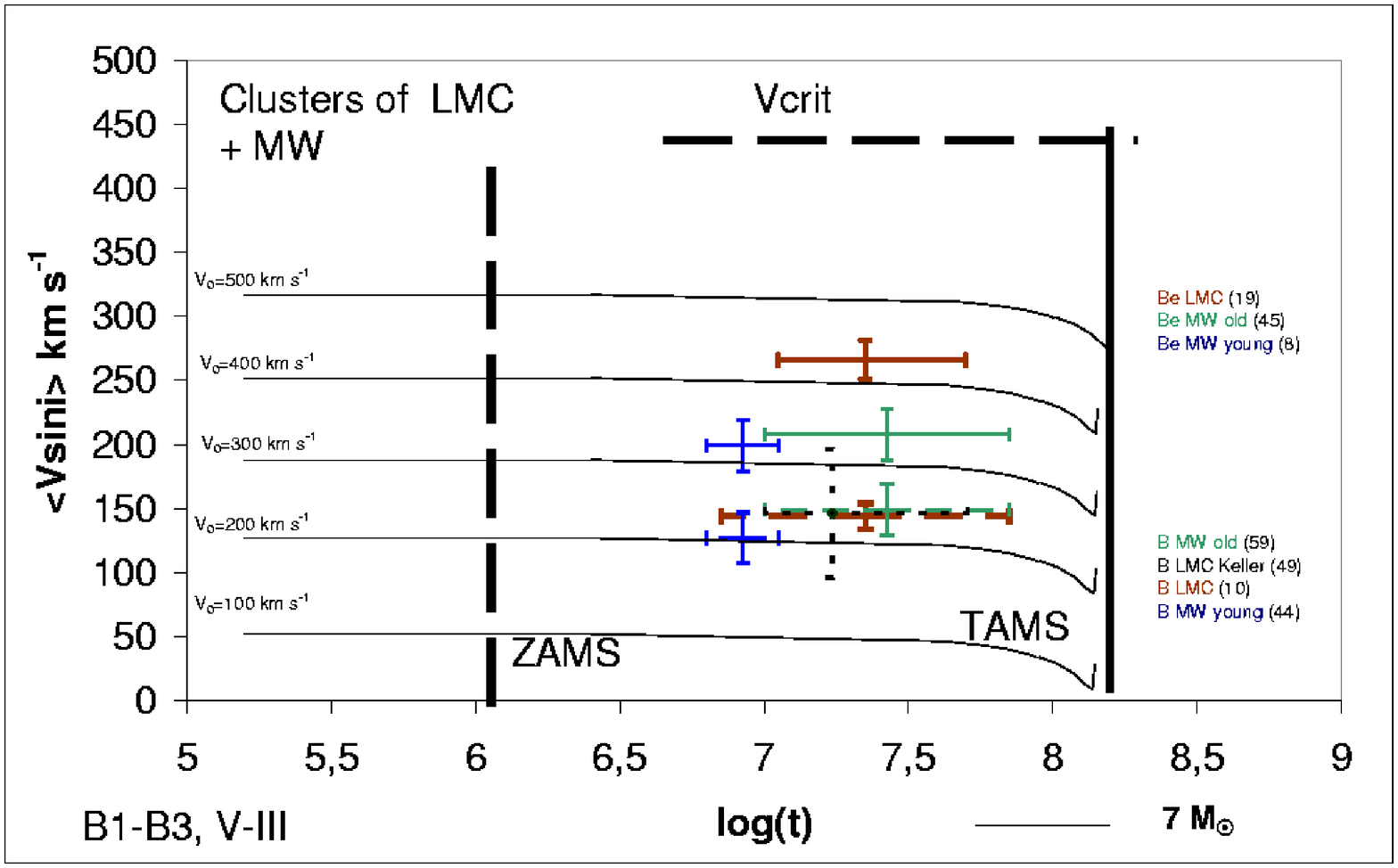}}

\caption{Top: Comparison of mean \vsini~in the LMC and the MW for field B and Be
stars. Evolutionary tracks of rotational velocity during the Main Sequence life
are given for different initial velocities for a 7 M$_{\odot}$ star. The ZAMS
and TAMS are indicated by vertical lines and the critical \vsini~by a
horizontal dotted line. The number of stars for each study is given in
brackets. The dispersion in ages corresponds to the range of individual stellar ages
in the samples, when these ages are known, or to the Main Sequence lifetime.
The considered studies are: for the LMC, this paper and Keller (2004); for the
MW, Cha = Chauville et al. (2001), Yudin (2001), Gle = Glebocki \& Stawikowski
(2000), and Lev = Levato \& Grosso (2004).
Bottom: same figure but for clusters. The considered
studies are: for the LMC, this paper and Keller (2004), and for the clusters in
the MW, the WEBDA database.}
\label{agevsinifield}
\end{figure*}

The mean \vsini~obtained for B stars in the field of the LMC
closely agrees with Keller's (2004) results in the same age range
(see Table~\ref{Vsinifield} and Fig.~\ref{agevsinifield}, upper panel). To
compare the rotational velocity of B stars in the LMC with the
MW, we use the studies of Levato \& Grosso (2004) and Glebocki \&
Stawikowski (2000). In these studies, we select stars
with spectral types ranging from B1 to B3 and luminosity classes
from V to III, because ages and masses were not
determined. To compare the rotational velocity of Be stars in the
LMC with the MW, we use Chauville et al. (2001), Glebocki \&
Stawikowski (2000) and Yudin (2001) with  the same selection
criteria as for B stars. The comparison of \vsini~in the LMC and
the MW for B and Be stars in the field is presented in
Table~\ref{Vsinifield} and Fig.~\ref{agevsinifield}, upper panel. 
The range of
stellar ages is reported as the dispersion in age in the figure.
For samples with an unknown age, we adopt as error bar the
duration of the Main Sequence for a 7 M$_{\odot}$ star, which
overestimates the age uncertainty. The curves in
Fig.~\ref{agevsinifield}, upper panel, show the evolutionary tracks of
rotational velocity during the Main Sequence for different
initial velocity for a 7 M$_{\odot}$ star, which corresponds to
the maximum of the mass function of the B-stars sample.

In order to know if the samples contain a sufficient number of
elements for the statistic to be relevant and give an average
\vsini~not biased by inclination effects, we calculate
averages for samples with different number of elements. The
deviation of the averages gives the statistical error. If this
statistical error is smaller than the error on the data, the
value determined for the data is statistically significant and
does not represent an effect of inclination. Thanks to this
test, we find that our samples in the LMC and in the
MW are statistically significant even in the cases when the
number of stars does not exceed 10. For example, in
Table~\ref{Vsinifield}, we have found no difference between 
the cluster B stars (\vsini=144 \kms, 10 stars) following our
results  and according to Keller (2004) (\vsini=146 \kms, 49
stars).

We complete the statistical studies by a Student's t--test (see
Table~\ref{testStudent1}) in order to know  whether the
differences observed in the samples are significant. The Student test gives:\\
(i) for field B stars: There is no significant difference
between the LMC studies carried-on in this paper and Keller's
(2004). The test further gives no significant difference between
LMC and MW studies, when the data used for the MW are those of
Glebocki \& Stawikowski (2000). There are, however, significant
differences when the comparison is based on data taken from
Levato \& Grosso (2004). It is therefore difficult to conclude whether B
stars have similar
rotational velocities in the LMC and the MW fields.\\
(ii) for field Be stars: A significant difference between studies in the LMC (this
paper) and the MW.  Field Be stars in the LMC have a rotational velocity higher than in
the MW.

\subsubsection{B and Be stars in clusters}
\label{BBecl}

As for B stars in the field, we find that the mean \vsini~of B
stars in the LMC clusters ($\log(t)$ $>$ 7) determined from our
observations closely agrees with Keller's (2004) results (see
Table~\ref{Vsinifield} and Fig.~\ref{agevsinifield}, lower panel). In the
MW, we selected young clusters with $\log(t)$ $<$ 8 (some of them
were observed by GH04) in the WEBDA database for which published
\vsini~and MK classification exist. We distinguish two groups:
the younger clusters with $\log(t)$ $<$ 7  and older clusters
with $\log(t)$ $>$ 7. The younger clusters (given by increasing
age) are: IC\,1805, Trumpler\,14, IC\,2944, NGC\,6193, NGC\,2362,
NGC\,2244, NGC\,6611, NGC\,2384, NGC\,3293, and NGC\,1502. The
older clusters (given by increasing age) are: NGC\,869, NGC\,884,
NGC\,4755, IC\,2395, NGC\,7160,  and NGC\,2422. Their difference
in age gives the age--dispersion reported in
Fig~\ref{agevsinifield}, lower panel. The ages are those given by GH04 and
in WEBDA.

The results concerning B stars in the LMC clusters and MW
Be stars with $\log(t)$ $<$ 7 must be taken cautiously.
In fact, the samples are not numerous enough to have the inclination angle effects, 
in \vsini~averages, entirely removed so as to reflect the
<V>-dependent information properly.
For the stars in these clusters, the Student's t--test (see Table~\ref{testStudent1}) gives:\\
(i) for B stars: A slight difference between younger ($\log(t)$ $<$ 7) and older
($\log(t)$ $>$ 7) clusters in the MW, which may be explained by the effect of evolution
on rotational velocity, but no significant difference between the LMC and the MW
clusters with $\log(t)$ $>$ 7. B stars in the LMC and MW clusters seem to have a similar
rotational velocity when intervals of similar ages are compared. \\
(ii) for Be stars: A significant difference between the LMC and the MW clusters. Be
stars in the LMC clusters have a rotational velocity higher than in the MW clusters.
However, the number of Be stars observed in LMC clusters, as well as the number of Be
stars identified in young MW clusters, is poor and may affect the statistics.

Fig.~\ref{agevsinifield}, lower panel, illustrates the comparison of the mean \vsini~in the LMC and
the MW for B and Be stars in clusters. As in Fig~\ref{agevsinifield}, upper panel, the curves show
the evolutionary tracks of rotational velocity during the Main Sequence for different
initial velocity for a 7 M$_{\odot}$ star.

\subsubsection{Comparison between field and clusters}

Results on rotational velocity of B and Be stars in the field and clusters
presented in Sect.~\ref{BBefield} and ~\ref{BBecl} (see also
Table~\ref{Vsinifield} and Table~\ref{testStudent1}) lead to the following conclusions:\\
(i) According to Meynet \& Maeder (2000), the lower the mass-loss in massive
stars is, the lower the metallicity is. We can the expect that stars
in low-metallicity regions may lose less angular momentum and then
better preserve the initial high rotational rates.
%
Such a metallicity effect seems to be present in Be stars, as we see that they rotate faster
in the field and clusters of the LMC than in the MW. However, we were not able to detect such
an effect in B stars.\\
(ii) Be stars rotate more rapidly than B stars in the field as
well as in clusters, in the LMC, and the MW. Be stars would begin
their life on the MS with an initial rotational velocity higher
than the one of B stars. The lower the metallicity environment
is, the higher the initial rotational velocity of Be stars would
be. Moreover, we note that these objects would require an initial
rotational velocity of at least $\sim$250 \kms.\\
(iii) No significant differences can be found between the
rotational
velocity of field and cluster Be stars, neither in the LMC nor in the MW.\\
(iv) No significant differences can be found between
rotational velocities of young field and cluster B stars in the
LMC and in the MW. However, there is a significant difference
between the rotational velocity of older field and cluster B
stars in the MW. This fact can be explained in terms of evolution
of rotational velocities.

\subsubsection{Be stars: mass and rotation}
\label{Bemasses}

\begin{figure*}[]
\centering
\resizebox{\hsize}{!}{\includegraphics[angle=-90]{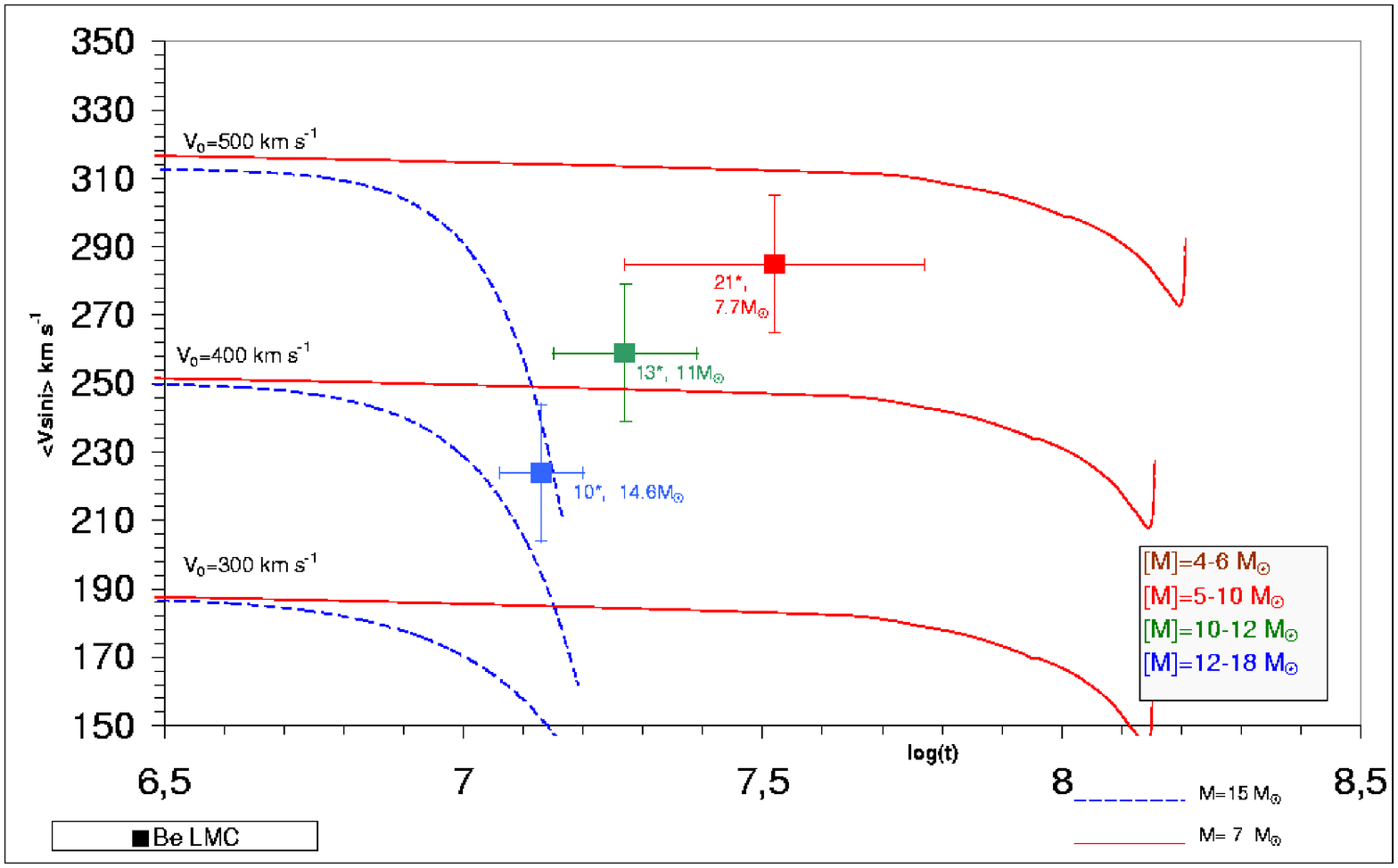}}
\caption{Comparison of mean \vsini~for Be stars in the study for three
different intervals of masses: $ 5 M_{\odot} < M < 10 M_{\odot}$, $10 M_{\odot} < M < 12
M_{\odot}$ and $ 12 M_{\odot} < M < 18 M_{\odot}$. The numbers given, 
towards the crosses, are the number of stars and the mean mass.
The curves show the evolutionary tracks of
rotational velocity during the Main Sequence for different initial velocity for
stars with 7 M$_{\odot}$ (solid curve) and 15 M$_{\odot}$ (dotted curve). The
dispersion in ages corresponds to the range of individual stellar ages in the
sub-samples.}
\label{MBecomp}
\end{figure*}

The number of Be stars is too low to make a statistical study by mass range on
\vsini~in clusters and in the field of the LMC separately. As mentioned in the
previous section, we did not find significant differences in mean
\vsini~values for Be stars between clusters and the field. Therefore we compare
Be stars only by intervals of mass, regardless of their location.\\
(i) for $5 < M < 10 M_{\odot}$, the sub-sample contains 21 stars. The determined mean parameters are:
$<M>$ = 7.7 $M_{\odot}$, $<R>$ = 5.8 $R_{\odot}$, and  $<$\vsini$>$ = 285 \kms, which correspond
to a mean ratio \omc $\simeq$ 85\% $\pm$ 9 \%\\
(ii) for $10 < M < 12 M_{\odot}$, the sub-sample contains 13 stars. The determined mean parameters are:
$<M>$ = 11.0 $M_{\odot}$, $<R>$ = 9.3 $R_{\odot}$, and $<$\vsini$>$ = 259 \kms, which correspond
to a mean ratio \omc $\simeq$ 83\% $\pm$ 9 \%;\\
(iii) for $18 > M > 12 M_{\odot}$, the sub-sample contains 10 stars. The determined mean parameters are:
$<M>$ = 14.6 $M_{\odot}$ close to 15 $M_{\odot}$, $<R>$ = 12.7 $R_{\odot}$, and $<$\vsini$>$ = 224\kms,
which correspond to a mean ratio \omc $\simeq$ 73\% $\pm$ 9 \%.

These results are plotted in Fig.~\ref{MBecomp} and compared to the theoretical
evolutionary tracks for a 7 and 15 M$_{\odot}$ star for different values of
initial rotational velocity (Maeder \& Meynet 2001). These curves can be
considered as envelopes of evolutionary tracks of our sample of Be stars in the
LMC. This figure shows that stars follow the theoretical rotational velocity
evolution in a low metallicity environment: the mean \vsini~decreases as the
mass increases. This trend can be explained by a difference in mass loss
between massive and less  massive stars.

\subsubsection{Star formation conditions and magnetic field}

According to Stepi\'en (2002), the sufficient condition for a
star to rotate rapidly on the ZAMS is the presence  of a weak or
moderate stellar magnetic field and the existence of an accretion
disk for at least 10\% of its pre-Main Sequence (PMS) phase. The
magnetic field and its interactions with the disk or wind and
other phenomena such as accretion have an impact on
rotational velocity during the PMS and affect the initial
rotational velocity (spin down for strong values of magnetic
field) in the Main Sequence (MS). For stars in the MW, the
progenitors of all Be stars would possess a fossil magnetic field
with a surface intensity between 40 and 400 G and, due to the
short PMS phase for the early types, they would conserve their
strong rotational velocity during the MS. On the opposite, stars
with a magnetic field stronger than 400 G would become slowly
rotating magnetic B stars. In the LMC and other environments of
low metallicity, the magnetic field has less braking impact on
the velocity  as explained by Penny et al. (2004) due to
the lower abundances of metals. It may explain why Be
stars in the LMC can rotate initially with higher velocities than
in the MW, as shown in Fig.~\ref{agevsinifield}, upper panel. 
Note that a weak magnetic field is
suspected in the classical Be star \object{$\omega$ Ori} (Neiner
et al. 2003).

\begin{figure*}[]
\centering
\resizebox{\hsize}{!}{\includegraphics[angle=-90]{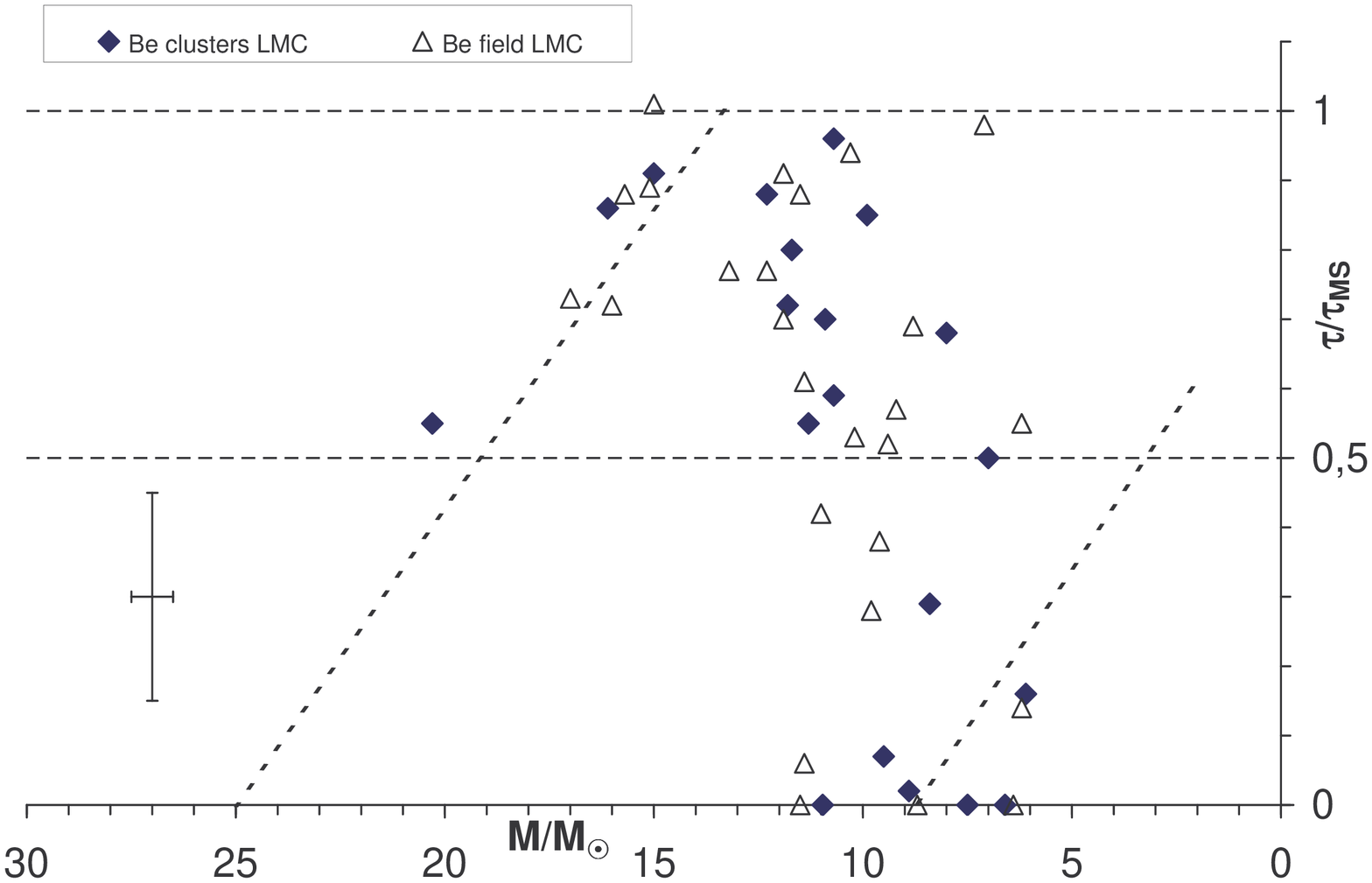}}
\caption{Evolutionary status of Be stars in the sample. The parameters of Be
stars are corrected with FASTROT with \omc~= 85\% and ages are determined in HR
diagrams with $V_{0}$ = 300 \kms. The typical errors are shown in the lower
left corner. The dotted diagonals, which show the area of existing Be stars in
the Milky Way, come from Zorec et al. (2005).}
\label{evolBeLMCveil}
\end{figure*}

\subsection{Evolutionary status of Be stars in the LMC}
\label{evol}

Using the same approach as the one described by Zorec et al.
(2005), we studied the evolutionary status of the LMC Be stars in
our sample. We used evolutionary tracks with an initial velocity
V$_0$ = 300 \kms~provided by Maeder \& Meynet (2000). These
evolutionary tracks for rotating stars are only available for
stars in the MW. They show a slight shift towards lower
temperatures and an extension of the time a star may spend on the
MS ($\tau_{MS}$), compared to evolutionary tracks of
non-rotating stars. Therefore, a star placed in a HR diagram for
non-rotating stars has a different age ($\tau$) and a different
evolutionary status \ttms than if it would be placed in an HR
diagram for rotating stars. Fig.~\ref{evolBeLMCveil} shows the
evolutionary status of Be stars in the sample. It appears that
more massive Be stars in our sample in the LMC seem to be
evolved, since they  are localized mainly in the second part of
the MS.  Contrary to previous similar studies (Zorec et
al. 2005, Fr\'emat et al. 2005), in our Be star sample, 
massive stars  (M$\ga$10M$_{\odot}$) by the end of the MS
evolutionary phase represent a  high fraction of the total
number of the studied stars (60\%). The  distribution obtained
cannot correspond only to differences in the  mass-dependent
evolutionary sampling, but it could reflect some star formation
history in the region: stars with M$\ga$10M$_{\odot}$ in
$\tau/\tau_{\rm  MS}\ga0.5$ have an average age
$<\!\!\tau\!\!>\sim(1.5\pm0.4)\times10^7$ yr,  stars with
M$\la$10M$_{\odot}$ in $\tau/\tau_{\rm MS}\ga0.5$ have $<\!\!\tau
\!\!>\sim(2.9\pm0.2)\times10^8$ yr, while those with
M$\la$10M$_{\odot}$ and  $\tau/\tau_{\rm MS}\la0.5$ have
$<\!\!\tau\!\!>\sim(0.7\pm0.6)\times10^7$ yr.

The observed trend is only indicative, because evolutionary tracks for
massive stars are mass loss dependent and the initial rotational velocities of
Be stars are higher than 300 \kms. Nevertheless, according to Zorec et al.
(2005), changes of initial velocities from  V$_0$ = 300 \kms~to higher values
do not seem to strongly affect evolutionary tracks.

However, several Be stars with lower mass seem to be close to the
ZAMS, which is inconsistent with the assumption that the Be star
phenomenon occurs preferentially in the second half of the MS
life. Those objects, for which spectra were mostly
obtained with low S/N ratio, need to be reobserved to clearly
confirm their fundamental parameters.

\section{Conclusions}

With the VLT-GIRAFFE spectrograph, we obtained spectra of a large sample of B
and Be stars in the LMC-NGC\,2004 and in its surrounding field. We determined
fundamental parameters for B stars in the sample, and apparent and parent
non-rotating counterpart (pnrc) fundamental parameters for fast rotators such
as Be stars.

From the \vsini~ study for B and Be stars in the LMC and its comparison with
the MW, we conclude that Be stars begin their life on the MS with a stronger
initial velocity than B stars. Moreover, this initial velocity is sensitive to
the metallicity. Consequently, only a fraction of B stars can become Be stars.
This result may explain the differences in the proportion of Be stars in
clusters with similar age.

Our results support Stepien's scenario (2002): massive stars with a weak or
moderate magnetic field and with an accretion disk during at least 10\% of
their PMS lifetime would reach the ZAMS with sufficiently high initial
rotational velocity to become Be stars.

We find no clear influence of the metallicity on rotational velocity in B-type
stars.  The low metallicity may favour the PMS evolution of high velocity stars
by minimizing the braking due to magnetic field interactions with the disk, but
the influence of metallicity during the life of B-type stars in the MS is not
preponderant. As Be stars are not critical rotators, an additional process,
such as magnetic field by transfering momentum to the surface or non-radial
pulsations (see Rivinius et al. 1998), must provide additional angular momentum
to eject material from the star.

The effects of metallicity, the star formation conditions and the evolutionary
status of B and Be stars discussed in this paper will be investigated in a
forthcoming paper in the Small Magellanic Cloud, which has a lower metallicity
than the Large Magellanic Cloud, in order to enlarge the results presented
here.

\begin{acknowledgements}
We would like to thank Dr H. Flores for performing the observing run in
November 2003 with success and good quality. We thank Drs M.R. Cioni and J.
Smoker for their help during the observing run in April 2004.
We also thank the referee Dr S. J. Smartt for his constructive remarks.
This research has made use of the Simbad database and Vizier
database maintained at CDS, Strasbourg, France, as well as of the WEBDA
database.
\end{acknowledgements}

\end{document}